\documentclass{llncs}
\usepackage{graphicx}
\usepackage{amsmath}
\usepackage{float}
\usepackage[caption=false]{subfig}
\usepackage{makecell}
\usepackage{comment}
\usepackage[title]{appendix}
\usepackage[linesnumbered, ruled]{algorithm2e}

\SetKwRepeat{Do}{do}{while}
\SetKwBlock{Repeat}{repeat}{end}
\begin{document}
\pagenumbering{gobble}
\title{Shape-Faithful Graph Drawings\thanks{This work is supported by ARC grant DP190103301.}}
%
%
\author{Amyra Meidiana \and
Seok-Hee Hong \and
Peter Eades}

\authorrunning{A. Meidiana et al.}

\institute{University of Sydney, Australia \\
\email{amei2916@uni.sydney.edu.au, \{seokhee.hong, peter.eades\}@sydney.edu.au}}
\maketitle              

\begin{abstract}
\textit{Shape-based metrics} measure how faithfully a drawing $D$ represents the structure of a graph $G$, using the \textit{proximity graph} $S$ of $D$. 
While some limited graph classes admit proximity drawings (i.e., optimally shape-faithful drawings, where $S = G$), algorithms for shape-faithful drawings of general graphs have not been investigated.

In this paper, we present the first study for shape-faithful drawings of {\em general graphs}.
First, we conduct extensive comparison experiments for popular graph layouts using the shape-based metrics, and examine the properties of highly shape-faithful drawings.
Then, we present $ShFR$ and $ShSM$, algorithms for shape-faithful drawings based on force-directed and stress-based algorithms, by introducing new  {\em proximity forces/stress}. 
Experiments show that $ShFR$ and $ShSM$ obtain significant improvement over $FR$ (Fruchterman-Reingold) and $SM$ (Stress Majorization), on average 12\% and 35\% respectively, on shape-based metrics.

\end{abstract}                                                                                                                 

\section{Introduction}

Recently, \textit{shape-based metrics}~\cite{eades2017shape} have been introduced for evaluating the quality of large graph drawing. It measures how faithfully the ``shape'' of a drawing $D$ represents the ground truth structure of a graph $G$, by comparing the similarity between the \textit{proximity graph} $S$ of the vertex point set of $D$ and the graph $G$.

For a point set $P$ in the plane, proximity graphs are defined as: two points are connected by an edge if they are ``close enough''. 
Specifically, a \textit{proximity region} is defined for each pair of points, and if the proximity region is empty, the points are connected by an edge in the proximity graph~\cite{preparata2012computational}.

Some limited graph classes always admit a {\em proximity drawing} $D$, where the graph $G$ is realized as a proximity graph $S$ in $D$.
For such proximity drawable graph classes,
some characterizations are known, and algorithms to construct such proximity drawings are available~\cite{battista1994proximity,bose1996characterizing}. 
Consequently, such proximity drawings are optimally shape-faithful (i.e, shape-based metric of 1), since $S=G$.

However, such optimally shape-faithful drawings are only applicable for very limited graph classes. Algorithms to optimize shape-based metrics for {\em general graphs} (i.e., not proximity drawable graphs) have not been studied yet.

In this paper, we present the first study for shape-faithful drawings of general graphs. 
Specifically, our main contributions can be summarized as follows:

\begin{enumerate}
\item We evaluate the shape-faithfulness of popular graph drawing algorithms for various proximity drawable graph classes, including \textit{strong} proximity drawable graphs (i.e., the best possible shape-based metric is 1),  \textit{almost} proximity drawable graphs with some forbidden subgraphs, {\em weak} proximity drawable graphs, and mesh graphs.

Experiments show that $tsNET$~\cite{kruiger2017graph} obtains the highest shape-faithfulness on most large graph instances, for strong and almost proximity drawable graphs, and stress-based layouts~\cite{gansner2004graph} achieve good results on mesh graphs.

\item We present $ShFR$ and $ShSM$, algorithms for shape-faithful drawings for general graphs, based on the force-directed and stress-based layouts, by introducing new  {\em proximity forces/stress}. 

Experiments with strong proximity drawable graphs, scale-free graphs and benchmark graphs show that  $ShFR$ and $ShSM$ obtain significant improvement (on average, 12\% and 35\%) on the shape-based metrics over $FR$ (Fruchterman-Reingold)~\cite{fruchterman1991graph} and $SM$ (Stress Majorization)~\cite{gansner2004graph}.

\end{enumerate}

\section{Related Work}
\label{sec:litreview}

\subsection{Shape-Based Metrics}

Shape-based metrics measure how faithfully the ``shape'' of a drawing $D$ represents the ground truth structure of a graph $G$, by comparing the similarity between the \textit{proximity graph} $S$ of the vertex point set of $D$ and the graph $G$~\cite{eades2017shape}. 

Specifically, the shape-based metrics use proximity graphs such as the Gabriel Graph ($GG$) and Relative Neighborhood Graph ($RNG$) (defined in  Section \ref{sec:litreview_proxgraph}). 
To compute the similarity between $G$ and $S$, both with vertex set $V$, the shape-based metrics use the \textit{Jaccard Similarity (JS)}~\cite{jaccard1912distribution} as follows:  
$JS (G, S)= \frac{1}{|V|}\sum_{v \in V} \frac{N_G(v) \cap N_S(v)}{N_G(v) \cup N_S(v)}$, 
where $N_G(v)$ (resp., $N_S(v)$) is the set of neighbors of vertex $v$ in $G$ (resp., $S$). 
We denote the shape-based metrics computed with this formula using $RNG$ (resp., $GG$) as $Q_{RNG}$ (resp., $Q_{GG}$), having values between 0 and 1 where 1 means perfectly shape-faithful.


\subsection{Proximity Graphs}
\label{sec:litreview_proxgraph}

For a point set $P$ in the plane,
a proximity graph $S$ of $P$ is roughly defined as follows: two points are connected by an edge if and only if they are ``close enough''.
Namely, the \textit{proximity region} defined for the two points should be empty (i.e., contains no other points)~\cite{preparata2012computational,toth2017handbook}.
For example,  \textit{Gabriel Graph (GG)}~\cite{gabriel1969new} (resp., \textit{Relative Neighborhood Graph (RNG)}~\cite{toussaint1980relative}) is a proximity graph where two points $x$ and $y$ are connected by an edge if and only if the closed disk (resp., open lens) having line segment $xy$ as its diameter contains no other points.

For {\em strong proximity}, two conditions must be fulfilled: (a) two points are connected by an edge only if their proximity region is empty, and (b) two points are not connected by an edge only if their proximity region is not empty~\cite{battista1995strength}.

A relaxation of condition (b) gives rise to the definition of {\em weak proximity}, where the proximity graph may omit an edge between points $x$ and $y$ even if their proximity region is empty~\cite{battista1995strength}. 
Namely, while points need to be ``close enough'' to be connected by an edge in the proximity graph $S$, points can be made to be not connected by an edge in $S$ even if they are ``close enough''.

\subsection{Proximity Graph Drawing}

Characterizations of {\em strong proximity drawable graphs} (i.e., graphs that admit a proximity drawing  $D$, where the graph $G$ is realized as a proximity graph $S=G$ in $D$) are known for $RNG$ and $GG$~\cite{bose1996characterizing,lenhart1996proximity}:

\begin{itemize}
    \item $RNG$-drawable graphs: trees with maximum degree 5, maximal outerplanar graphs, biconnected outerplanar graphs
    \item $GG$-drawable graphs: trees with maximum degree 4 and no degree 4 vertex with all ``wide'' subtrees, maximal and biconnected outerplanar graphs
\end{itemize}

Moreover, forbidden subgraphs have also been characterized: no $GG$- and $RNG$-drawable graphs may contain $K_4$ and $K_{2,3}$ as subgraphs~\cite{gabriel1969new}.

Characterizations of {\em weak  proximity drawable graphs} include 
wider classes: 

\begin{itemize}
 \item trees (regardless of maximum degree): weak $GG$- and $RNG$-drawable~\cite{battista1995strength}

\item 1-connected outerplanar graphs with no vertex of degree 1:  weak $GG$-drawable~\cite{evans2011approximate}.
 \end{itemize}

Algorithms to construct proximity drawings of both strong and weak proximity drawable graphs are 
available~\cite{battista1995strength,bose1996characterizing,lenhart1996proximity}, although implementations are unavailable and challenging due to requiring precise geometric computations. 
For details on proximity graph drawing, see a survey~\cite{Liotta04proximitydrawings}.




\section{Graph Layout Comparison Experiments}
\label{sec:layoutcomp}

\subsection{Experiment Design and Data Sets}

In this Section, we present extensive experiments using the shape-based metrics $Q_{RNG}$ and $Q_{GG}$ to compare popular graph drawing algorithms:

\begin{itemize}
	\item Force-directed layouts: Fruchterman-Reingold ($FR$)~\cite{fruchterman1991graph}, Organic ($OR$)~\cite{wiese2004yfiles}.
	
	\item Multi-level force-directed layouts: $FM^3$~\cite{hachul2004drawing}, $sfdp$~\cite{hu2005efficient}.
	
	\item Backbone layout  ($BB$)~\cite{nocaj2015untangling}, which untangles hairballs in a drawing.
	
	\item LinLog layout ($LL$)~\cite{noack2003energy}, a  force-directed algorithm displaying clusters.
	
	\item Stress-based layouts to minimize the \textit{stress}: Stress Majorization ($SM$)~\cite{gansner2004graph},  Stochastic Gradient Descent ($SGD$)~\cite{zheng2018graph}.
	
	\item $tsNET$ layout~\cite{kruiger2017graph}, based on the t-SNE dimension reduction~\cite{Laurens:2008:tSNE}.
	
	\item Walker's level drawing algorithm ($W$) for trees~\cite{walker1990node}.
	
	\item Chrobak and Kant algorithm ($CK$)~\cite{chrobak1997convex} for convex grid drawings of triconnected planar graphs in quadratic area.
\end{itemize}

For data sets, we generate graphs with various  sizes: {\em small} graphs with 50-250 vertices, {\em medium} graphs with 250-500 vertices, and {\em large} graphs with 500-1000 vertices.
Furthermore, we consider graph types based on proximity drawability characterization: 
{\em strong} proximity drawable graphs, 
{\em almost} proximity drawable graphs, 
and {\em weak} proximity drawable graphs. 
We also use {\em mesh} graphs, which do not fall into known proximity drawability characterizations.
For each graph type and size, we generate ten graph instances.

\medskip
\noindent {\bf Strong Proximity Drawable Graphs: }
We generate strong proximity drawable graphs 
based on known characterizations~\cite{bose1996characterizing,lenhart1996proximity}:

\begin{itemize}
    \item \textit{Maximum outerplanar graphs}, generated using the connected planar graph generator of OGDF~\cite{chimani2013open}. 
    \item \textit{Biconnected outerplanar graphs}:  We start $G$ as a cycle of random length $\leq$ the target size $n$.
    Then, select an edge $(u,v)$ in $G$ that is only involved in one cycle. Select a cycle length $x < n$, create a path $p$ of length $x - 2$, and add an edge between $u$ and the first vertex of $p$, and between $v$ and the last vertex of $p$. Repeat while the number of vertices in $G$ is less than $n$.
    \item Proximity drawable \textit{trees}, generated  using the random tree generator of OGDF:  
    For $RNG$-drawable trees, we set the maximum vertex degree as 5; 
    for $GG$-drawable trees, we set the maximum vertex degree as 4, and then prune forbidden subtrees until the tree contains no more forbidden subtrees.
\end{itemize}

\noindent {\bf Almost Proximity Drawable Graphs with Forbidden Subgraphs:}
We start with a strong proximity drawable graph $G$, and then add a few edges and/or vertices  to create a forbidden subgraph. 
The number of edges (resp., vertices) added are limited to at most 10 (resp., 5). 
Specifically, we perform two types of forbidden subgraph augmentation:

\begin{itemize}
    \item \texttt{L-AUG} (\textit{Local} Augmentation) graphs: We choose a vertex $v$ of $G$ and add new vertices and edges around $v$ to create a  forbidden subgraph $F$.
    \item \texttt{F-AUG} (\textit{Global} Augmentation) graphs: We select a subset of vertices of $G$, all separated by a shortest path length above a predefined threshold, and add edges between the selected vertices to create a forbidden subgraph $F$. 
\end{itemize}

\noindent {\bf Weak Proximity Drawable Graphs:}
We also use \textit{weak} proximity drawable graphs
based on the weak proximity drawability characterization~\cite{battista1995strength}:

\begin{itemize}
    \item \textit{1-connected} outerplanar graph with a minimum degree of 2, which are weak $GG$-drawable~\cite{evans2011approximate}: 
    We generate the graphs in a similar way to the biconnected outerplanar graphs, however  alternately appending the new cycle to a random vertex rather than a random edge.
\end{itemize}

\noindent {\bf Mesh Graphs:}
We use simple \textit{mesh} graphs containing no chordless cycles of length $> 3$,  
from the $jagmesh$ set of the SuiteSparse Matrix collection~\cite{davis2011university}.
These graphs are not part of known proximity drawability characterizations,
but can be drawn as an $RNG$ drawing, by drawing each 3-cycle as an equilateral triangle. 

\subsection{Results}

\subsubsection{Strong Proximity Drawable Graphs}

On strong proximity drawable trees, all the drawing algorithms used fail to obtain shape-based metrics close to optimal. 

\begin{figure}[t]
\centering
\subfloat[Small]{
\includegraphics[width=0.28\columnwidth]{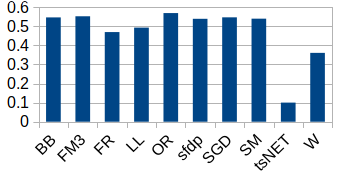}
}
\subfloat[Medium]{
\includegraphics[width=0.28\columnwidth]{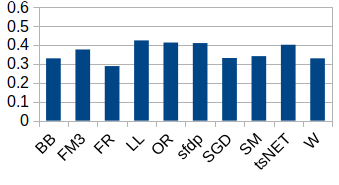}
}
\subfloat[Large]{
\includegraphics[width=0.28\columnwidth]{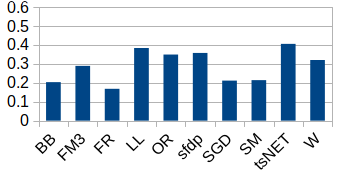}
}
\caption{Average $Q_{RNG}$ for trees. $LL$, $OR$, and $sfdp$ consistently perform well, with $tsNET$ performing much better on large trees. Even the highest-performing layouts are still far from optimal shape-faithfulness ($Q_{RNG}$ = 1).}
\label{fig:layoutcompmetrics_rngtree}
\end{figure}

Figure \ref{fig:layoutcompmetrics_rngtree} shows the average $Q_{RNG}$ for $RNG$-drawable trees.
On small trees, the best performing layouts, $OR$, $BB$, multi-level layouts, and stress-based layouts, only obtain $Q_{RNG}$ of 0.5-0.6 on average. 
$tsNET$ becomes the best performing layout on medium and large trees, with $Q_{RNG}$ of about 0.4 on average. On large trees, the differences in $Q_{RNG}$ between layouts are more pronounced, with $tsNET$ and $LL$ performing the best, followed by $sfdp$ and $OR$.

\begin{figure}[t]
\centering
\subfloat[Small]{
\includegraphics[width=0.28\columnwidth]{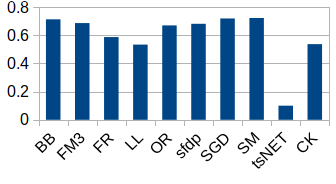}
}
\subfloat[Medium]{
\includegraphics[width=0.28\columnwidth]{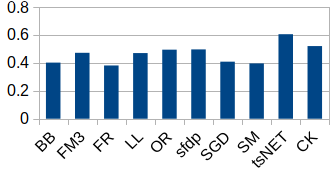}
}
\subfloat[Large]{
\includegraphics[width=0.28\columnwidth]{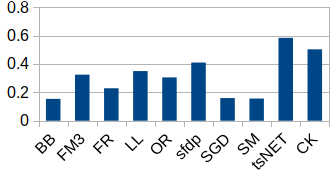}
}
\caption{Average $Q_{RNG}$ for maximum outerplanar graphs. $tsNET$ and $CK$ are the top performing layouts on medium and large graphs. For highest-performing layouts, $Q_{RNG}$ is slightly closer to optimal compared to $RNG$-drawable trees.}
\label{fig:layoutcompmetrics_maxouterplanar_rng}
\end{figure}

For small proximity drawable outerplanar graphs (both $GG$- and $RNG$- drawable), the best performing layouts, stress-based layouts and $BB$, obtain $Q_{RNG}$ of around 0.7 (see Figure \ref{fig:layoutcompmetrics_maxouterplanar_rng}). This is notably closer to optimal compared to  $RNG$-drawable trees, where all layouts obtain average $Q_{RNG}$ of at most 0.6.

\begin{table}[h!]
\centering
\caption{Example layout comparison for a large $RNG$-drawable tree.}
\label{table:layoutcomp_rngtree}
\begin{tabular}{|c|c|c|c|c|}
\hline
$BB$ & $FM^3$ & $FR$ & $LL$ & $OR$ \\
\hline
\includegraphics[width=0.19\textwidth]{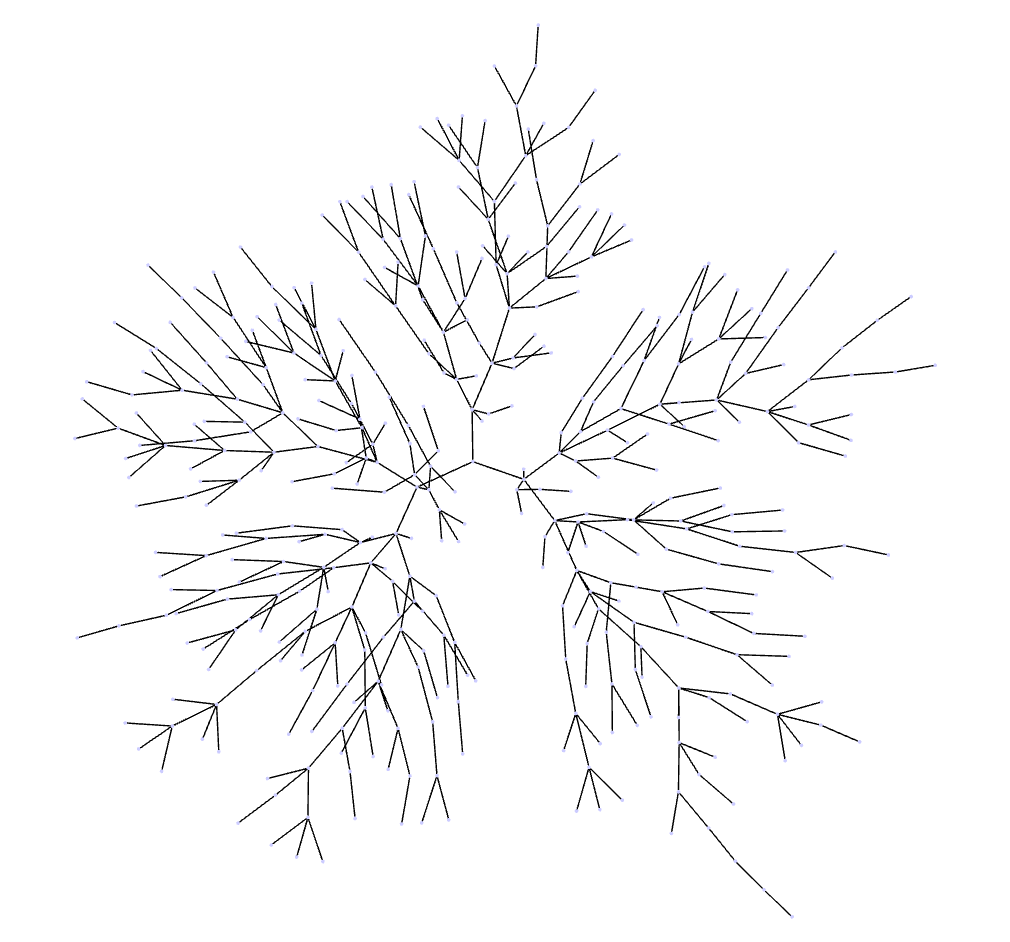} &
\includegraphics[width=0.19\textwidth]{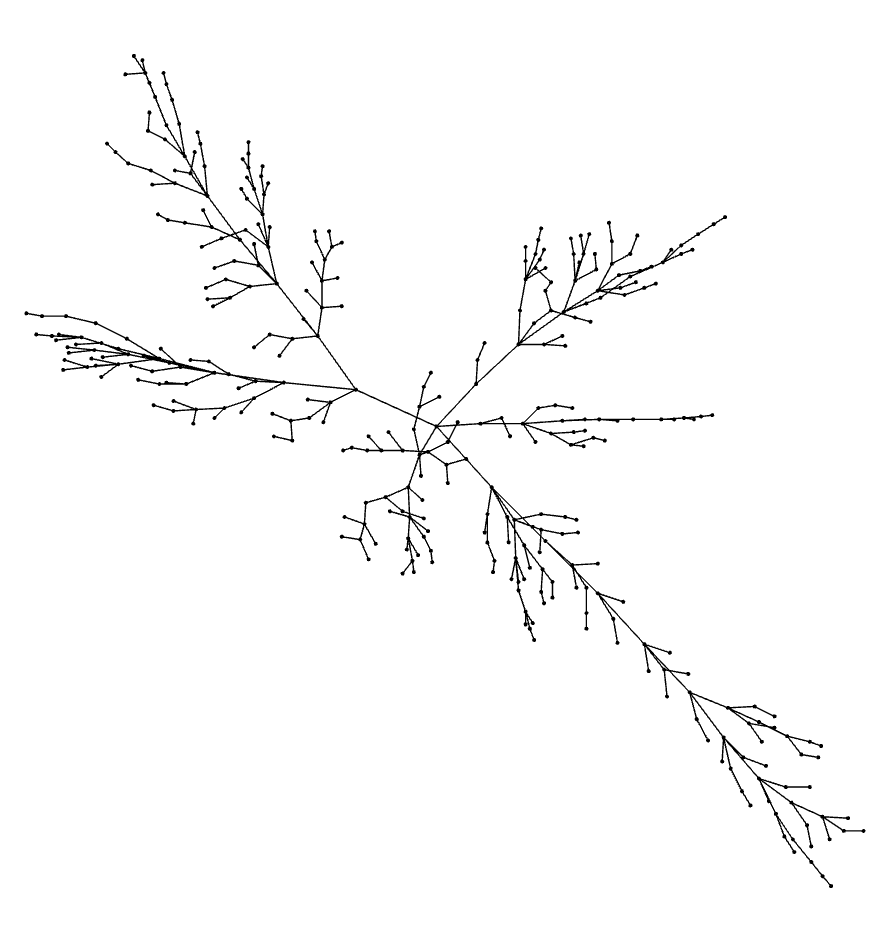} &
\includegraphics[width=0.19\textwidth]{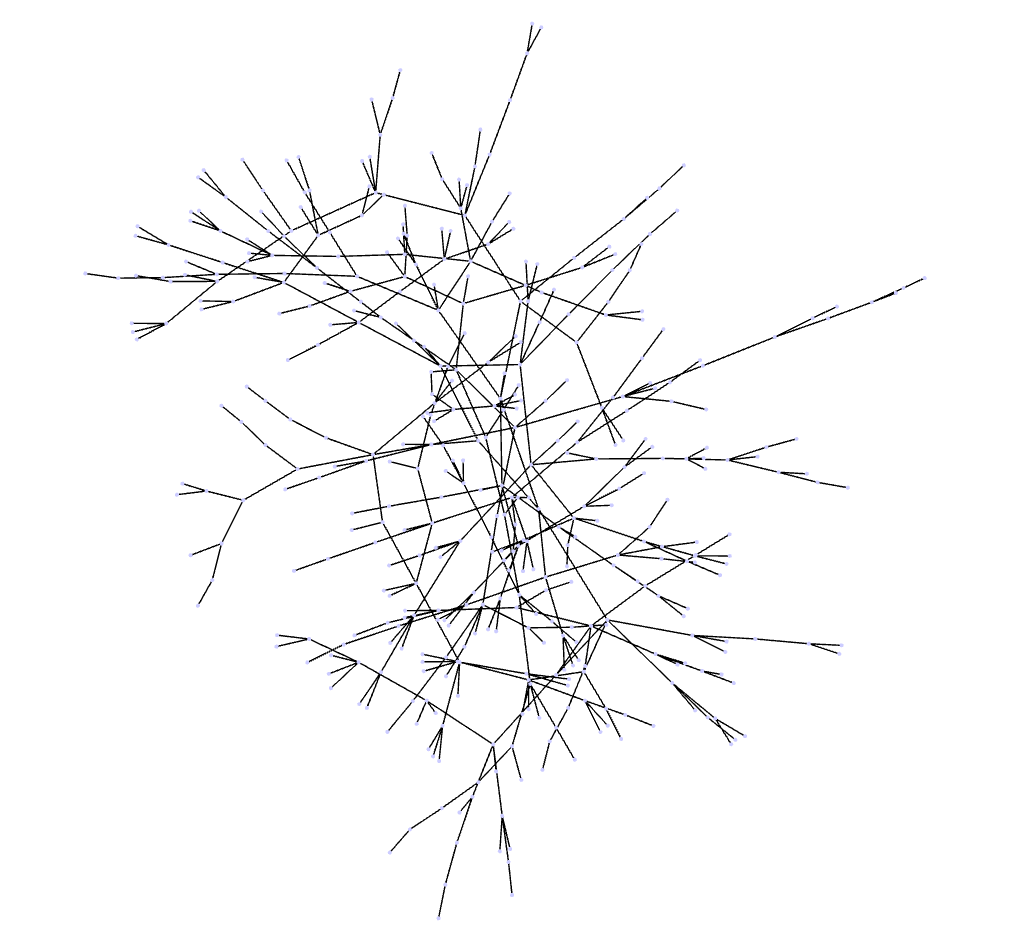} &
\includegraphics[width=0.19\textwidth]{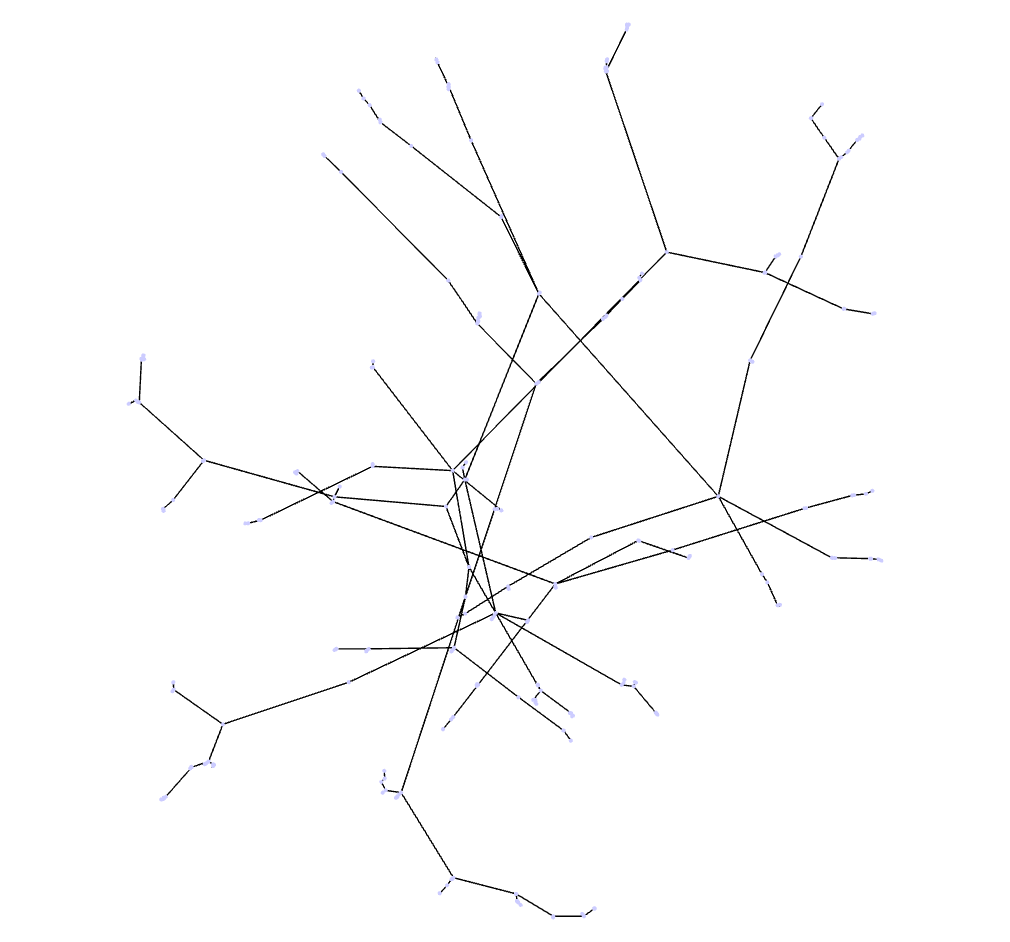} &
\includegraphics[width=0.19\textwidth]{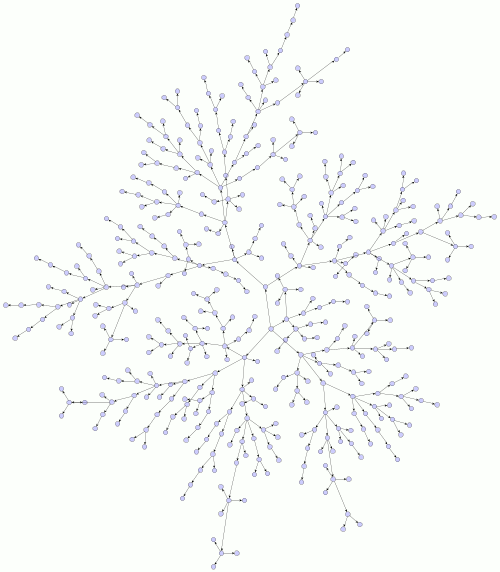} \\
\hline
$sfdp$ & $SGD$ & $SM$ & $tsNET$ & $W$ \\
\hline
\includegraphics[width=0.19\textwidth]{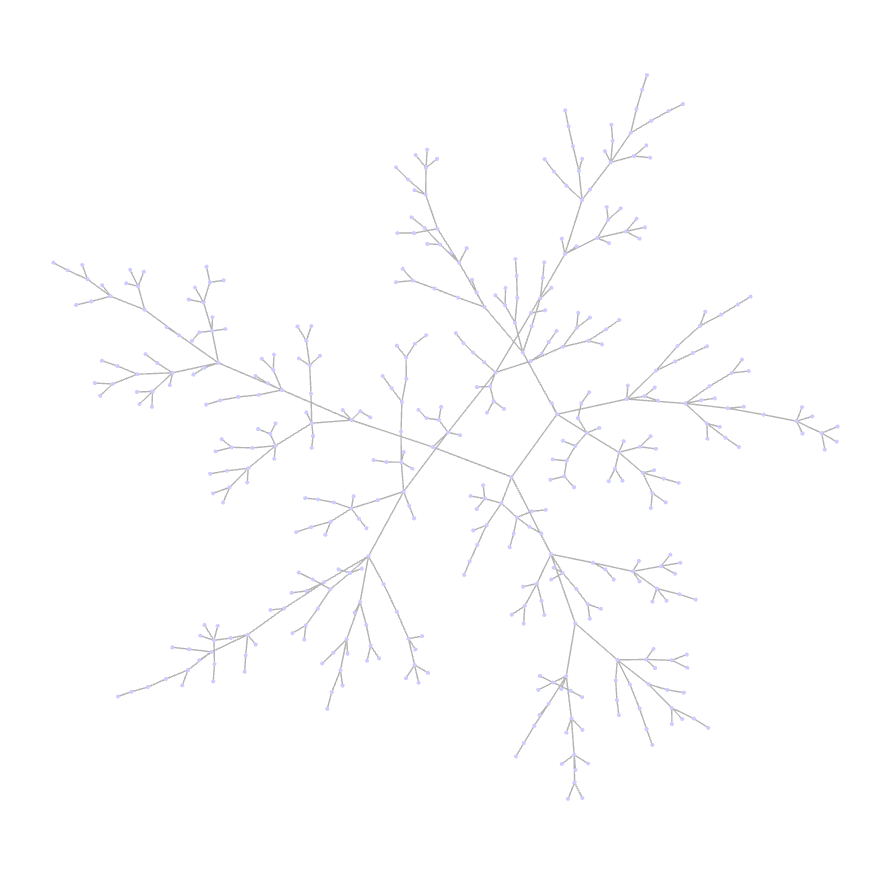} &
\includegraphics[width=0.19\textwidth]{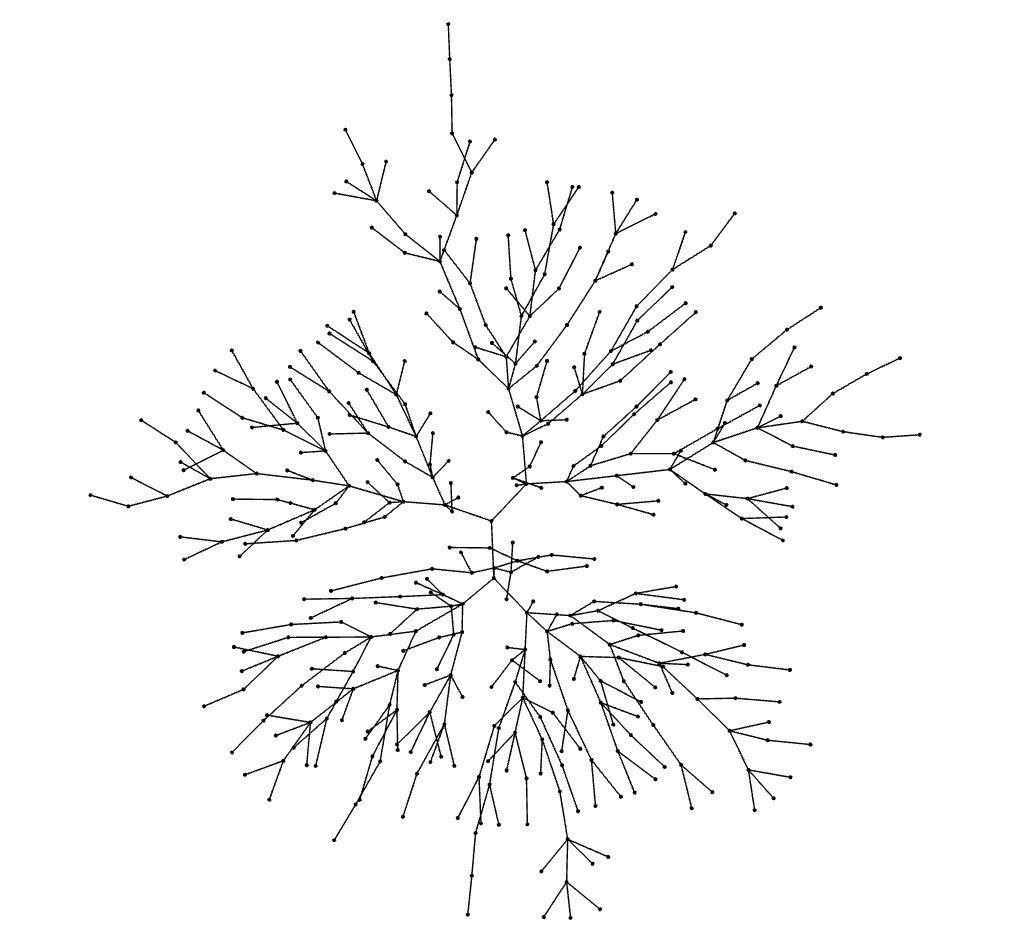} &
\includegraphics[width=0.19\textwidth]{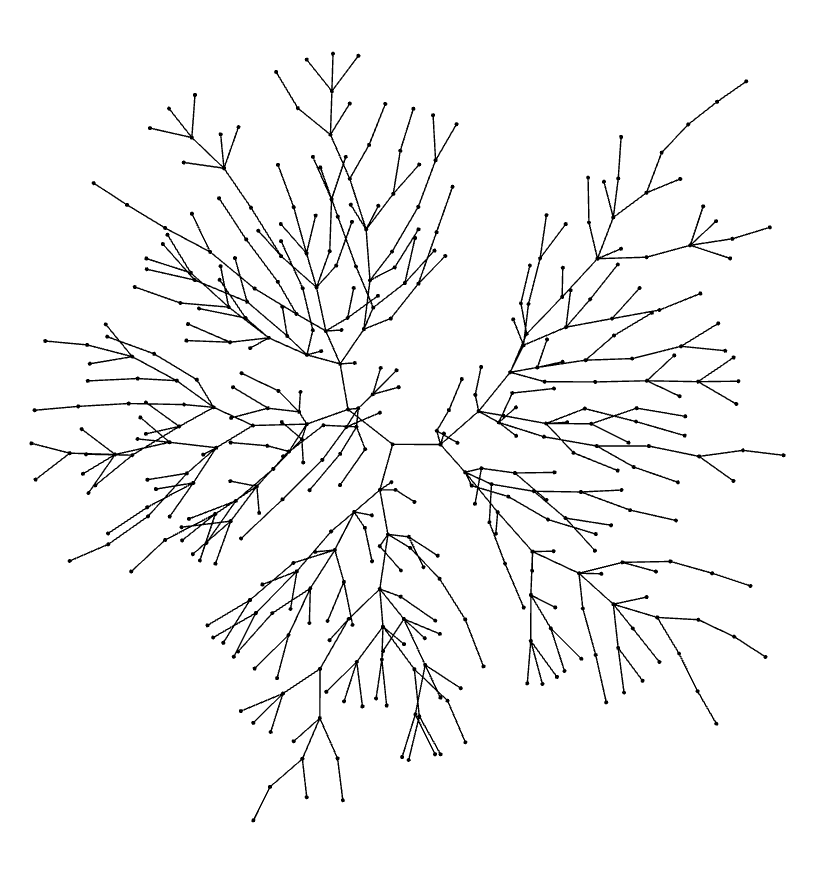} &
\includegraphics[width=0.19\textwidth]{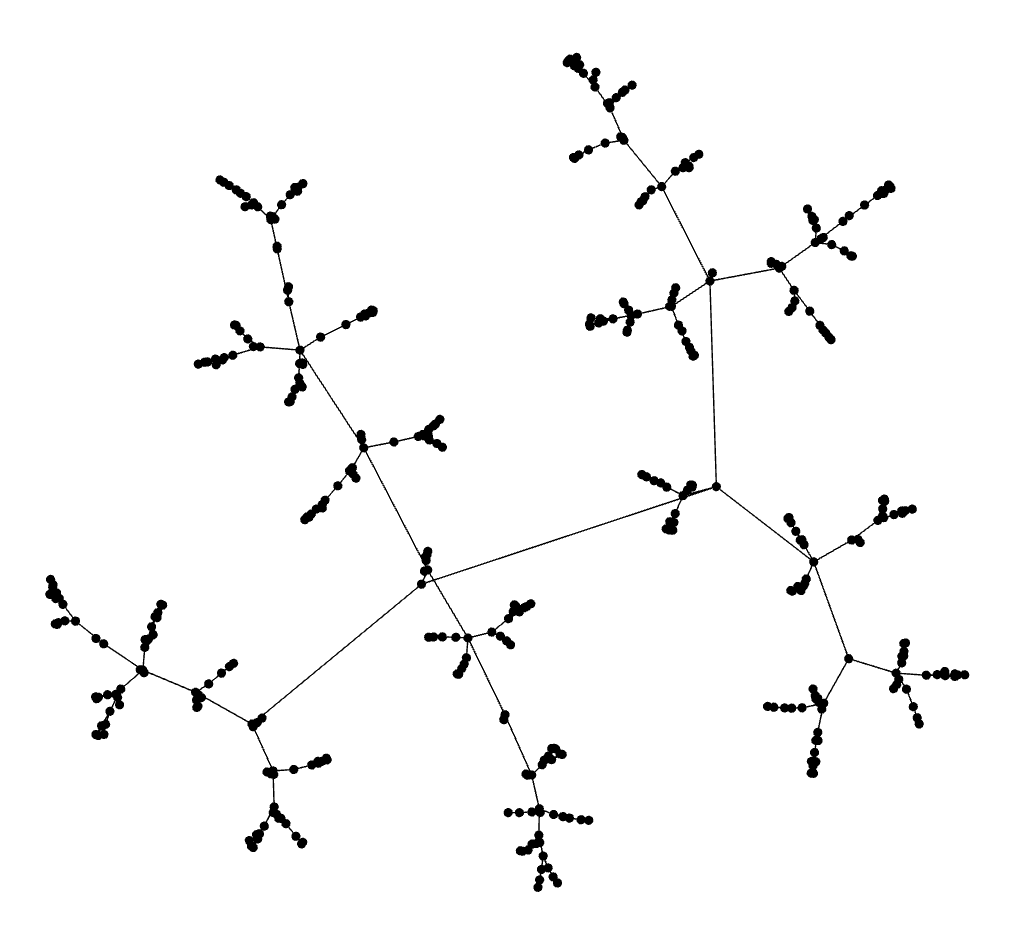} &
\includegraphics[width=0.19\textwidth]{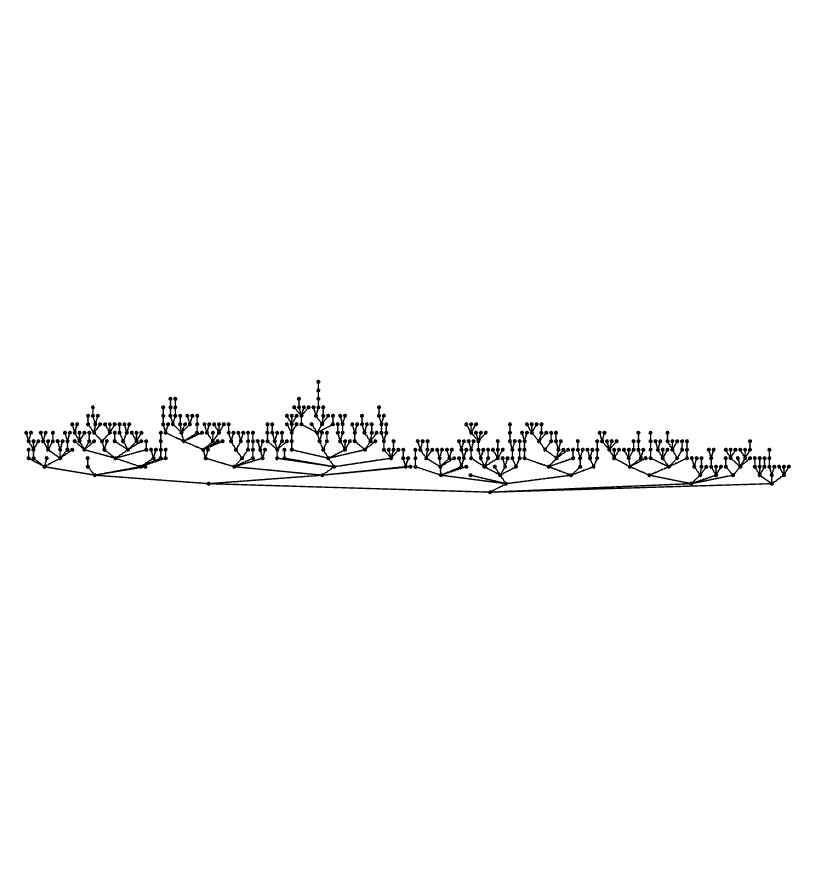} \\ \hline
\end{tabular}
\end{table}

\begin{table}[h!]
\centering
\caption{Example layout comparison for a large maximum outerplanar graph.}
\label{table:layoutcomp_maxouterplanar}
\begin{tabular}{|c|c|c|c|c|}
\hline
BB & FM\textsuperscript{3} & FR & LL& OR \\ \hline
\includegraphics[width=0.19\textwidth]{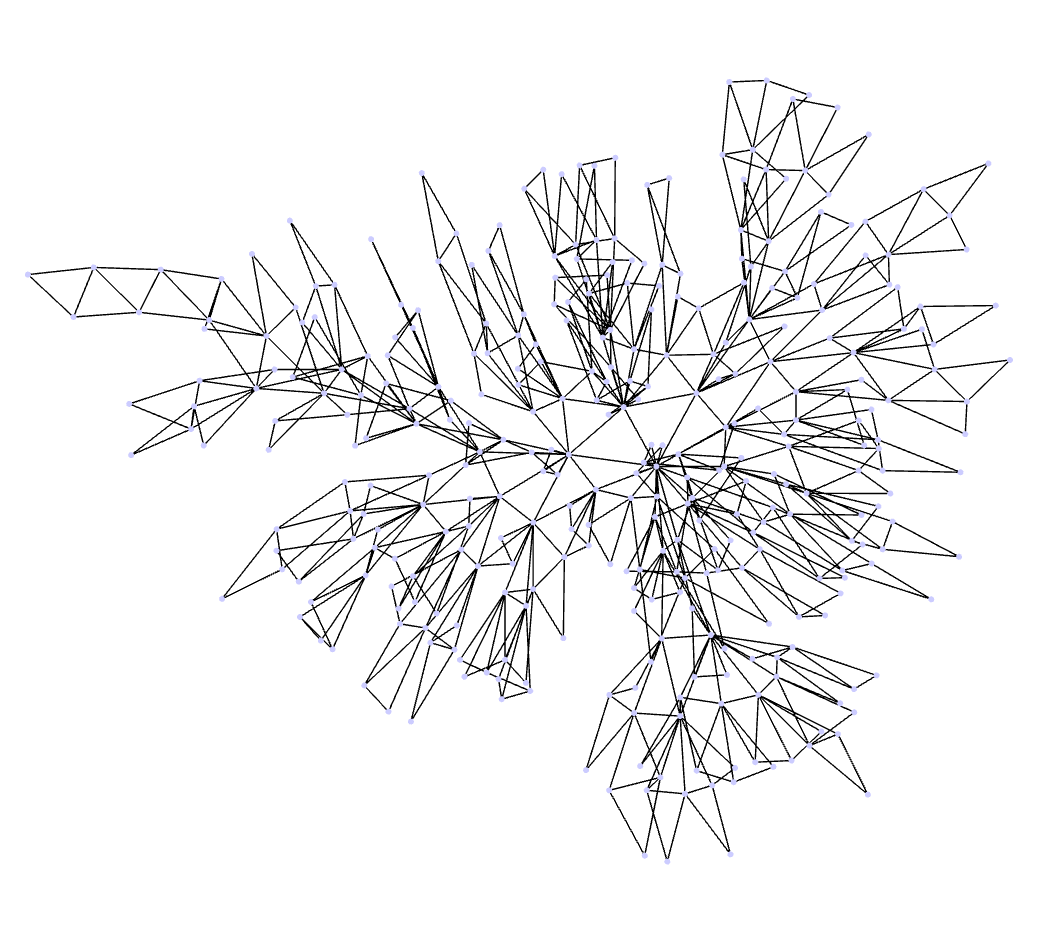} &
\includegraphics[width=0.19\textwidth]{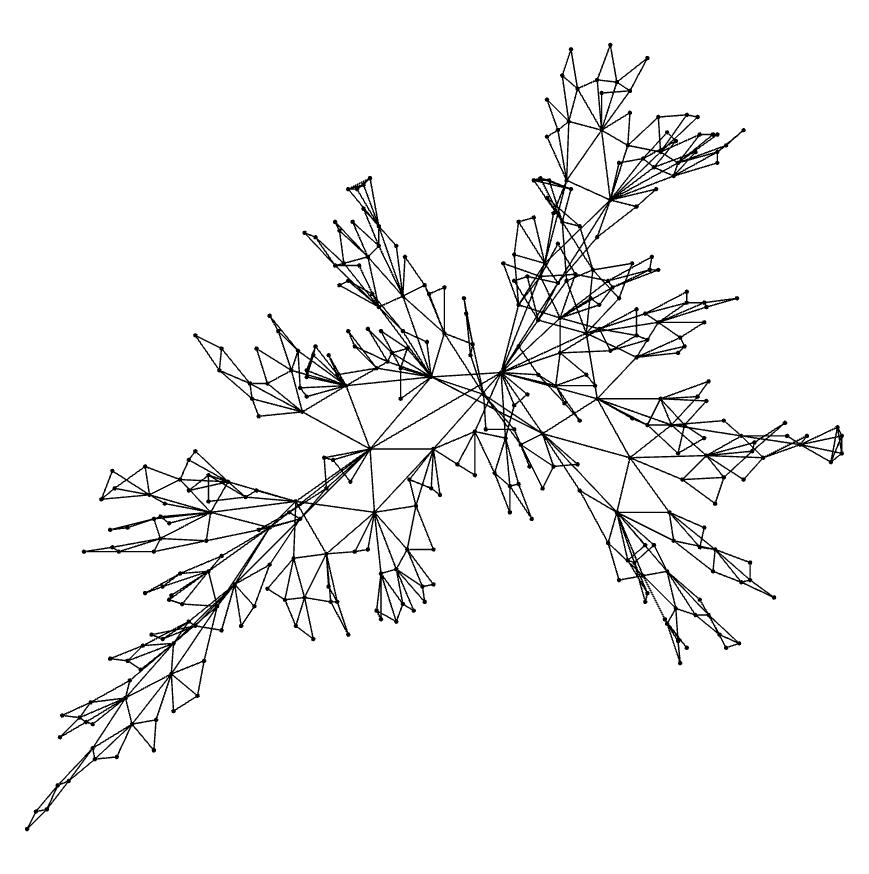} &
\includegraphics[width=0.19\textwidth]{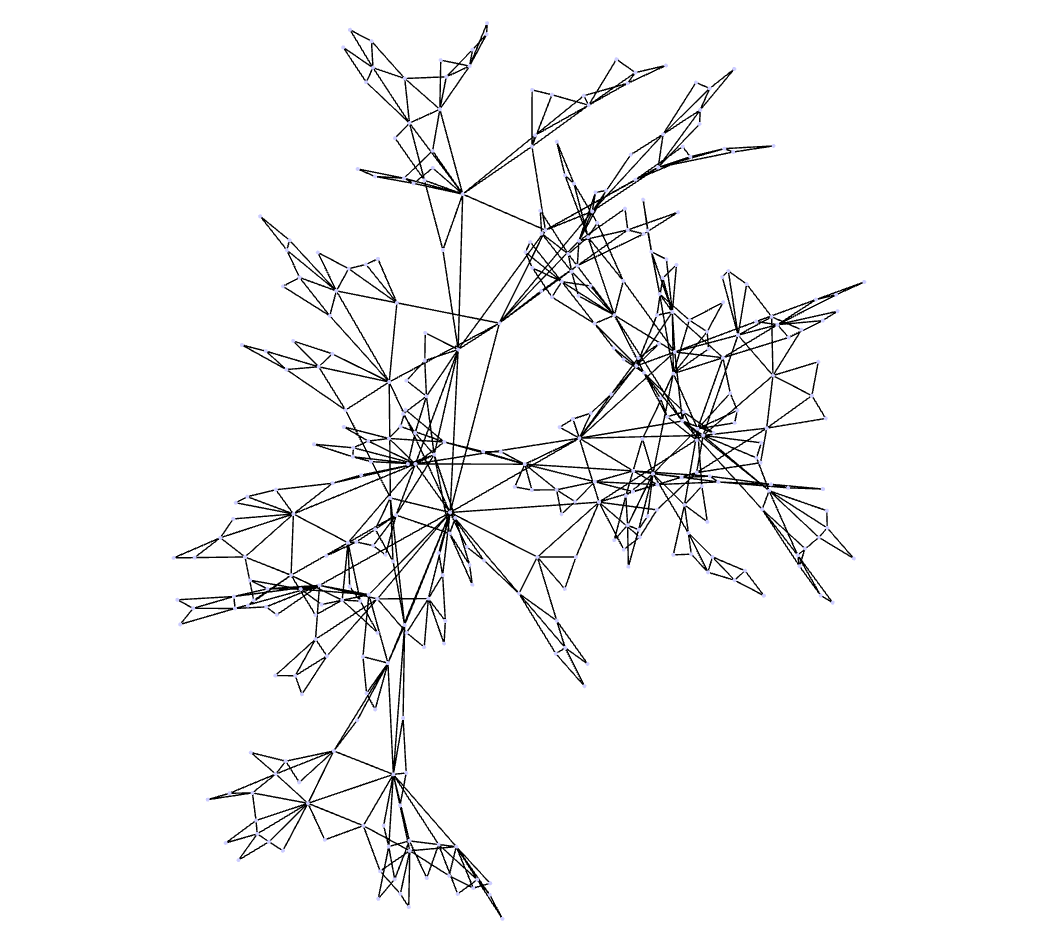} &
\includegraphics[width=0.19\textwidth]{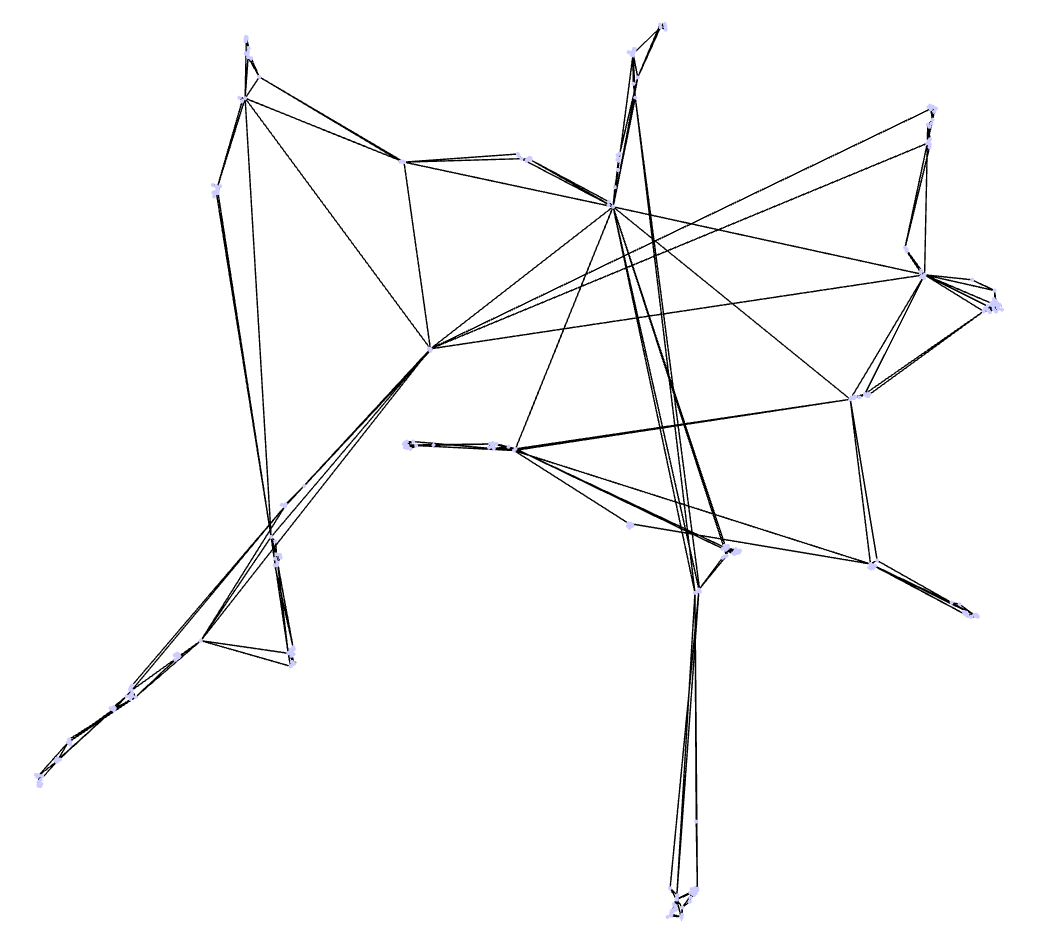} &
\includegraphics[width=0.19\textwidth]{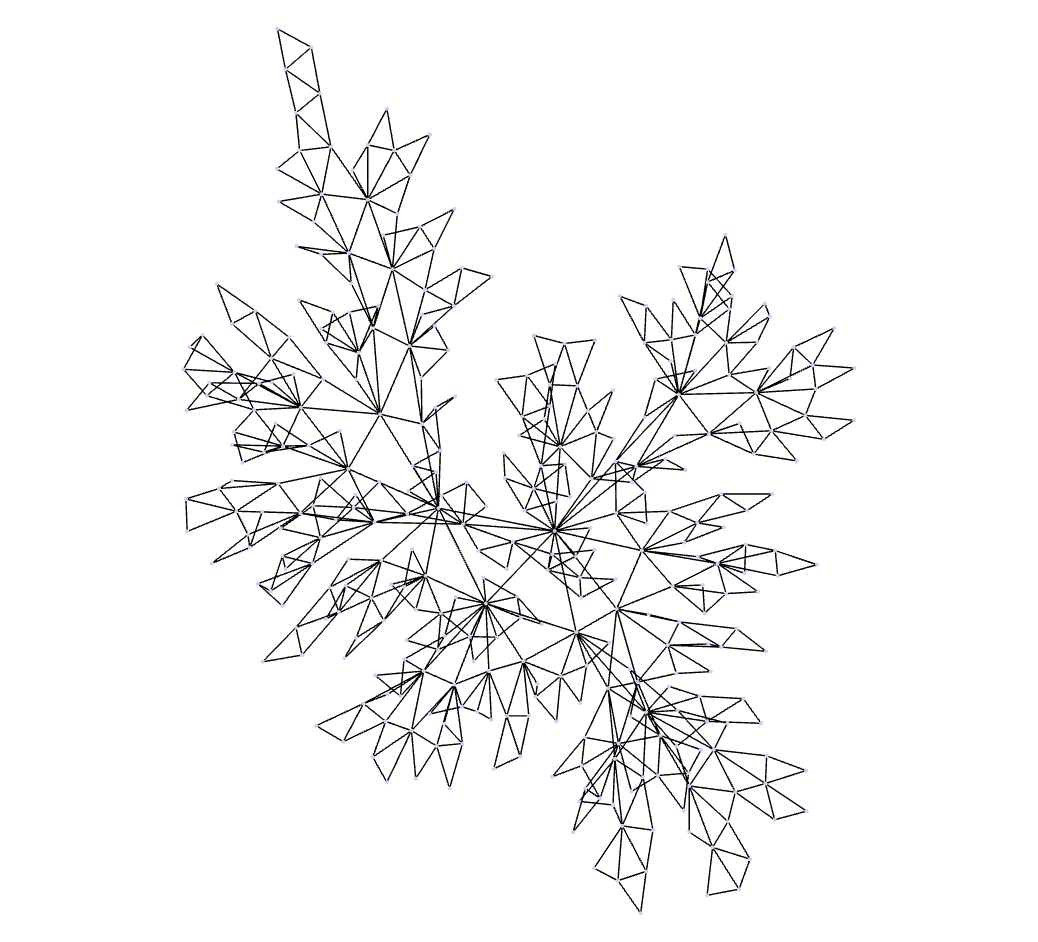} \\ \hline
$sfdp$ & $SGD$ & $SM$ & $tsNET$ & $CK$ \\ \hline
\includegraphics[width=0.19\textwidth]{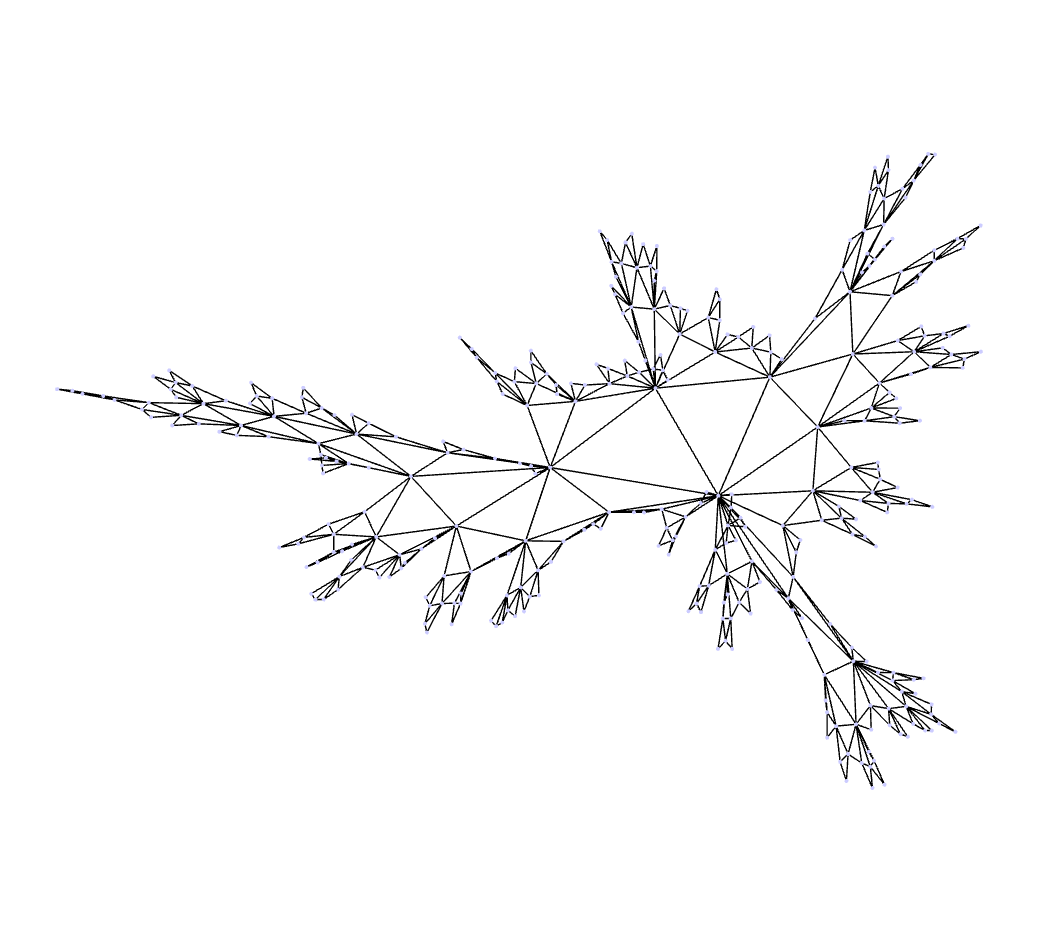} &
\includegraphics[width=0.19\textwidth]{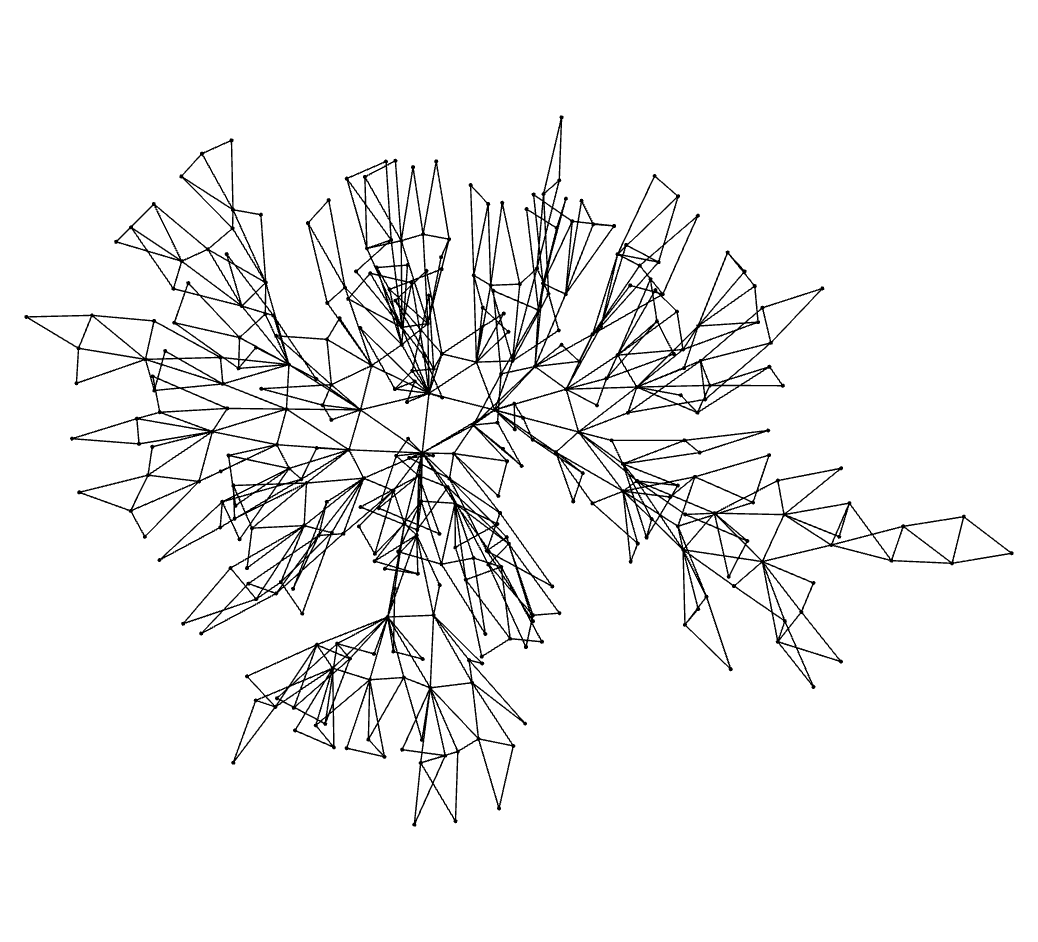} &
\includegraphics[width=0.19\textwidth]{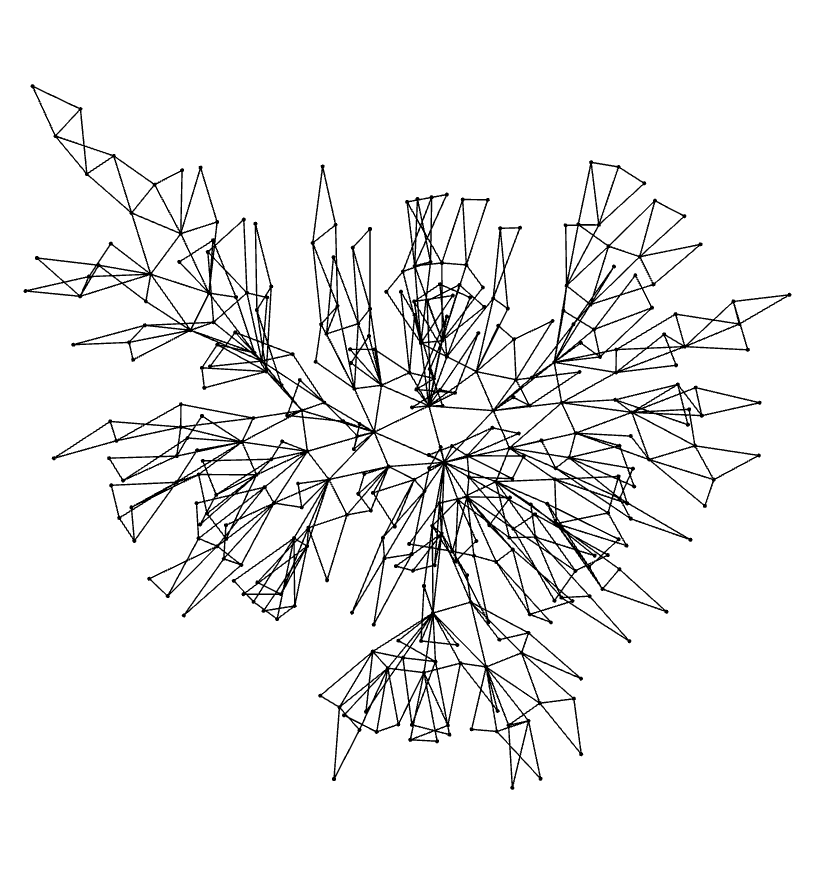} &
\includegraphics[width=0.19\textwidth]{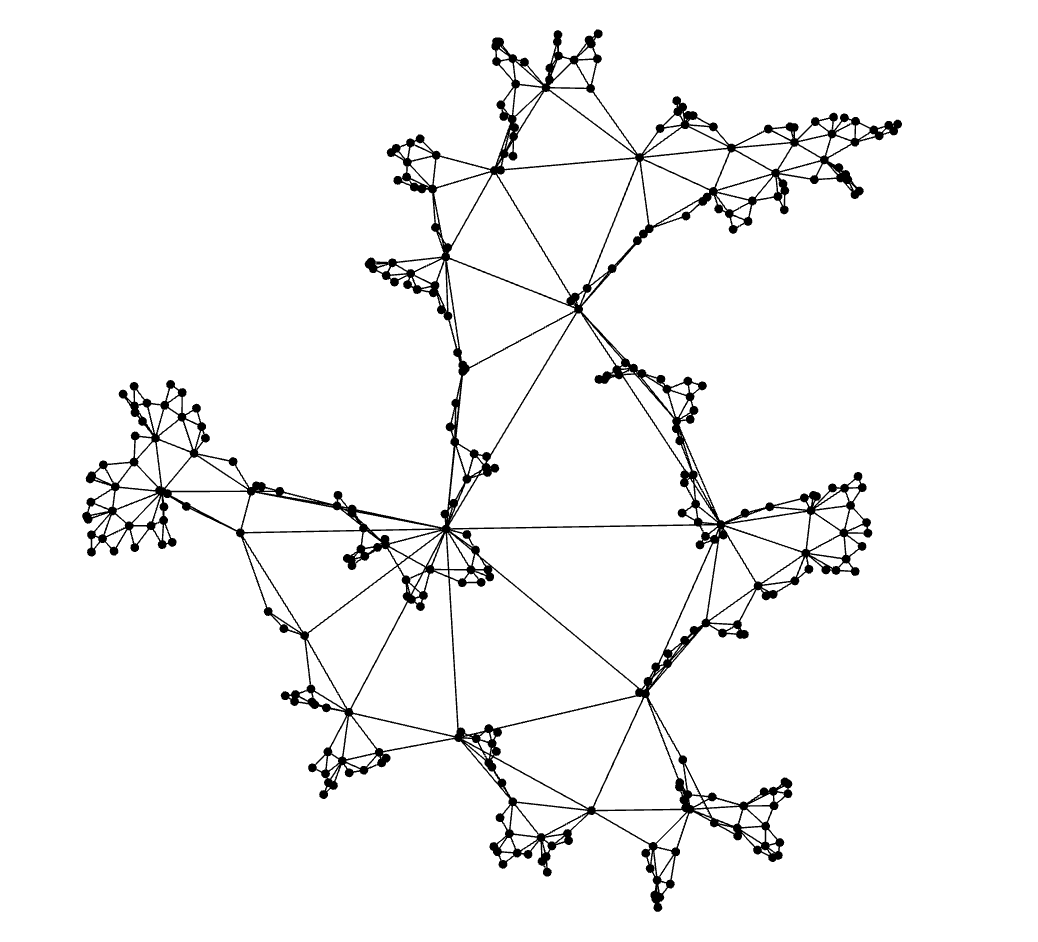} &
\includegraphics[width=0.19\textwidth]{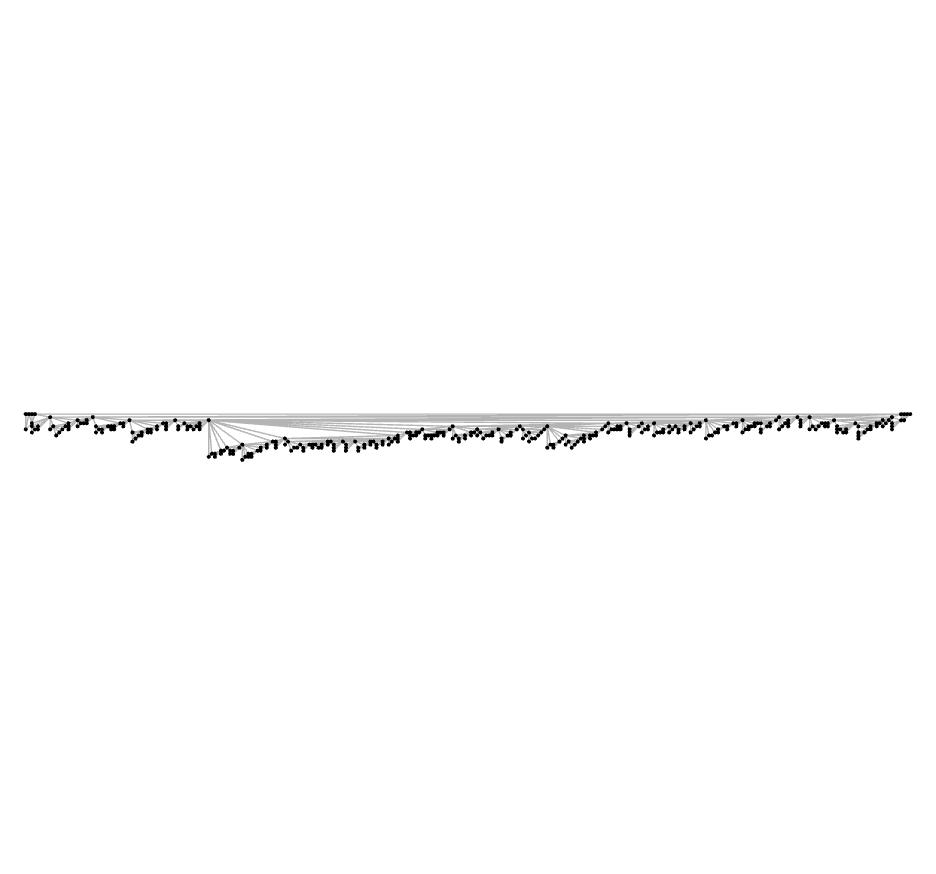} \\ \hline
\end{tabular}
\end{table}

$tsNET$ and $CK$ are the top performing layouts on medium and large outerplanar graphs, despite lower performance on small graphs: on medium and large maximum outerplanar graphs, $tsNET$ (resp., $CK$) obtains $Q_{RNG}$ of 0.6 (resp., 0.5) on average. This is closer to optimal compared to large $RNG$-drawable trees, where $tsNET$, the best performing layout, only obtains average $Q_{RNG}$ of 0.4.


For $GG$-drawable trees, the results on $Q_{GG}$ are mostly similar to $Q_{RNG}$;
similarly, for $GG$-drawable outerplanar graphs (same set of graphs as $RNG$-drawable outerplanar graphs),
the results on $Q_{GG}$ are similar to $Q_{RNG}$. For details, see Figures \ref{fig:layoutcompmetrics_ggtree} and \ref{fig:layoutcompmetrics_maxouterplanar_gg} in Appendix \ref{sec:app_layoutcomp}.

Table \ref{table:layoutcomp_rngtree} shows a visual comparison of graph layouts on a large $RNG$-drawable tree. 
For the best performing layouts $tsNET$ and $LL$, subtrees closer to the leaves are often ``compacted'' together, compared to the second best performing layouts such as OR and sfdp, where all branches are more ``opened'' up.

Table \ref{table:layoutcomp_maxouterplanar} shows a visual comparison of layouts on a maximal outerplanar graph. 
The best performing layout, $tsNET$, collapses the faces on the periphery,  compared to the faces in the middle of the drawing. The ``long'' drawing of $CK$ may have obtained a comparable effect, producing high shape-based metrics.

\begin{figure}[h]
\centering
\subfloat[\texttt{L-AUG}]{
\includegraphics[width=0.4\columnwidth]{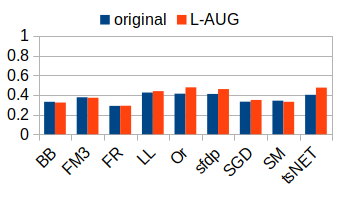}
}
\subfloat[\texttt{F-AUG}]{
\includegraphics[width=0.4\columnwidth]{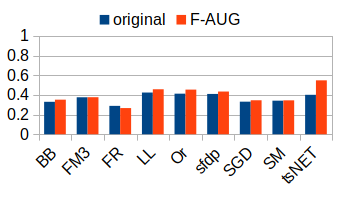}
}
\caption{Average $Q_{RNG}$ for \texttt{L-AUG} and \texttt{F-AUG} graphs, compared to $RNG$-drawable trees. $tsNET$ obtains the highest shape-based metrics; surprisingly, $Q_{RNG}$ is sometimes higher on \texttt{L-AUG} and \texttt{F-AUG} than on the strong proximity drawable graphs.}
\label{fig:layoutcompmetrics_rngtreeaug}
\end{figure}

\begin{table}[h!]
\centering
\caption{Example layout comparison for a medium \texttt{F-AUG} graph.}
\label{table:layoutcomp_rngtreeaug}
\begin{tabular}{|c|c|c|c|c|}
\hline
$BB$ & $FM^3$ & $FR$ & $LL$ & $OR$ \\ \hline
\includegraphics[width=0.19\textwidth]{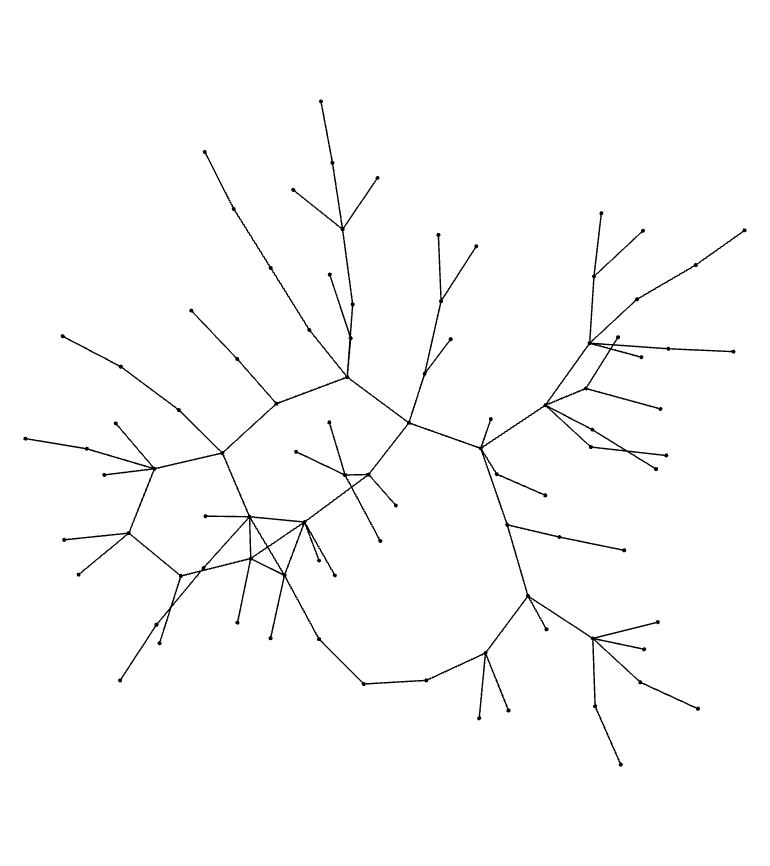} &
\includegraphics[width=0.19\textwidth]{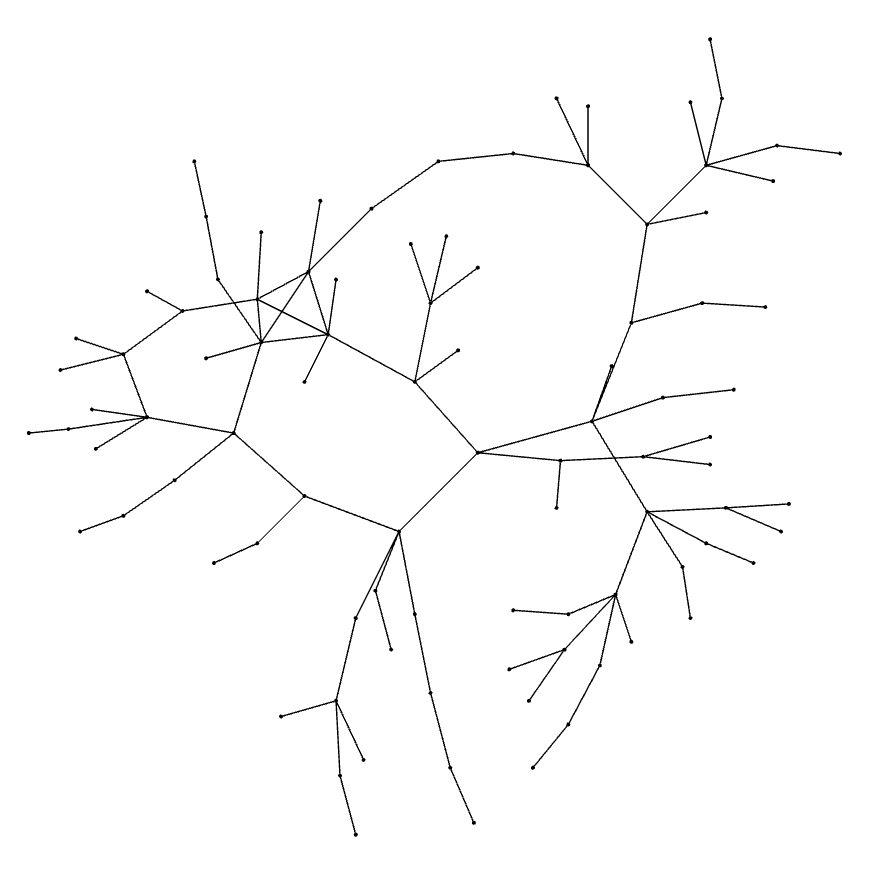} &
\includegraphics[width=0.19\textwidth]{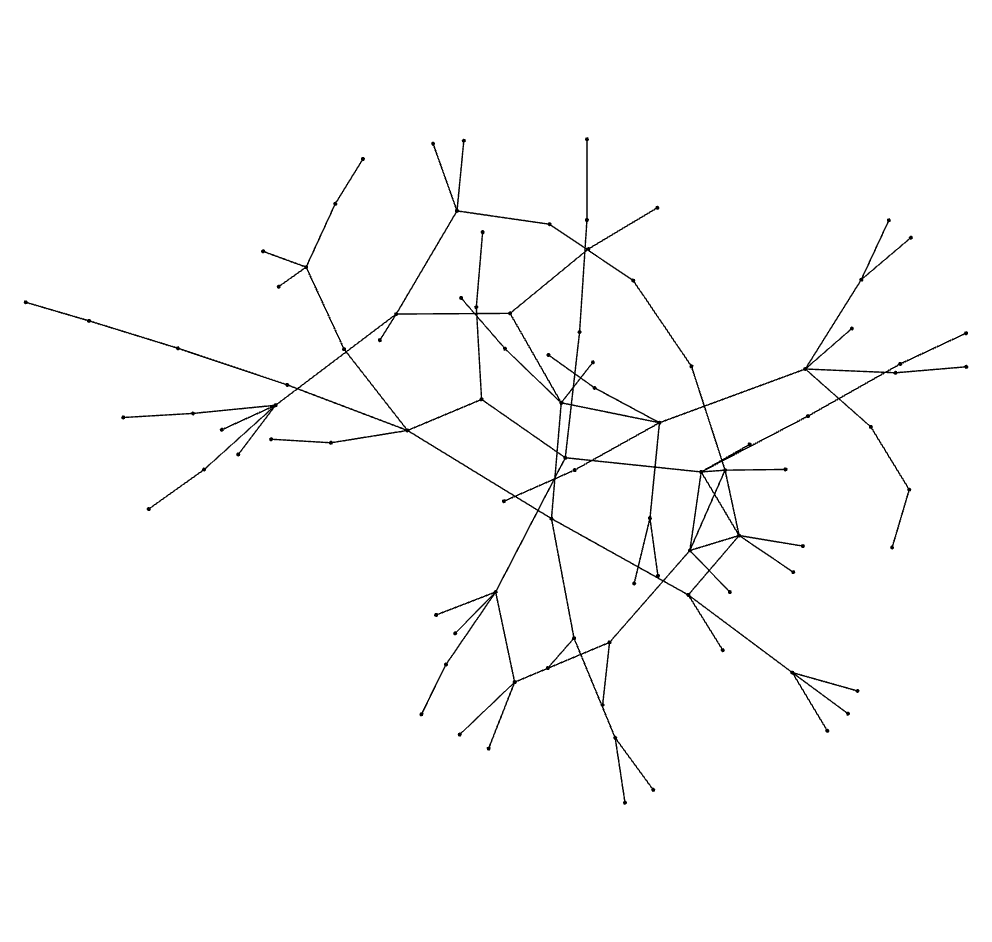} &
\includegraphics[width=0.19\textwidth]{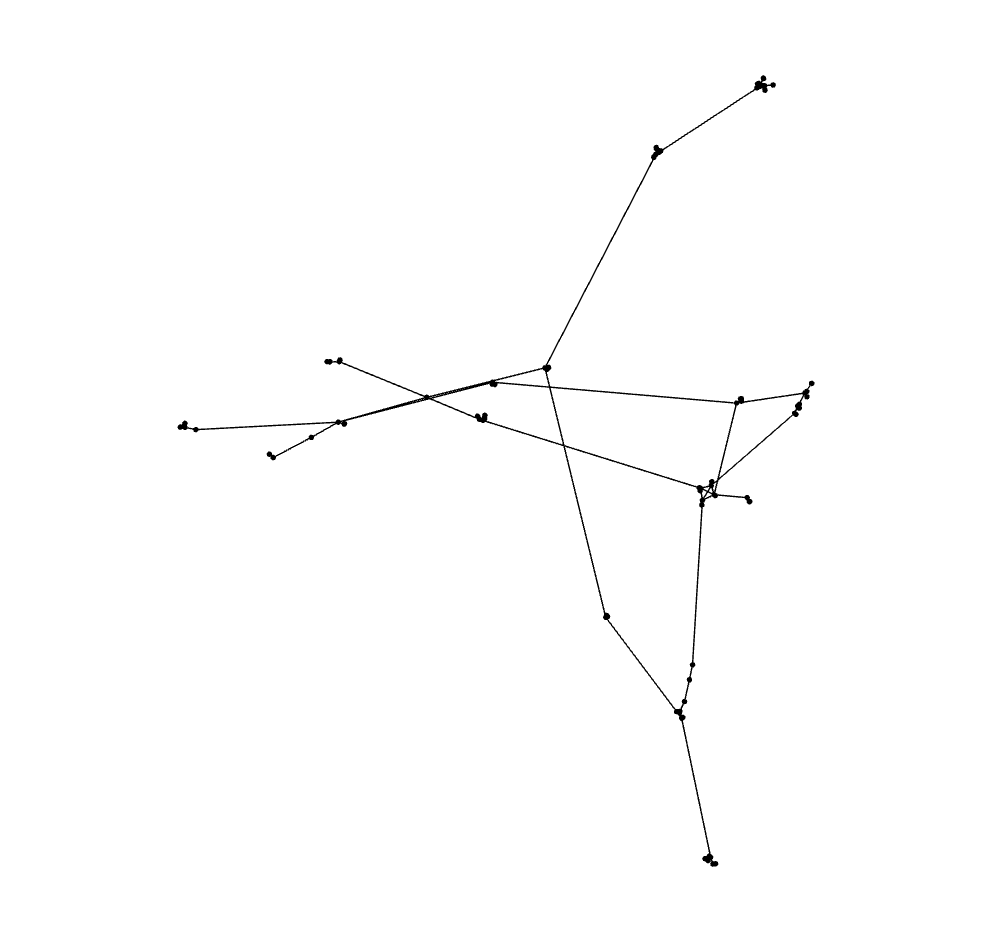} &
\includegraphics[width=0.19\textwidth]{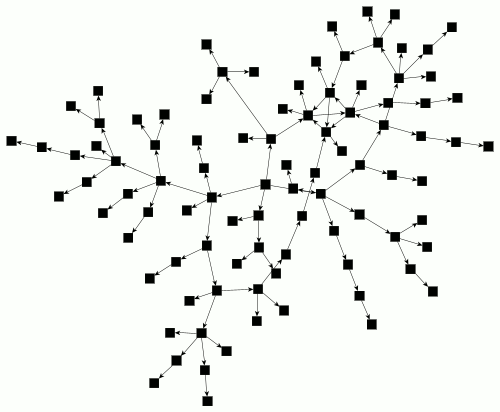} \\ \hline
$sfdp$ & $SGD$ & $SM$ & $tsNET$ & \\ \hline
\includegraphics[width=0.17\textwidth]{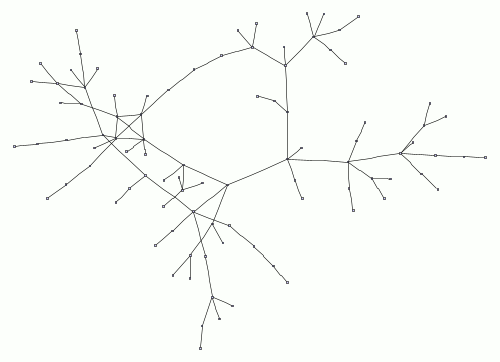} &
\includegraphics[width=0.17\textwidth]{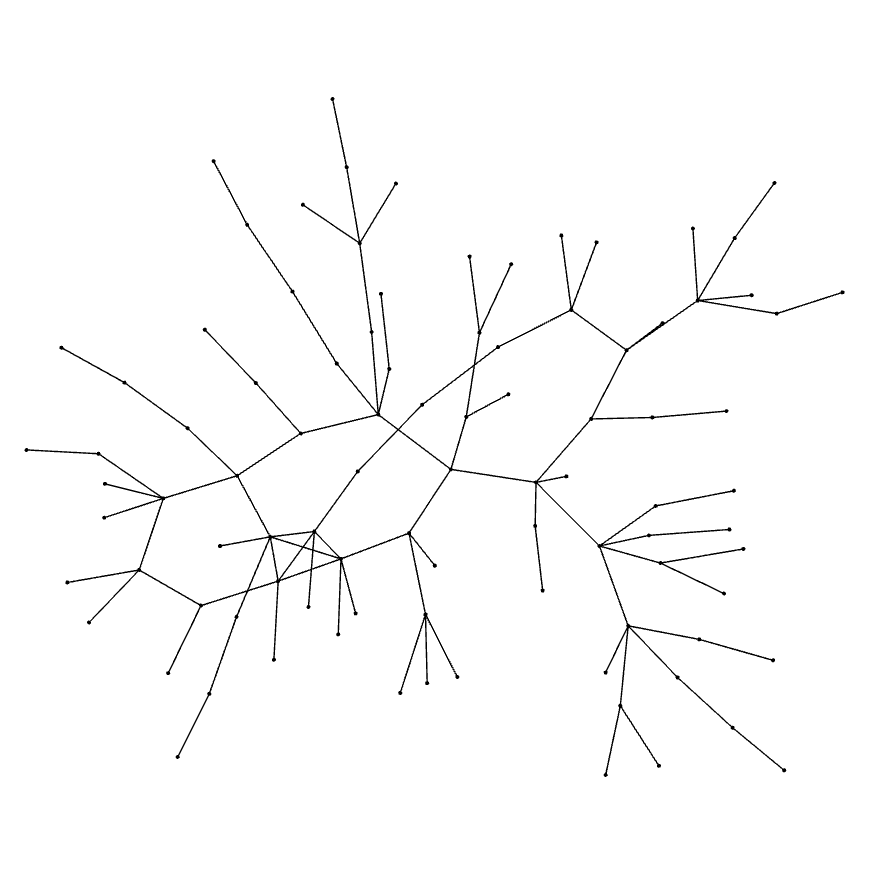} &
\includegraphics[width=0.17\textwidth]{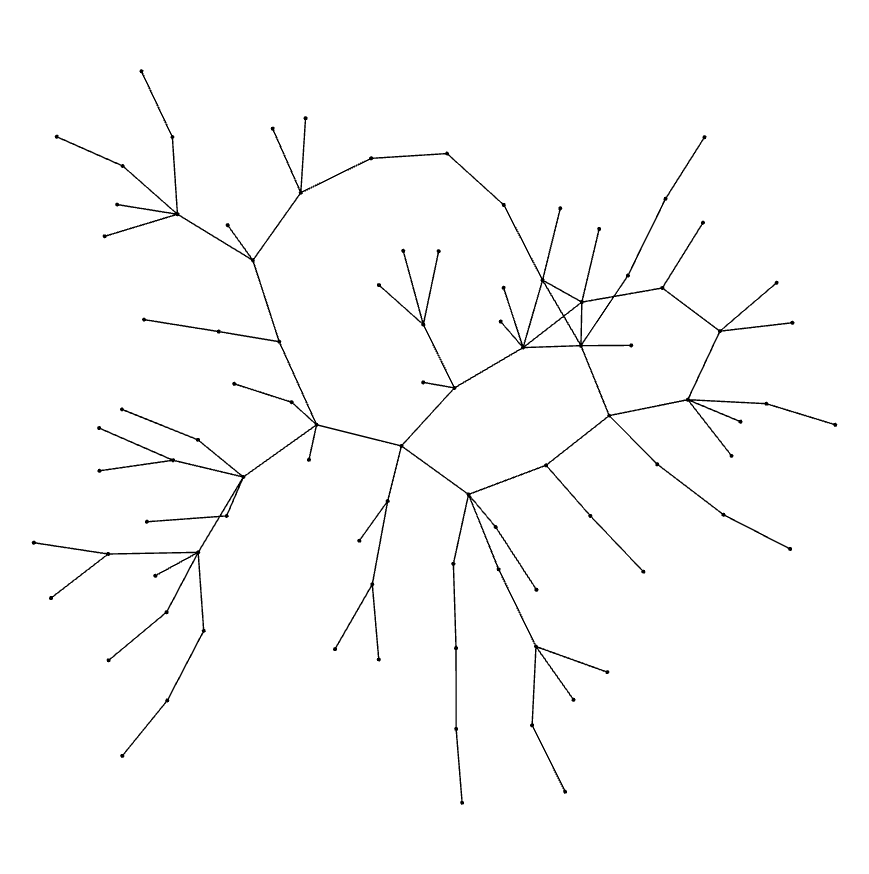} &
\includegraphics[width=0.17\textwidth]{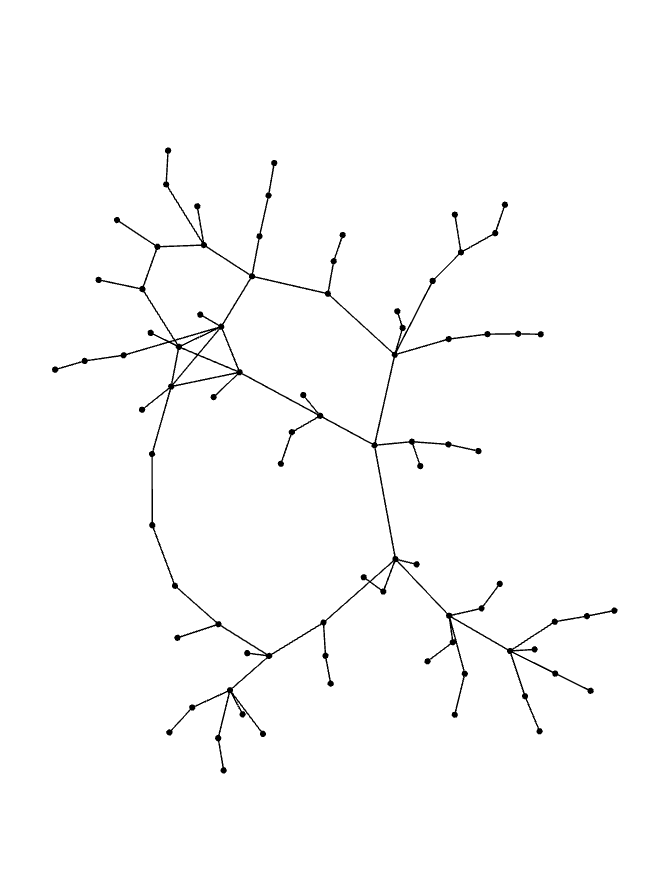} & \\ \hline
\end{tabular}
\end{table}

\subsubsection{Almost Proximity Drawable Graphs}

In general, the ranking of the graph drawing algorithms on the shape-based metrics do not change much between strong proximity drawable graphs and almost proximity drawable graphs.

Figure \ref{fig:layoutcompmetrics_rngtreeaug} shows comparisons on $Q_{RNG}$ for the  base $RNG$-drawable trees and the \texttt{L-AUG} and \texttt{F-AUG} graphs, where $tsNET$ still obtains the highest $Q_{RNG}$. 
$LL$ also obtains the second highest $Q_{RNG}$, although with a smaller difference to the next best performing layouts $OR$ and $sfdp$, compared to  $RNG$-drawable trees.

Table \ref{table:layoutcomp_rngtreeaug} shows a visual comparison on a \texttt{F-AUG} graph, where the layouts with highest shape-based metrics, such as tsNET and LL, draw the ``branches'' in the periphery of the drawing in a more compact way, than other layout.
This observation is consistent with the pattern also seen in the visual comparison for strong proximity drawable trees and outerplanar graphs.

\begin{figure}[h!]
\centering
\subfloat[Small]{
\includegraphics[width=0.28\columnwidth]{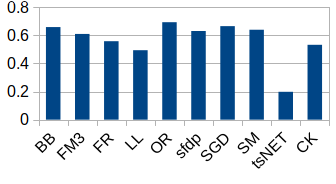}
}
\subfloat[Medium]{
\includegraphics[width=0.28\columnwidth]{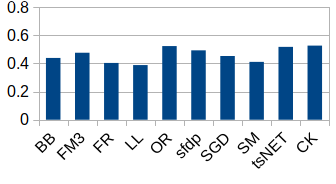}
}
\subfloat[Large]{
\includegraphics[width=0.28\columnwidth]{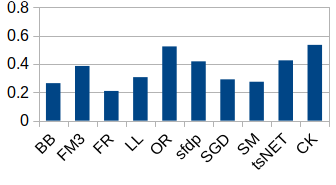}
}
\caption{Average $Q_{GG}$ for 1-connected outerplanar graphs. $OR$ and $CK$ performs the best on large 1-connected outerplanar graphs.}
\label{fig:layoutcompmetrics_connouterplanar_gg}
\end{figure}

\begin{table}[h!]
\centering
\caption{Example layout comparison for a large 1-connected outerplanar graph.}
\label{table:layoutcomp_connouterplanar}
\begin{tabular}{|c|c|c|c|c|}
\hline
$BB$ & $FM^3$ & $FR$ & $LL$ & $OR$ \\ \hline
\includegraphics[width=0.19\textwidth]{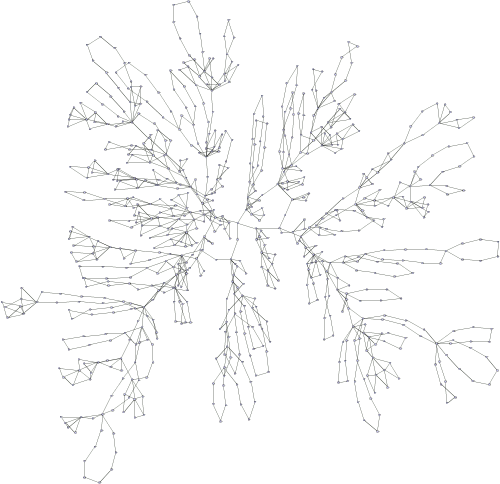} &
\includegraphics[width=0.19\textwidth]{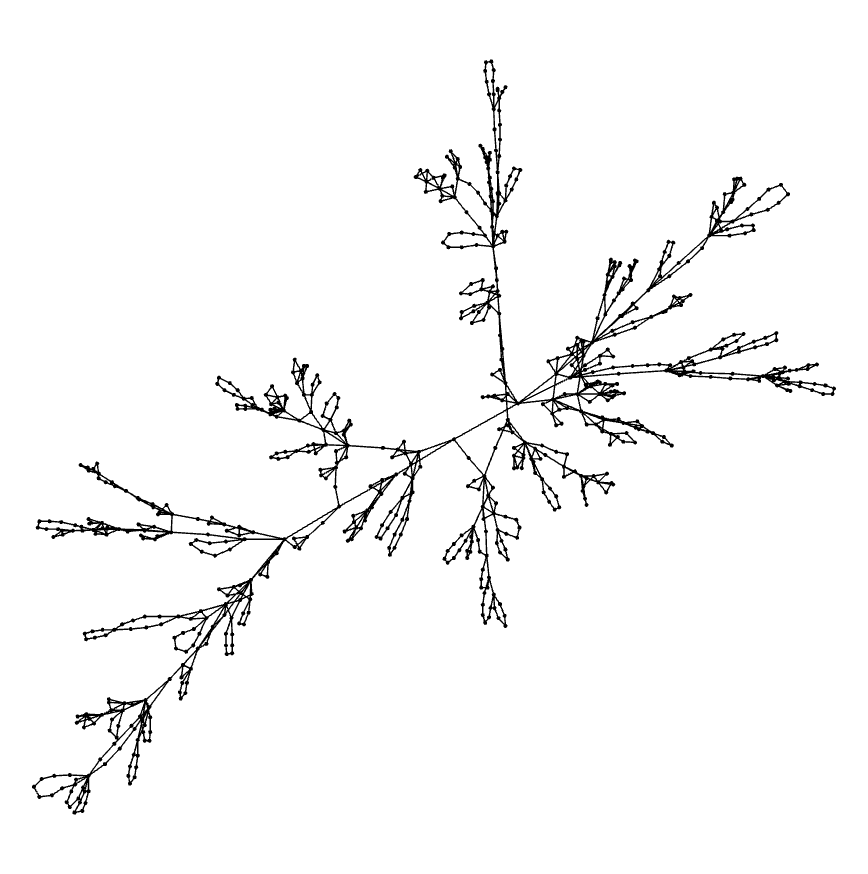} &
\includegraphics[width=0.19\textwidth]{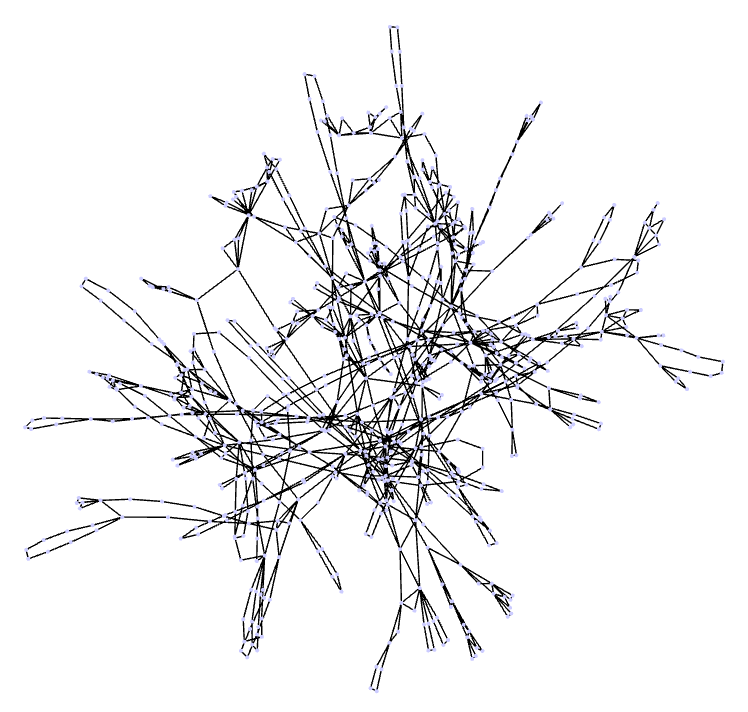} &
\includegraphics[width=0.19\textwidth]{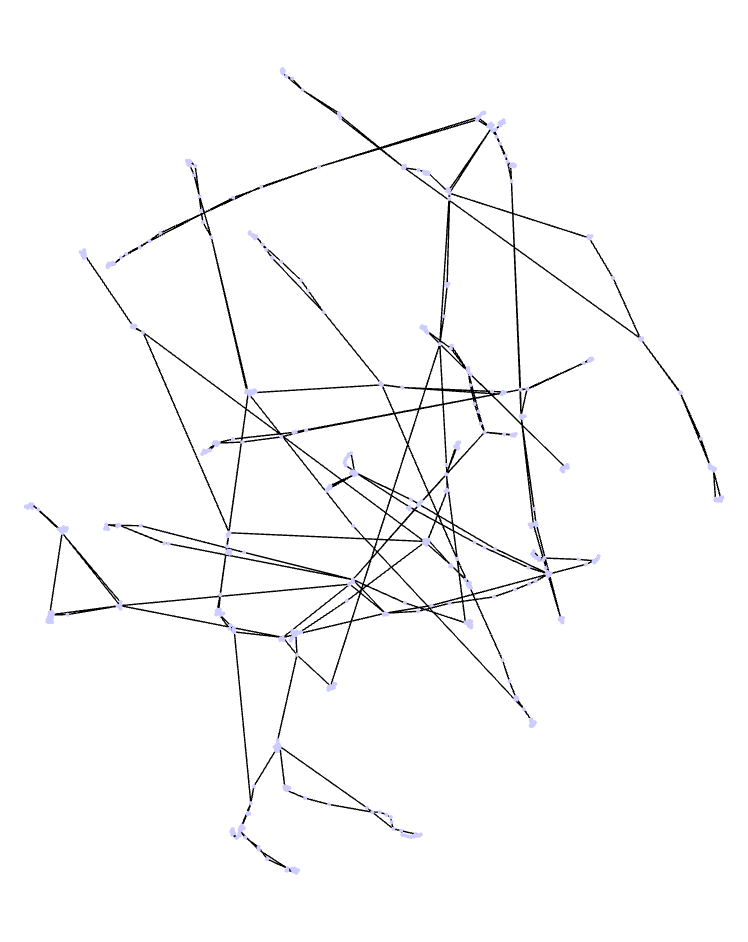} &
\includegraphics[width=0.19\textwidth]{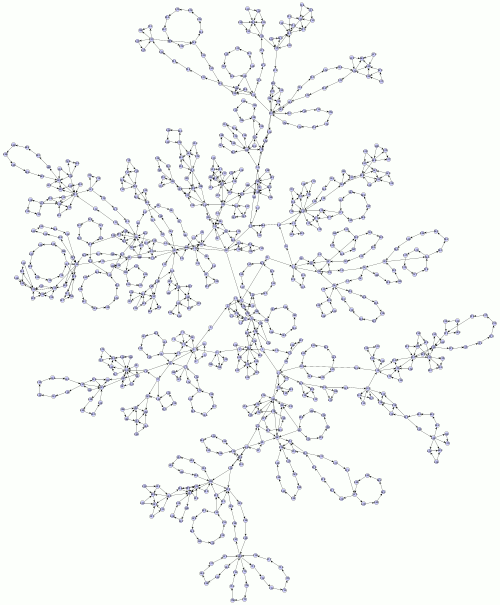} \\ \hline
$sfdp$ & $SGD$ & $SM$ & $tsNET$ & $CK$ \\ \hline
\includegraphics[width=0.19\textwidth]{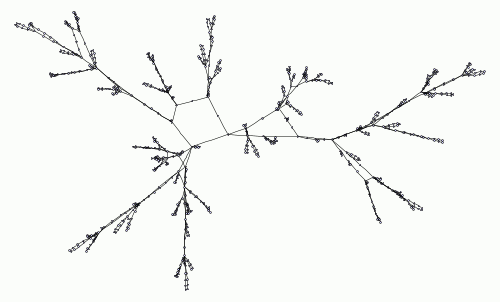} &
\includegraphics[width=0.19\textwidth]{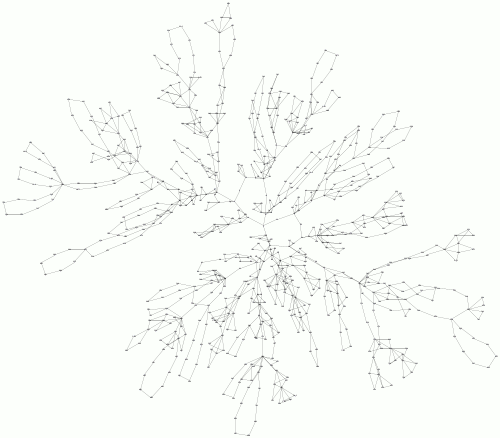} &
\includegraphics[width=0.19\textwidth]{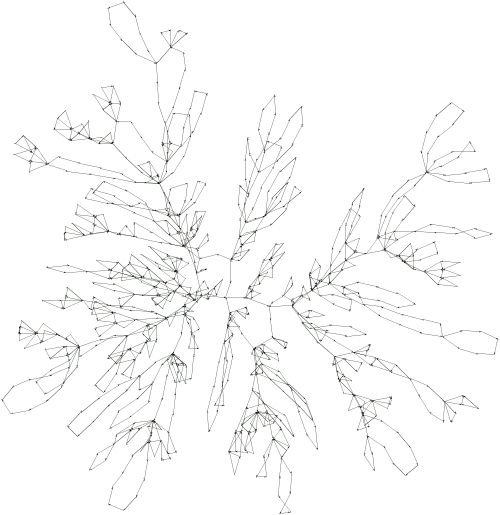} &
\includegraphics[width=0.19\textwidth]{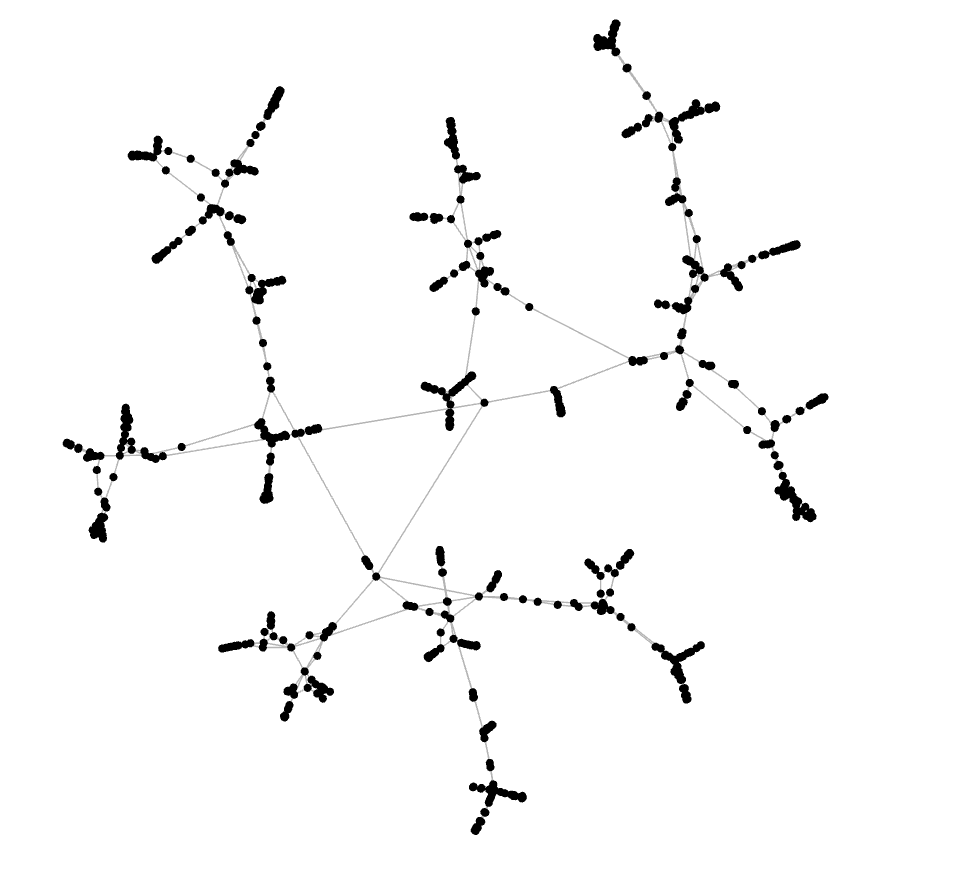} &
\includegraphics[width=0.19\textwidth]{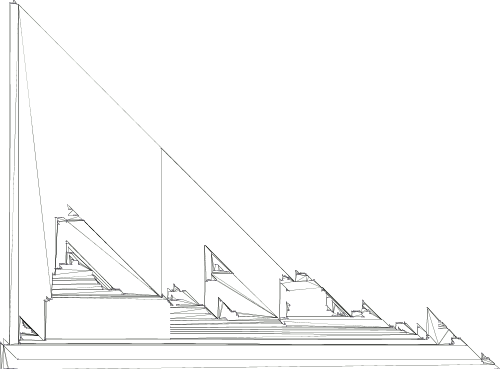} \\ \hline
\end{tabular}
\end{table}

\subsubsection{Weak Proximity Drawable Graphs}

For weak $GG$-drawable 1-connected outerplanar graphs, $OR$ surprisingly obtains the highest $Q_{GG}$ on large 1-connected outerplanar graphs, followed by CK and tsNET; see Figure \ref{fig:layoutcompmetrics_connouterplanar_gg}.

Table \ref{table:layoutcomp_connouterplanar} shows a  visual comparison, where $OR$ draws a number of chordless cycles with their vertices in a regular polygon configuration. 
In fact, this is the correct way to draw such cycles as $GG$, resulting in high $Q_{GG}$.

\subsubsection{Mesh Graphs}

\begin{figure}[h!]
\centering
\includegraphics[width=0.5\textwidth]{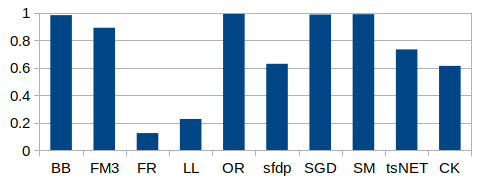}
\caption{Average $Q_{RNG}$ for mesh graphs. Stress-based layouts obtain the best shape-based metrics, at almost perfect.}
\label{fig:layoutcompmetrics_mesh}
\end{figure}

On mesh graphs, the best performing layouts, stress-based layouts, obtain on average much higher shape-based metrics than on other strong proximity drawable graphs, see Figure \ref{fig:layoutcompmetrics_mesh}. 
In particular, $SGD$ and $SM$ obtain near-perfect shape-based metrics ($Q_{RNG} = 0.99$ on average), and $OR$ and $BB$ also obtain very high shape-based metrics ($Q_{RNG} = 0.98$ on average). 
On the other hand, $tsNET$ obtains comparatively lower shape-based metrics.

\begin{table}[h!]
\centering
\caption{Example layout comparison for mesh.}
\label{table:layoutcomp_mesh}
\begin{tabular}{|c|c|c|c|c|}
\hline
$BB$ & $FM^3$ & $FR$ & $LL$ & $OR$ \\ \hline
\includegraphics[width=0.19\textwidth]{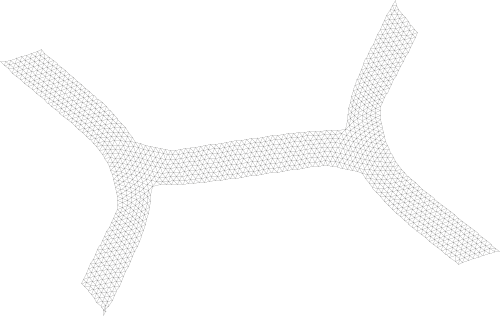} &
\includegraphics[width=0.19\textwidth]{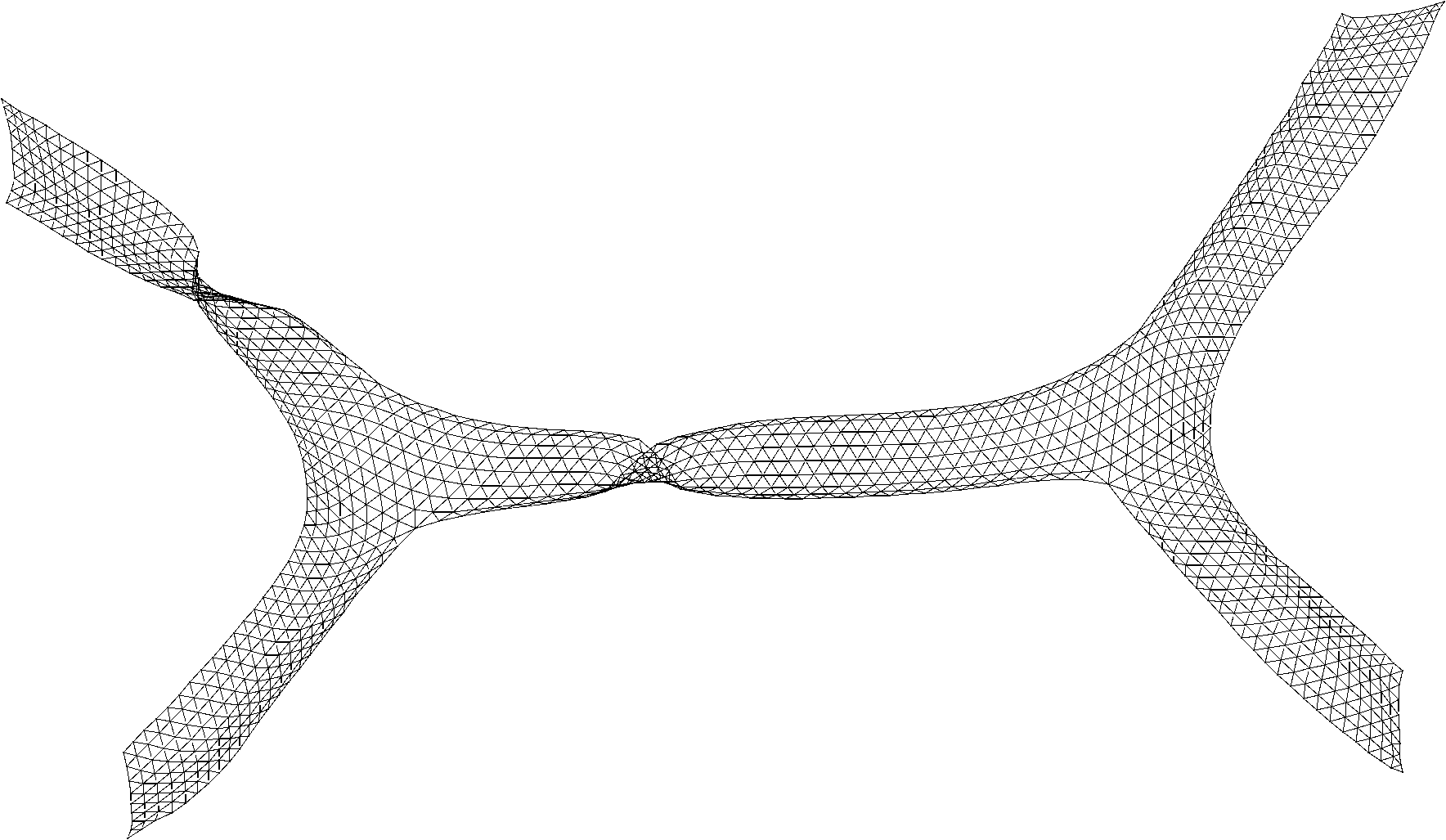} &
\includegraphics[width=0.19\textwidth]{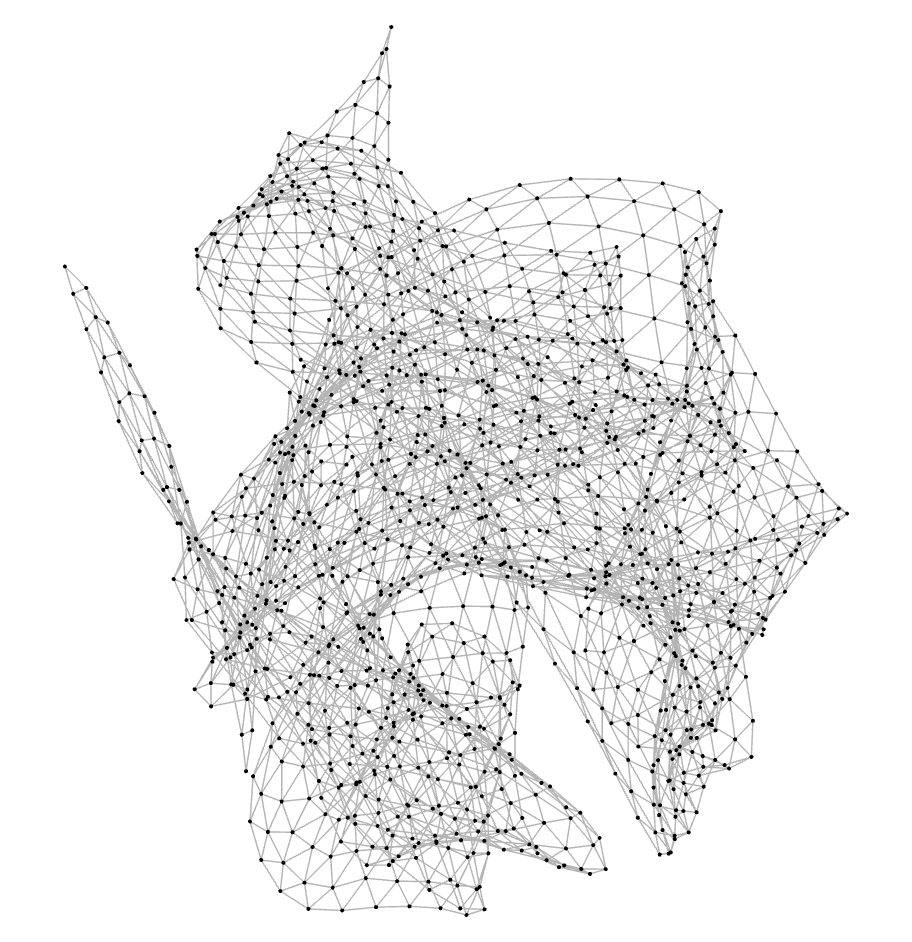} &
\includegraphics[width=0.19\textwidth]{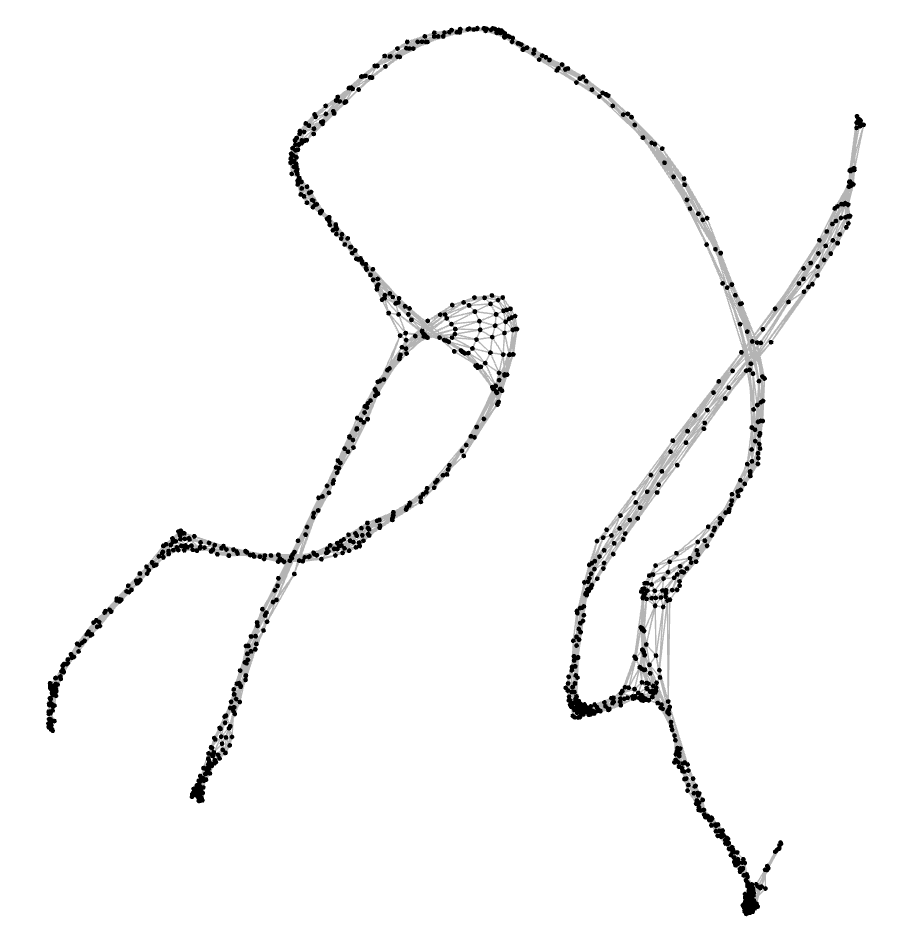} &
\includegraphics[width=0.19\textwidth]{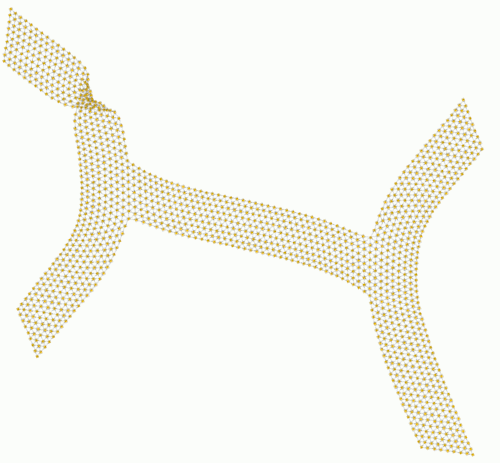} \\ \hline
$sfdp$ & $SGD$ & $SM$ & $tsNET$ & $CK$ \\ \hline
\includegraphics[width=0.19\textwidth]{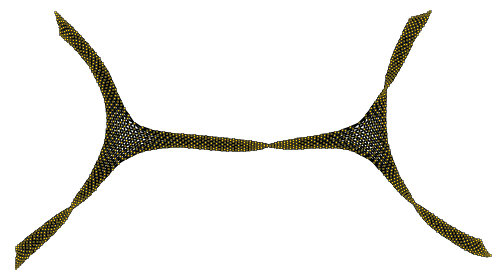} &
\includegraphics[width=0.19\textwidth]{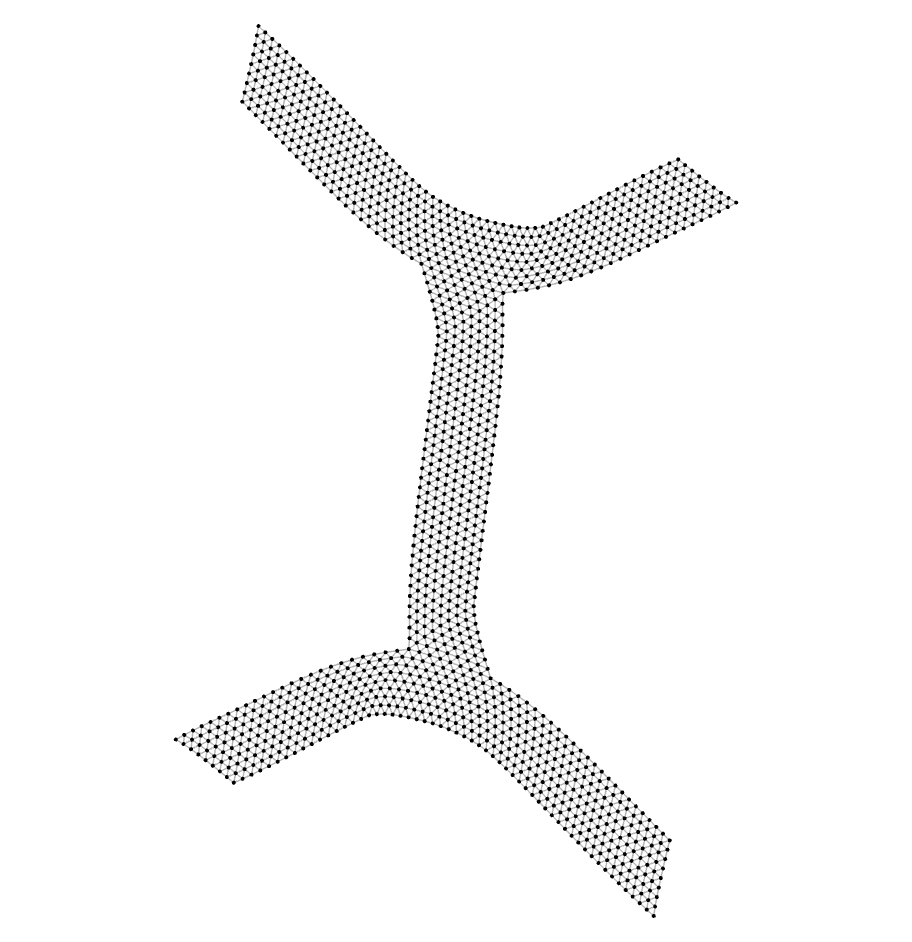} &
\includegraphics[width=0.19\textwidth]{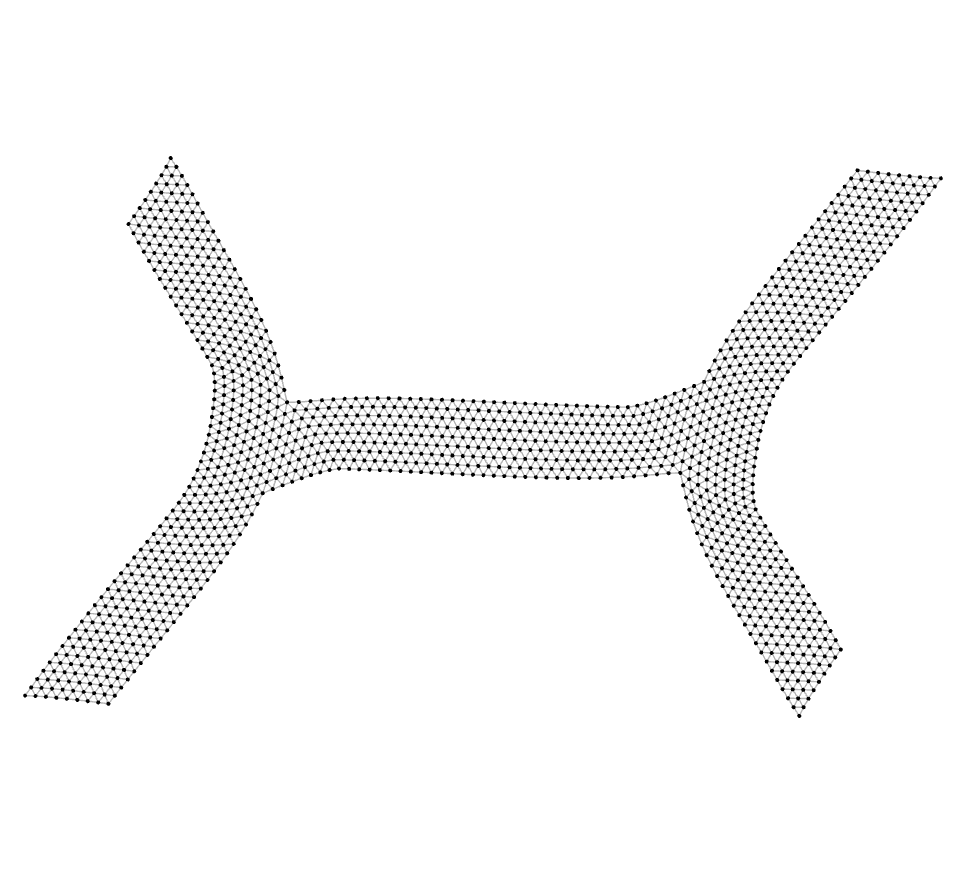} &
\includegraphics[width=0.19\textwidth]{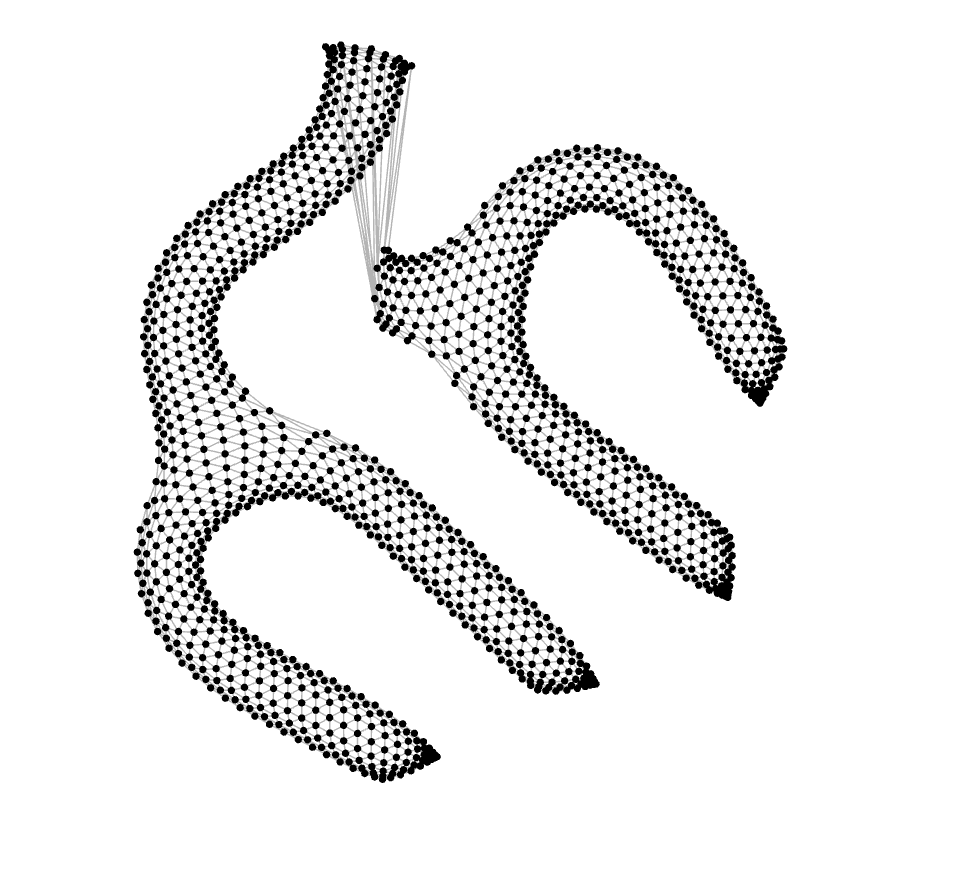} &
\includegraphics[width=0.19\textwidth]{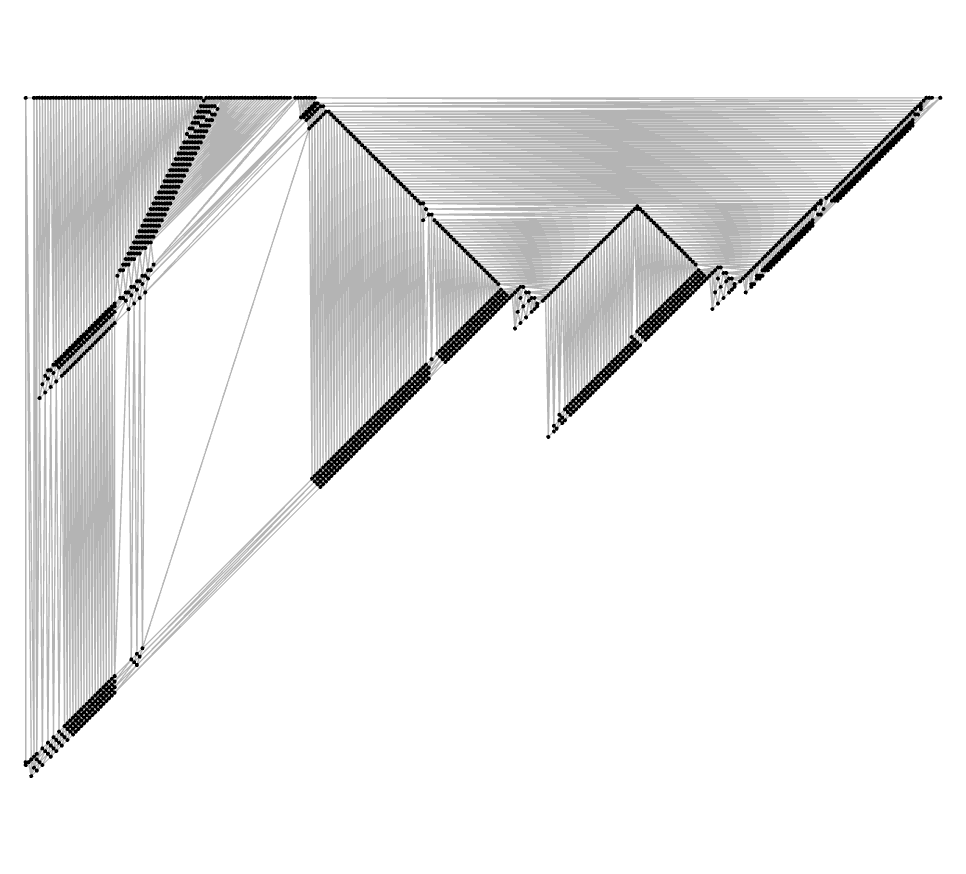} \\ \hline
\end{tabular}
\end{table}

Table \ref{table:layoutcomp_mesh} shows a visual comparison on a mesh graph; most layouts manage to untangle the mesh. Furthermore, $SGD$ and $SM$ manage to untangle without twists or ``distortions'', where triangles in the periphery are more ``squashed'' compared to the triangles in the middle, as seen in $sfdp$ or $tsNET$ layouts.

\subsection{Discussion and Summary}

Overall, $tsNET$ performs the best on large strong proximity drawable graphs, followed by $LL$. 
Looking at the visual comparison, these layouts often ``collapse'' subgraphs on the periphery. 
This may have lead to fewer non-adjacent vertices being close to each other, leading to better shape-based metrics. Moreover, this improvement compared to other layouts is more apparent in larger graphs, where the larger number of vertices means more non-adjacent vertices being close to each other in drawings where subgraphs on the periphery are not ``collapsed''.

Most layout algorithms are better at computing  drawings closer to optimal shape-faithfulness for  {\em dense} strong proximity drawable graphs: 
the best-performing layouts, $tsNET$ and $LL$, obtain much higher average shape-based metrics on outerplanar graphs compared to trees. 
Lower density means more pairs of vertices are not adjacent in $G$, i.e., more proximity regions need to be non-empty in $D$.

The mesh graphs are drawn as $RNG$ by drawing each face as equilateral triangles, i.e., having uniform edge lengths, a readability metric which is often used as a goal for a number of layout algorithms. 
This may be why more layout algorithms, especially stress-based layouts which emphasize distance faithfulness, are able to produce almost-perfect shape-faithful drawings for the mesh graphs.

\section{Algorithms for Shape-Faithful  Graph Drawings}
\label{sec:shapemod}

In this Section, we present algorithms for shape-faithful drawings. 
Based on the qualitative observations from the layout comparison experiments in Section \ref{sec:layoutcomp}, 
high shape-based metrics are obtained often by ``collapsing'' subgraphs on the drawing's periphery  - this keeps non-adjacent vertices in $G$ distant from each other, and adjacent vertices in the collapsed subgraphs within close proximity. 
Therefore, our main idea for shape-faithful graph drawings is to ``drive away'' non-adjacent vertices in $G$ that are geometrically too close in the drawing $D$.
 
Specifically, we present two algorithms $ShFR$ and $ShSM$ based on two popular graph drawing algorithms, force-directed and stress minimization algorithms. 
$ShFR$ and $ShSM$  aim to improve shape-based metrics by introducing two new types of {\em proximity forces/stress}. 
For a pair of adjacent vertices $v$ and $u$ in $G$ and another vertex of $t$ currently located in the proximity region of $v$ and $u$ in $D$:

\begin{itemize}
    \item {\em proximity repulsion  force/stress}: push $t$ out of the proximity region of $u,v$; 
    \item {\em proximity attraction  force/stress}: pull $v$ and $u$ closer together.
\end{itemize}

\subsection{$ShFR$: Force-Directed Layout for  Shape-Faithful Drawings}

We present $ShFR$, a force-directed layout for shape-faithful drawing, incorporating {\em proximity forces} with Fruchterman-Reingold ($FR$)~\cite{fruchterman1991graph}.

To explain the design rationale for $ShFR$, consider the following case: for a pair of adjacent vertices $u$ and $v$ in a graph $G=(V,E)$, the edge $(u,v)$ does not exist in the proximity graph $S=(V,E')$ of a drawing $D$ of $G$, due to a vertex $t$ located inside the proximity region of $u$ and $v$ in $D$. 
For such a case, to add back the edge $(u,v)$ in the proximity graph $S$ to achieve $S=G$,
we introduce two new {\em proximity forces}:
(1) repulsion force to repel $t$ out of the proximity region of $u$ and $v$; (2) attraction force on $u$ and $v$ to shrink the proximity region.

We first add a {\em proximity repulsion force} to drive $t$ out of the proximity region of $u$ and $v$ in $D$. 
From the midpoint $m$ between $u$ and $v$, we add a repulsion force acting on $t$, with a magnitude proportional to how far $t$ needs to be away from $m$ in order to be driven out of the proximity region of $u$ and $v$. 
Specifically, the $x$-displacement of $t$ induced by the repulsion force can be computed as: $\frac{x_t - x_m}{|| X_t - X_m ||^2} fl^2 \frac{|| X_v - X_u ||}{|| X_t - X_m ||}$, where $x_t$ is the $x$-coordinate of $t$, $|| X_t - X_m ||$ is the Euclidean distance between $t$ and $m$, $l$ is the parameter for ideal spring length (i.e., target edge length), and $f$ is the parameter for spring stiffness.

Next, we add a {\em proximity attraction force} for a pair of adjacent vertices $u$ and $v$ in $G$ with non-empty proximity regions.
Specifically, we add an attraction force acting between $u$ and $v$: $(x_u - x_v)(|| X_v - X_u ||) l^{-1}$.

The new proximity forces can be added to any force-directed algorithms.
For our specific implementation, we add the proximity forces in conjunction with $FR$, where the proximity force computations are added to each force computation iteration of $FR$. 
For details, see Algorithm \ref{alg:shfr} in Appendix \ref{sec:app_shfr}.

$GG$ and $RNG$ are subgraphs of the Delaunay Triangulation, which can be computed in $O(n \log n)$ time~\cite{preparata2012computational}. The original $FR$ algorithm runs in $O(n^2)$ time. Therefore, the total runtime of $ShFR$ is $O(n^2)$.


\subsection{$ShSM$: Stress-Based Layout for  Shape-Faithful Drawings}

We now present $ShSM$ for shape-faithful drawing, incorporating {\em proximity stress} with Stress Majorization ($SM$)~\cite{gansner2004graph}.
Similar to the force-directed case, for each case where in drawing $D$ a vertex $t$ lies in the proximity region of two neighboring vertices $v$ and $u$, i.e. $(u,v) \in E$ but $(u,v) \notin E'$, 
we add two new types of stress: (1) repulsion stress to push $t$ out of the proximity region; (2) attraction stress to pull $v$ and $u$ closer together.

We first add the \textit{proximity repulsion stress} by exerting stress on $t$ from the midpoint $m$ of $u$ and $v$.
Specifically, we compute the $x$-displacement of $t$ due to the stress between $t$ and $m$ as $w_{uv} (x_m) + d_{uv} (x_m - x_t) || X_v - X_u || / || X_t - X_m ||)$, where $d_{uv}$ is the shortest path distance between $u$ and $v$ and $w_{uv}$ is the weight computed for the vertex pair $u$ and $v$, often computed as $(d_{uv})^k$ for a constant $k$. 
Since $m$ is not an actual vertex of $G$, there is no graph theoretic distance or weight between $m$ and $t$; we instead use $d_{uv}$ and $w_{uv}$, and then scale them based on the ratio of the Euclidean distances between $u,v$, and between $t,m$.

We next add the \textit{proximity attraction stress} which has a weight lower than the standard stress of $SM$, to attract $u$ and $v$ closer in order to to reduce the distance between $u$ and $v$. 
The $x$-displacement of $v$ due to this additional stress is computed as $w'_{uv} (x_u) + d_{uv} (x_v - x_u) / || X_v - X_u ||)$, where $w'_{uv} = w_{uv}^{k'}$ for $k' < 1$.

The new proximity stress can be added to any stress-based algorithms.
For our specific implementation, we add the proximity  stress in conjunction with $SM$, where the proximity stress computations are added to each stress computation iteration of $SM$. 
For details, see Algorithm \ref{alg:shsm} in Appendix \ref{sec:app_shsm}.

As with $ShFR$, $GG$ and $RNG$ can be computed in $O(n \log n)$ time and the original stress computation of $SM$ takes $O(n^2)$ time. The total runtime of $ShSM$ is therefore $O(n^2)$.

\section{$ShFR$ and $ShSM$ Experiments}

\subsection{Experiment Design and Data Sets}

In this experiment, we evaluate the effectiveness of  $ShFR$ and $ShSM$ over  $FR$ and $SM$ respectively, using shape-based metrics $Q_{RNG}$ and $Q_{GG}$. 

For data sets, we use {\em strong proximity drawable} graphs, as well as {\em scale-free} graphs and {\em benchmark} graphs: 

\begin{itemize}
    \item {\em strong proximity drawable} graphs, from Section \ref{sec:layoutcomp}.
    
    \item {\em scale-free} graphs: We generate synthetic scale-free graphs with density 2, 3, and 5, using the NetworkX~\cite{hagberg2008exploring} scale-free generator. 
    \item {\em benchmark} graphs, including  real-world scale-free graphs~\cite{davis2011university,snapnets,biosnapnets} with up to 6000 vertices and 15000 edges. For details, see Table \ref{tab:dataset_sh} in Appendix \ref{sec:app_dataset}.
\end{itemize}

To measure the improvement of the shape-based metrics, 
for example, on $Q_{RNG}$ by $ShFR$ over $FR$, 
we define the formula $I(Q_{RNG}) = \frac{Q_{RNG}(ShFR)-Q_{RNG}(FR)}{Q_{RNG}(FR)}$. We use the same formula for $Q_{GG}$, 
and for the improvement by $ShSM$ over $SM$.

\subsection{Results}

\begin{figure}[h!]
    \centering
    \subfloat[$ShFR$ improvement]{
        \includegraphics[width=0.47\columnwidth]{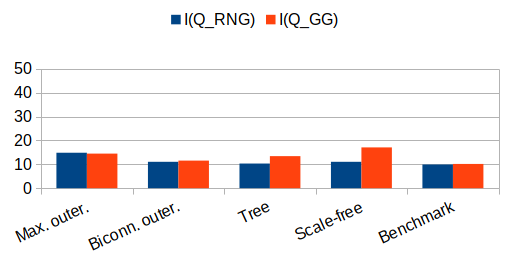}
    }
    \subfloat[$ShSM$ improvement]{
        \includegraphics[width=0.45\columnwidth]{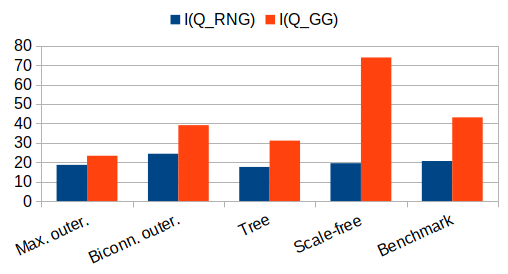}
    }
    \caption{Average shape-based metrics improvement (in percent) of $ShFR$ over $FR$ and $ShSM$ over $SM$ on $Q_{RNG}$ and $Q_{GG}$. $ShFR$ and $ShSM$ obtain significant improvement over $FR$ and $SM$ respectively on all data sets.}
    \label{fig:shfr_metrics}
\end{figure}

$ShFR$ obtains notable improvement over $FR$ on $Q_{RNG}$ and $Q_{GG}$ for large strong proximity drawable graphs, obtaining average improvement of 15\%, 12\%, and 12\% on maximum outerplanar graphs, biconnected outerplanar graphs, and trees respectively, see Figure \ref{fig:shfr_metrics} (a). 
$ShFR$ also obtains significant improvement over $FR$ on $Q_{GG}$ for scale-free graphs, at on average 18\%. 
For real-world benchmark graphs, the improvement on $Q_{RNG}$ and $Q_{GG}$ average at around 10\%.

\begin{table}[h!]
    \centering
    \caption{Visual comparison of $FR$ and $ShFR$, $SM$ and $ShSM$ on benchmark graphs. $ShFR$ often untangles the hairballs better than $FR$, and $ShSM$ expands faces that are ``collapsed'' by  $SM$.}
    \begin{tabular}{|c|c|}
    \hline
    $FR$& $ShFR$ \\ \hline
    \multicolumn{2}{|c|}{$G\_4$}  \\ \hline
    \includegraphics[width=0.4\textwidth]{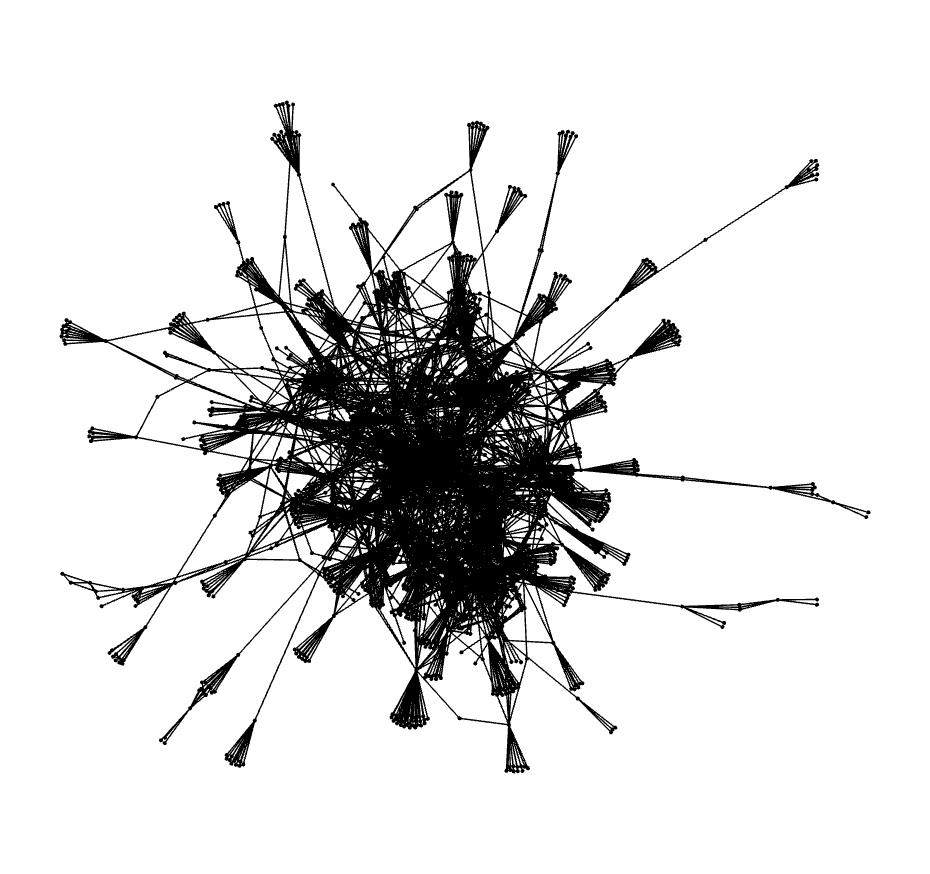} &
    \includegraphics[width=0.4\textwidth]{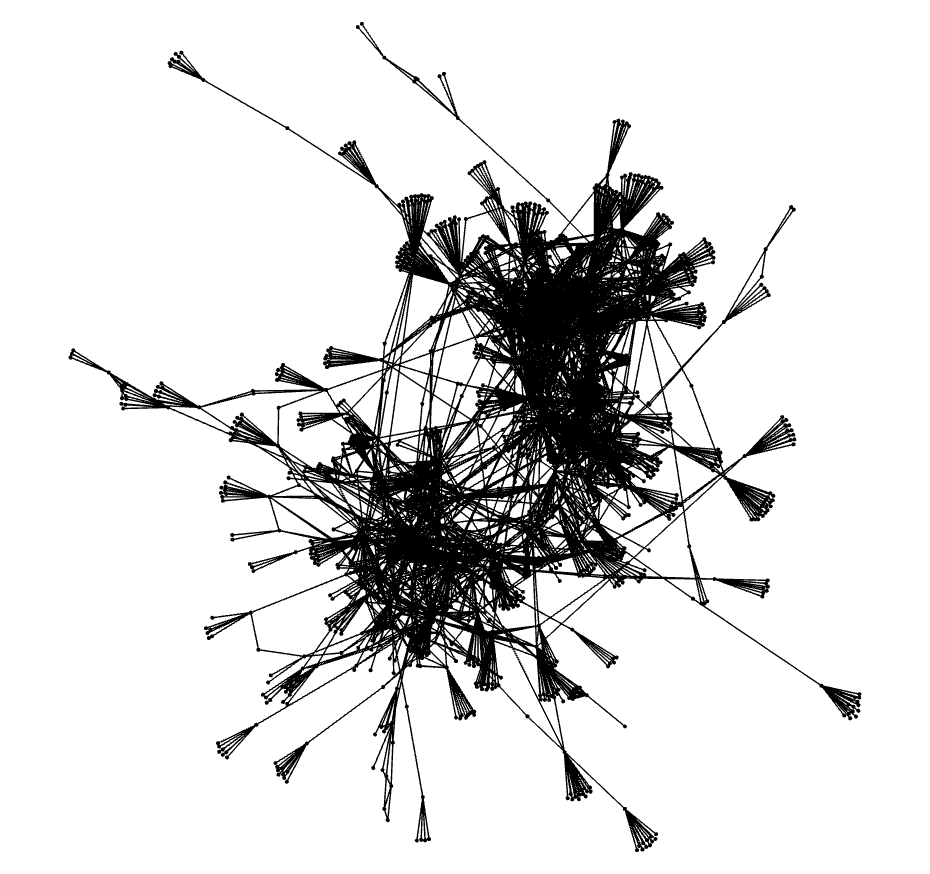} \\ \hline
    $SM$ & $ShSM$ \\ \hline
    \multicolumn{2}{|c|}{$netscience$}  \\ \hline
    \includegraphics[width=0.4\textwidth]{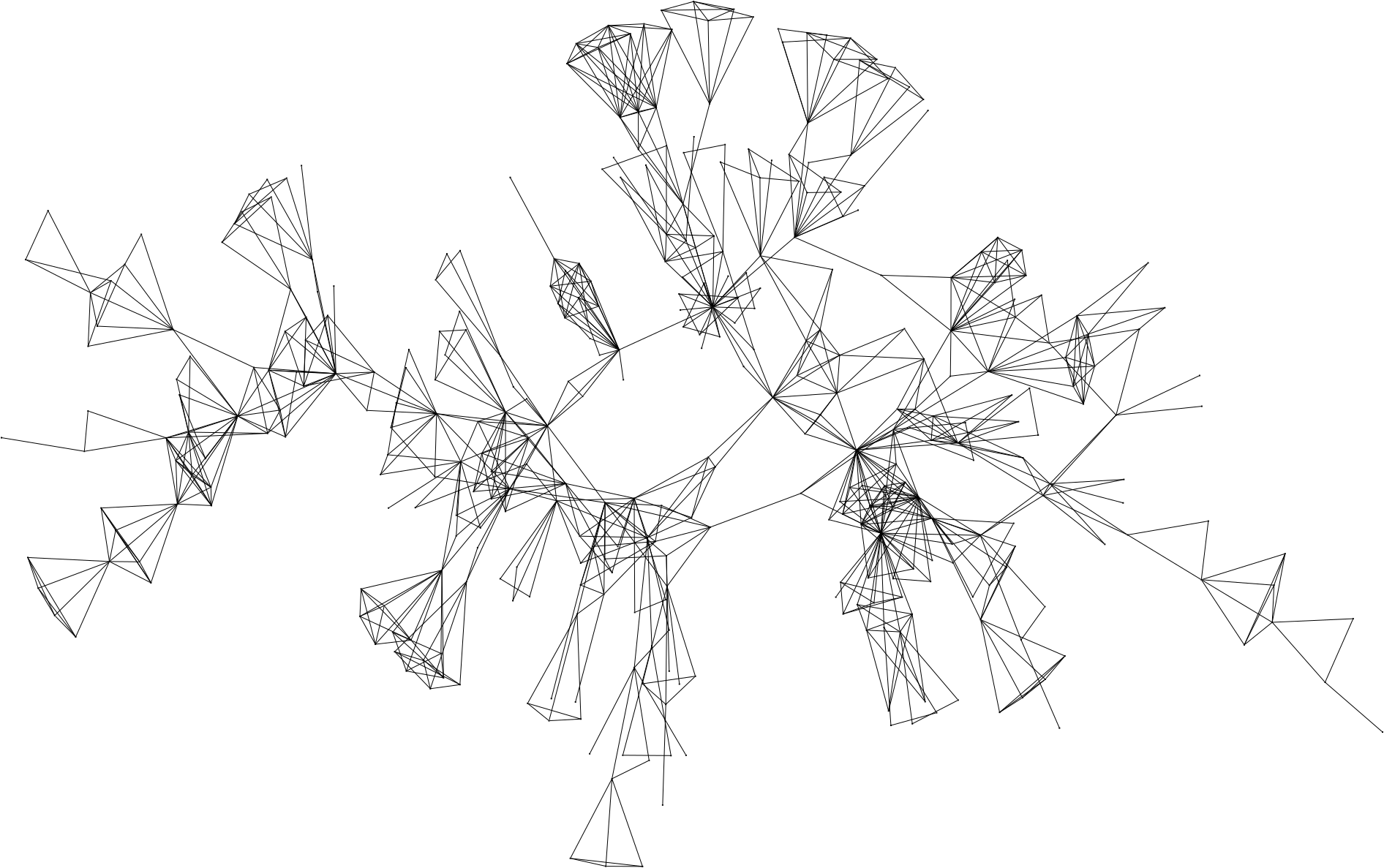} &
    \includegraphics[width=0.4\textwidth]{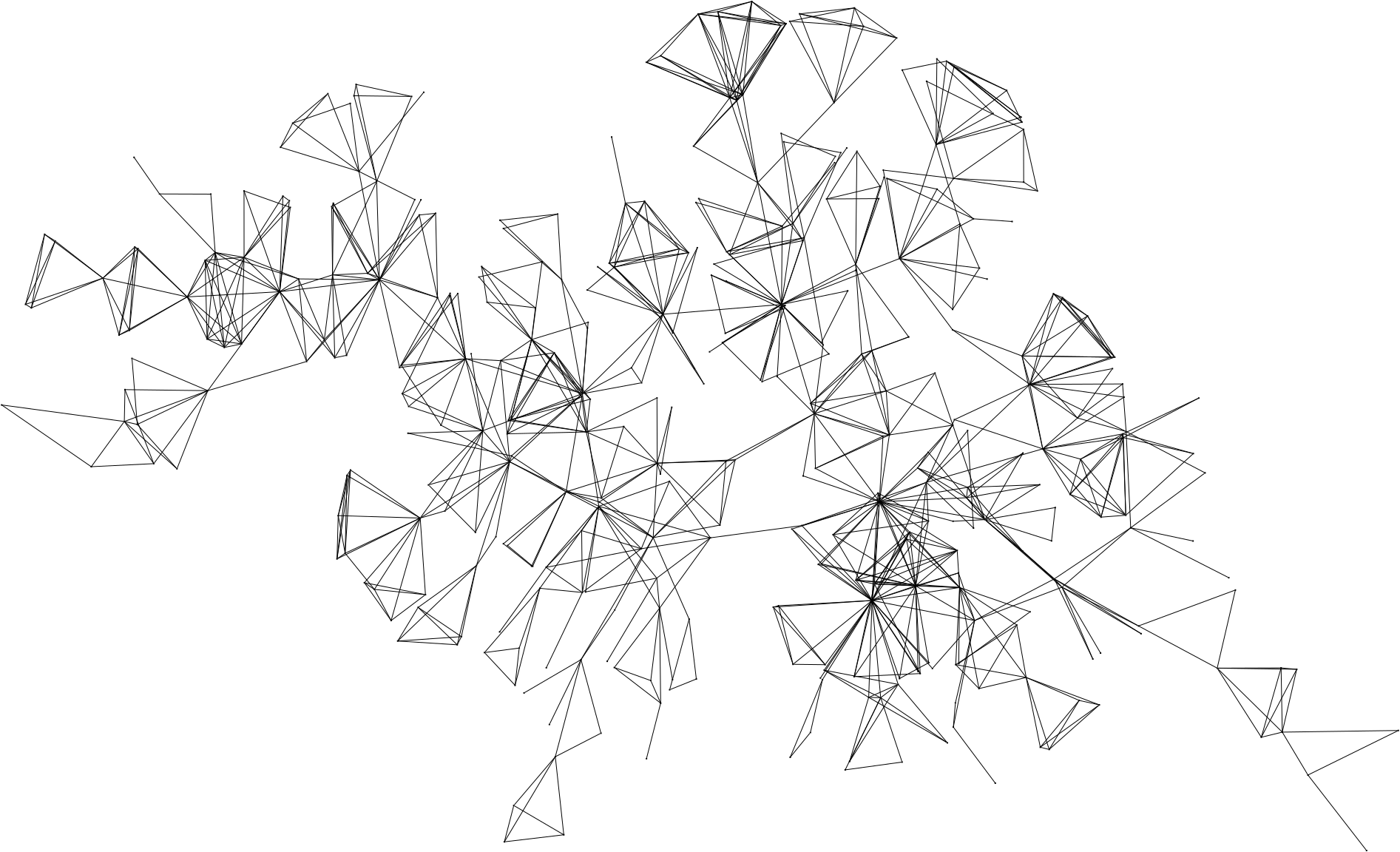} \\ \hline
    \end{tabular}
    \label{table:layoutcomp_shfr}
\end{table}

$ShSM$ obtains significant improvement over $SM$ for strong proximity drawable graphs, see Figure \ref{fig:shfr_metrics} (b). 
For maximum outerplanar graphs, $ShSM$ obtains significant improvement over $SM$ (average 20\% and 25\%) on $Q_{RNG}$ and $Q_{GG}$ respectively, which is much higher than the improvement by $ShFR$ over $FR$.
For biconnected outerplanar graphs,
an even larger improvement of on average 40\% is achieved on $Q_{GG}$.
For large trees, $ShSM$ also obtains significant improvement over $SM$, on average 18\% and 30\%  on $Q_{RNG}$ and $Q_{GG}$, respectively.

$ShSM$ also obtains significant improvement over $SM$ for scale-free graphs, on average 20\% improvement on $Q_{RNG}$. 
Notably, the largest improvement is obtained by $ShSM$ on $Q_{GG}$ for scale-free graphs, at over 70\%, on scale-free graphs.
Note that $ShSM$ obtains on average 20\% and 42\% improvement over $SM$ for real-world benchmark graphs, on $Q_{RNG}$ and $Q_{GG}$ respectively.

Table \ref{table:layoutcomp_shfr} shows a visual comparison of $FR$ and $ShFR$ on the benchmark scale-free graph $G\_4$. 
$ShFR$ untangles the ``hairball'' more clearly, compared to $FR$. 
Table \ref{table:layoutcomp_shfr} also shows a visual comparison of $SM$ and $ShSM$ on the benchmark scale-free graph $netscience$. 
$ShSM$ ``expands'' faces that are ``squashed'' in $SM$, showing the local neighborhood of some vertices more clearly. However, the expanded faces also leads to the drawing feeling more ``crowded'' compared to $SM$, thus increasing faithfulness but affecting readability. 
For more visual comparisons on other data sets, see Table \ref{table:layoutcomp_shfr_extra} in Appendix \ref{sec:app_shfrviscomp}.

\subsection{Discussion and Summary}

Our extensive experiments demonstrate the effectiveness of $ShFR$ and $ShSM$ for shape-faithful drawings.
$ShFR$ (resp., $ShSM$) obtains significant improvement over $FR$ (resp., $SM$) of 11\% and 13\% (resp., 20\% and 50\%) on $Q_{RNG}$ and $Q_{GG}$ respectively, averaged over all data sets.

For  strong proximity drawable graphs, $ShFR$ (resp., $ShSM$) obtains improvement over $FR$ (resp., $SM$) of on average 13\% and 13\% (resp., 20\% and 30\%) on $Q_{RNG}$ and $Q_{GG}$ respectively. 
For real-world benchmark graphs, $ShFR$ (resp., $ShSM$) obtains improvement over $FR$ (resp., $SM$) of on average 10\% and 10\% (resp., 20\% and 43\%) on $Q_{RNG}$ and $Q_{GG}$ respectively. 
For scale-free graphs, $ShFR$ (resp., $ShSM$) obtains improvement over $FR$ (resp., $SM$) of on average 10\% and 16\% (resp., 17\% and 70\%) on $Q_{RNG}$ and $Q_{GG}$ respectively.  Notably, the $Q_{GG}$ improvement of $ShSM$ over $SM$ on scale-free graphs at 70\% is the largest among all data sets.

The improvements are much higher for {\em large}  graphs.
In general, large graphs have many vertex pairs,
with a high ratio of non-adjacent vertices to adjacent pairs of vertices in $G$.
Therefore, there are potentially more vertices located in proximity region that should be empty, 
creating more instances for the proximity forces and stress to improve the shape-based metrics.

Furthermore, the best improvement is achieved by $ShSM$ over $SM$ on $Q_{GG}$, significantly higher than the improvement on $Q_{RNG}$ and the improvements of $ShFR$ over $FR$. 
Specifically, larger improvements are obtained on $Q_{GG}$ than $Q_{RNG}$ on scale-free and real-world benchmark graphs by $ShSM$.
Since the proximity region of $RNG$ (i.e., lens at points $u$ and $v$) is larger than the proximity region of $GG$ (i.e., disk with $uv$ as diameter), when applying proximity stress, it is harder to push all non-adjacent vertices out of the proximity region of $RNG$. In addition, the tendency for $ShSM$ to ``open up'' collapsed faces compared to $ShFR$ may have led to the better improvements obtained by $ShSM$.

\section{Conclusion and Future Work}
\label{sec:conclusion}

In this paper, we present the first study for the shape-faithful drawings of general graphs.
We first evaluate the shape-faithfulness of existing graph layouts and examine the properties of good shape-faithful drawings. 
In general, tsNET obtains the highest shape-faithfulness on medium-to-large graphs.

We then present $ShFR$ and $ShSM$, algorithms for shape-faithful drawings of general graphs, based on force-directed and stress-based layouts, introducing new proximity forces/stress.  
Extensive experiments show that $ShFR$ and $ShSM$ achieve significant improvement over $FR$ and $SM$, on average, 12\% and 35\% higher shape-based metrics respectively. 
Notably, $ShSM$ obtains a 70\% average improvement on $Q_{GG}$ over $SM$ for scale-free graphs.

Future work includes shape-faithful layouts based on various other layouts.

\newpage

\bibliographystyle{splncs04}
\bibliography{proxdraw_GD}

\begin{subappendices}
\renewcommand{\thesection}{\Alph{section}}

\newpage
\section{Data Sets}
\label{sec:app_dataset}
\begin{table}[]
    \centering
    \caption{Data set details for layout comparison experiments. Note that while the mesh graphs we use do not fall under known proximity drawability characterizations, they can be drawn as $RNG$ by drawing each 3-cycle as an equilateral triangle.}
    \begin{tabular}{|c|c|c|c|c|}
    \hline
        Name & $RNG$-drawable & $GG$-drawable  & Weak $GG$-drawable & Avg. density \\ \hline
        Tree (max deg. 5) & Y & N  & Y & 0.99 \\ \hline
        Tree (non-forbidden max deg. 4) & Y & Y & Y & 0.99 \\ \hline
        Max. outerplanar & Y & Y & Y & 2.01 \\ \hline
        Biconn. outerplanar & Y & Y & Y & 1.40 \\ \hline
        1-conn. outerplanar & N & N & Y & 1.50 \\ \hline
        \texttt{L-AUG} & N & N & N & 1.01 \\ \hline
        \texttt{F-AUG} & N & N & N & 1.01 \\ \hline
        Mesh & N* & N & N & 2.85 \\ \hline
    \end{tabular}
    \label{tab:dataset_layoutcomp}
\end{table}

\begin{table}[]
    \centering
    \caption{Benchmark data set details for $ShFR$ and $ShSM$ experiments}
    \begin{tabular}{|c|c|c|c|}
    \hline
    Name & $|V|$ & $|E|$ & density \\ \hline
    as19990606 & 5188 & 9930 & 1.91 \\ \hline
    migrations\_lcc & 6025 & 9378 & 1.56 \\ \hline
    netscience & 379 & 914 & 2.41 \\ \hline
    oflights\_lcc & 2905 & 15645 & 5.39 \\ \hline
    tvcg & 3213 & 10140 & 3.16 \\ \hline
    us\_powergrid & 4941 & 6594 & 1.33 \\ \hline
    yeastppi\_lcc & 2224 & 6609 & 2.97\\ \hline
    1138\_bus & 1138 & 2596 & 2.28 \\ \hline
    add32 & 4960 & 14422 & 2.91 \\ \hline
    eva & 4475 & 4652 & 1.04 \\ \hline
    G\_4 & 2075 & 4769 & 2.30 \\ \hline

    \end{tabular}
    \label{tab:dataset_sh}
\end{table}

\newpage
\section{Graph Layout Comparison Results}
\label{sec:app_layoutcomp}

\begin{figure}[h]
\centering
\subfloat[Small]{
\includegraphics[width=0.28\columnwidth]{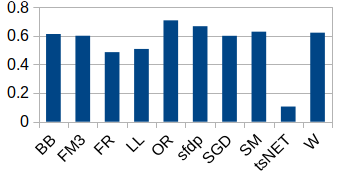}
}
\subfloat[Medium]{
\includegraphics[width=0.28\columnwidth]{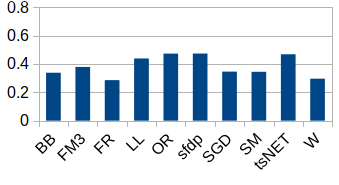}
}
\subfloat[Large]{
\includegraphics[width=0.28\columnwidth]{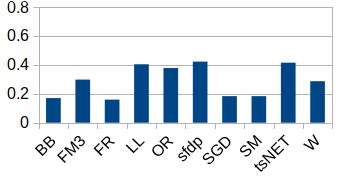}
}
\caption{Average $Q_{GG}$ for strong $GG$-drawable trees. The pattern in best-performing drawing algorithms are similar with $RNG$-drawable trees on medium and large trees. Even highest-performing layouts are still far from ideal ($Q_{GG}$ = 1).}
\label{fig:layoutcompmetrics_ggtree}
\end{figure}

\begin{figure}[h]
\centering
\subfloat[Small]{
\includegraphics[width=0.28\columnwidth]{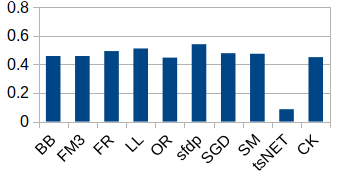}
}
\subfloat[Medium]{
\includegraphics[width=0.28\columnwidth]{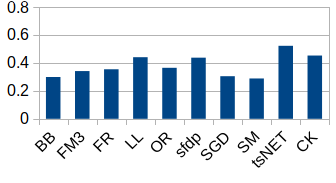}
}
\subfloat[Large]{
\includegraphics[width=0.28\columnwidth]{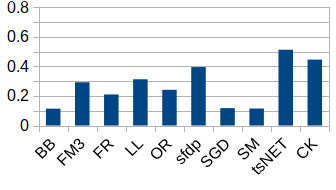}
}
\caption{Average $Q_{GG}$ for maximum outerplanar graphs. The ordering between layouts is mostly the same as with $Q_{RNG}$, with $tsNET$ and $CK$ obtaining the best $Q_{GG}$ on medium and large graphs.}
\label{fig:layoutcompmetrics_maxouterplanar_gg}
\end{figure}

\newpage
\section{$ShFR$ Algorithm}
\label{sec:app_shfr}

\begin{algorithm}
 \KwIn{Graph $G = (V,E)$, \# of iterations $k$}
    \Repeat ( $k$ times ){
    // Initialize displacement \\
    \For{$u \in V$}{
        $x'_u, y'_u = 0$\;
    }
    // $FR$ repulsion force \\
    \For{$u \in V$}{
        \For{$v \in V, v \neq u$}{
            $x'_u += \frac{x_v - x_u}{|| X_v - X_u ||^2} fl^2$\;
            $y'_u += \frac{x_v - x_u}{|| X_v - X_u ||^2} fl^2$\;
        }
    }
    // $FR$ attraction force \\
    \For{$e = (u,v) \in E$}{
        $x'_u -= (x_u - x_v)(|| X_v - X_u ||) / l$\;
        $y'_u -= (y_u - y_v)(|| X_v - X_u ||) / l$\;
        $x'_v += (x_u - x_v)(|| X_v - X_u ||) / l$\;
        $y'_v += (y_u - y_v)(|| X_v - X_u ||) / l$\;
    }
   // Update coordinates of vertices in $V$ according to the displacement \\
    \For{$u \in V$}{
        $x_u += x'_u$\;
        $y_u += y'_u$\;
    }
    Compute proximity graph $S = (V, E')$\;
    // Initialize displacement \\
    \For{$u \in V$}{
        $x'_u, y'_u = 0$\;
    }
    // $ShFR$ proximity forces \\
    \For{$e = (u,v) \in E \setminus E'$}{
        $m$: mid-point of $u$ and $v$\;
    
        // $ShFR$ proximity repulsion force \\
        \For{$t \in V$ where $(u,t) \in E' \setminus E$ or $(v,t) \in E' \setminus E$}{
        $x'_t += \frac{x_t - x_m}{|| X_t - X_m ||^2} fl^2 \frac{|| X_v - X_u ||}{|| X_t - X_m ||}$\;
        $y'_t += \frac{y_t - y_m}{|| X_t - X_m ||^2} fl^2 \frac{|| X_v - X_u ||}{|| X_t - X_m ||}$\;
        }
        // $ShFR$ proximity attraction force \\
        $x'_u -= (x_u - x_v)(|| X_v - X_u ||) / 2l$\;
        $y'_u -= (y_u - y_v)(|| X_v - X_u ||) / 2l$\;
        $x'_v += (x_u - x_v)(|| X_v - X_u ||) / 2l$\;
        $y'_v += (y_u - y_v)(|| X_v - X_u ||) / 2l$\;
        
    }
    // Update coordinates of vertices in $V$ according to the displacement \\ 
    \For{$u \in V$}{
        $x_u += x'_u$\;
        $y_u += y'_u$\;
    }
}
 \caption{\textbf{ShFR}}
 \label{alg:shfr}
\end{algorithm}

Before explaining our algorithm, we first provide an explanation of force-directed methods.
Force-directed algorithms model a graph as a system of bodies with two types of forces acting between them: a \textit{repulsion} force for each pair of vertices, and an \textit{attraction} force for each edge. 

For two vertices $u, v$, the \textit{x-displacement} $x'_v$ of $v$ induced by the repulsion force exerted by $u$ on $v$ can be computed as $\frac{x_v - x_u}{|| X_v - X_u ||^2} fl^2$, where $x_v$ is the $x$-coordinate of $v$, $|| X_v - X_u ||$ is the Euclidean distance between $u$ and $v$, $l$ is a parameter representing natural spring length (i.e., the target edge length), and $f$ is a parameter for spring stiffness. 

For two adjacent vertices $u, v$ (i.e., edge $(u,v)$ in $G$), the $x$-displacement of $v$ induced by the attraction force exerted by $u$ on $v$ can be computed as $(x_u - x_v)(|| X_v - X_u ||)  l^{-1}$.

\medskip
Algorithm \ref{alg:shfr} describes the details of $ShFR$ using the pseudo code.

\section{$ShSM$ Algorithm}
\label{sec:app_shsm}

Before explaining our algorithm, we first explain stress-based algorithms using Stress Majorization ($SM$), a popular stress-based algorithm. 
Stress-based algorithms aim to minimize the stress in a drawing, where low stress means that the Euclidean distances in the drawing are proportional to the graph-theoretic distances in a graph.  

For two vertices $v$ and $u$, the {\em stress} between the two vertices is defined by how proportional the Euclidean distances between the two vertices in $D$ is to the length of the shortest path between the two vertices in $G$. 
More precisely, the $x$-displacement $x'_v$ of $v$ induced by the stress between $v$ and $u$ can be computed as $w_{uv} (x_u) + d_{uv} (x_v - x_u) / || X_v - X_u ||)$, where $d_{uv}$ is the graph-theoretic distance between $v$ and $u$, and $w_{uv}$ is a weight for the pair $u$ and $v$, defined as $(d_{uv})^{-2}$ for $SM$.

\medskip
Algorithm \ref{alg:shsm} describes the details of $ShSM$ using the pseudo code.

\begin{algorithm}
 \KwIn{Graph $G = (V,E)$, \# of iterations $k$}
 $d_{uv} \leftarrow ShortestPaths(G)$\;
 Compute weights $w_{uv}$\;
 Compute initial layout $D$ of $G$ using PivotMDS\;

    \Repeat ( $k$ times ){
    // $SM$ stress \\
    \For{$u \in V$}{
        // Initialize weight sum and displacement \\
        $W_u = 0$\;
        $x'_u, y'_u = 0$\;
        
        // Stress minimization computation \\
        \For{$v \in V, v \neq u$}{
            $x'_u += w_{uv} (x_u) + d_{uv} (x_v - x_u) / || X_v - X_u ||)$\;
            $y'_u += w_{uv} (y_u) + d_{uv} (y_v - y_u) / || X_v - X_u ||)$\;
            $W_u += w_{uv}$\;
        }
        
        // Update coordinates of vertices in $V$ according to the displacement \\
        $x_u = x'_u / W_u$\;
        $y_u = y'_u / W_u$\;
    }
    Compute proximity graph $S=(V,E')$\;
    // Initialize weight sum and displacement \\
    \For{$u \in V$}{
        $W_u = 0$\;
        $x'_u, y'_u = 0$\;
    }

    // $ShSM$ proximity stress \\
    
    \For{$e = (u,v) \in E \setminus E'$}{
        $m$: mid-point of $u$ and $v$\;
    
                // $ShSM$ proximity repulsion stress \\
        \For{$t \in V$ where $(u,t) \in E' \setminus E$ or $(v,t) \in E' \setminus E$}{
                $x'_t += w_{uv} (x_m) + d_{uv} (x_m - x_t) || X_v - X_u || / || X_t - X_m ||)$\;
                $y'_t += w_{uv} (y_m) + d_{uv} (y_m - y_t) || X_v - X_u || / || X_t - X_m ||)$\;
                $W_t += w_{uv}$\;
        }
                // $ShSM$ proximity attraction stress \\
                $x'_u += (w_{uv})^k (x_u) + d_{uv} (x_v - x_u) / || X_v - X_u ||)$\;
                $y'_u += (w_{uv})^k (y_u) + d_{uv} (y_v - y_u) / || X_v - X_u ||)$\;
                $W_u += (w_{uv})^k$\;
                $x'_v += (w_{uv})^k (x_v) + d_{uv} (x_u - x_v) / || X_v - X_u ||)$\;
                $y'_v += (w_{uv})^k (y_v) + d_{uv} (y_u - y_v) / || X_v - X_u ||)$\;
                $W_v += (w_{uv})^k$\;
        
    }
    
                
    // Update coordinates of vertices in $V$ according to the displacement \\
    \For{$u \in V$}{
        $x_u = x'_u / W_u$\;
        $y_u = y'_u / W_u$\;
        
    }
}
 \caption{\textbf{ShSM}}
 \label{alg:shsm}
\end{algorithm}

\clearpage

\section{$ShFR$ and $ShSM$: Visual Comparison}
\label{sec:app_shfrviscomp}

\begin{table}[h]
\centering
\caption{Visual comparison of $FR$ and $ShFR$, $SM$ and $ShSM$ on strong proximity drawable graph classes and synthetic scale-free graphs. $ShFR$ manages to untangle strong proximity drawable graphs better than $FR$. Meanwhile, $ShSM$ manages to ``open up'' the faces collapsed by $SM$ and highlights dense areas in scale-free graphs better.}
\label{table:layoutcomp_shfr_extra}
\begin{tabular}{|c|c|c|c|c|c|}
\hline
$FR$& $ShFR$ & $FR$ & $ShFR$ & $FR$ & $ShFR$ \\ \hline
\multicolumn{2}{|c|}{Max. outerplanar} & \multicolumn{2}{|c|}{Biconn. outerplanar} & \multicolumn{2}{|c|}{Tree} \\ \hline
\includegraphics[width=0.16\textwidth]{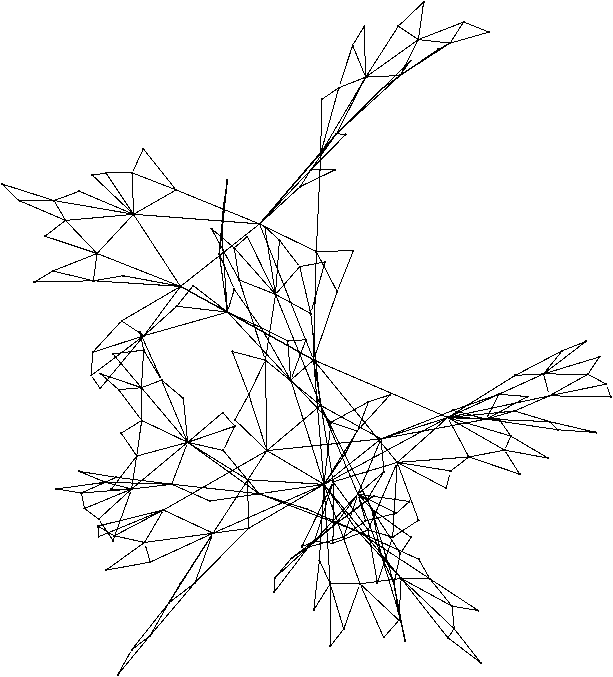} &
\includegraphics[width=0.16\textwidth]{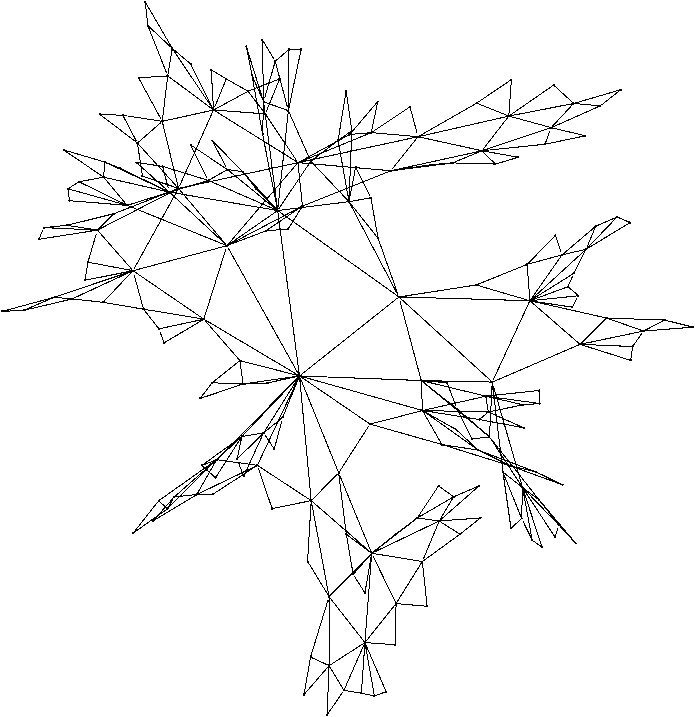} &
\includegraphics[width=0.16\textwidth]{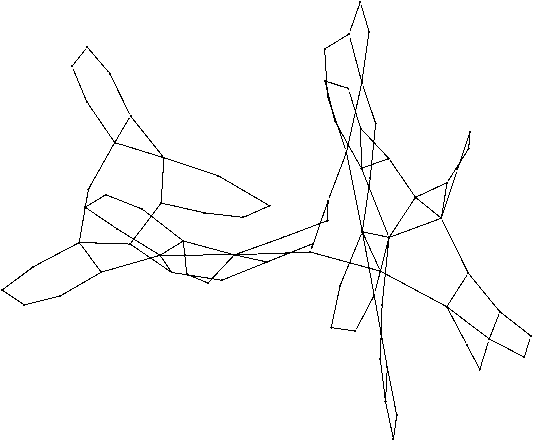} &
\includegraphics[width=0.16\textwidth]{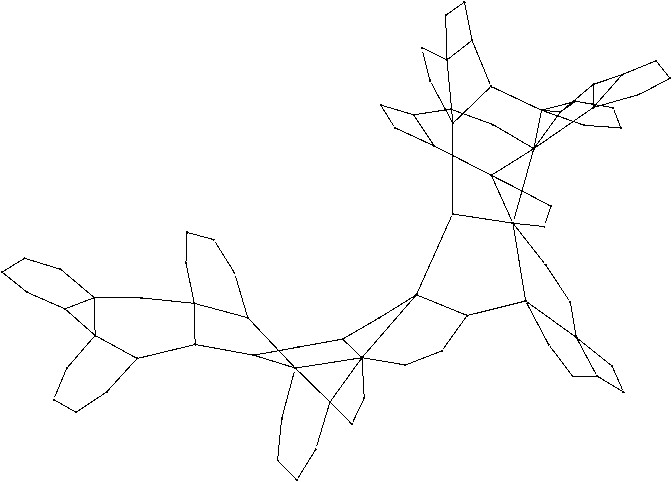} &
\includegraphics[width=0.16\textwidth]{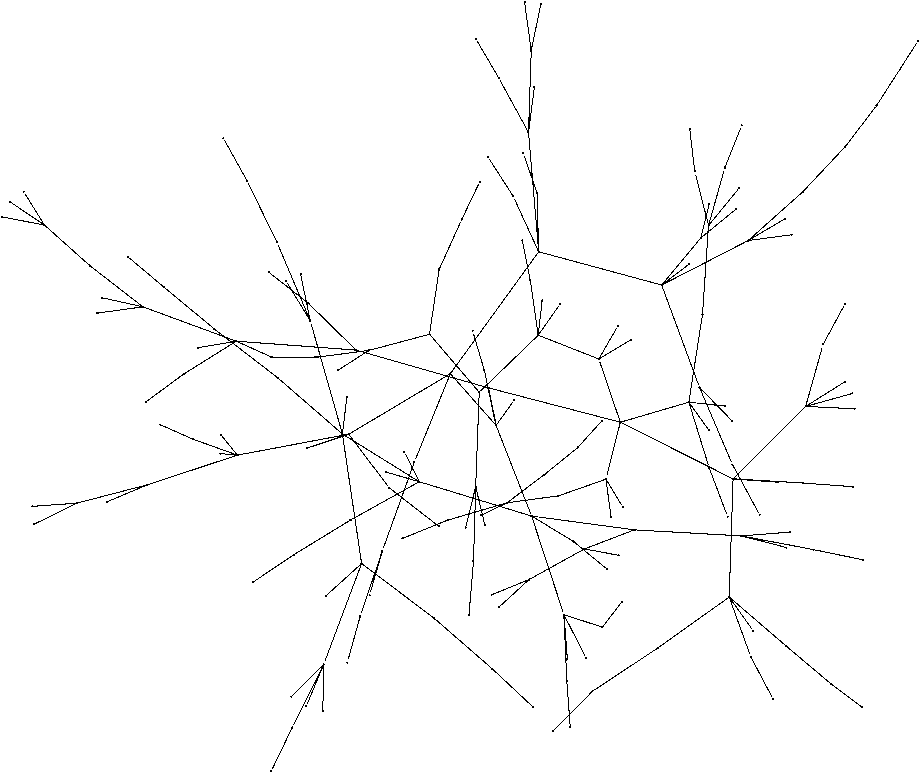} &
\includegraphics[width=0.16\textwidth]{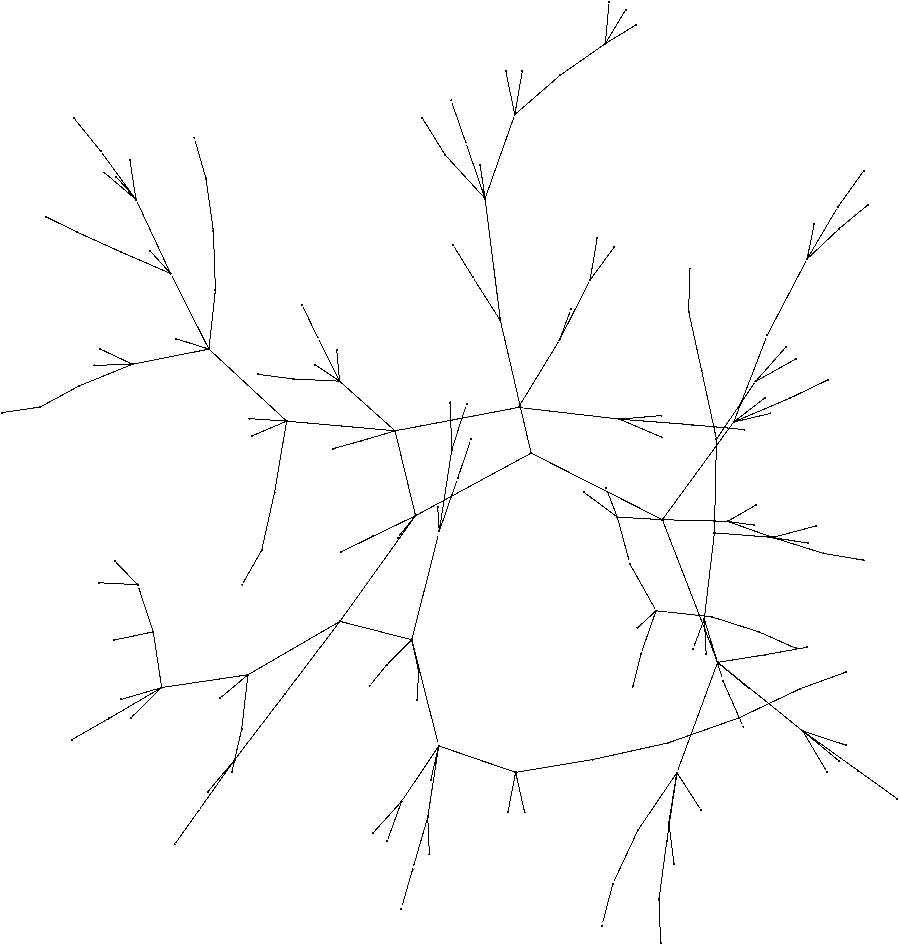} \\ \hline
\multicolumn{2}{|c|}{Scale-free ($d = 2$)} & \multicolumn{2}{|c|}{Scale-free ($d = 3$)} & \multicolumn{2}{|c|}{Scale-free ($d = 5$)} \\ \hline
\includegraphics[width=0.16\textwidth]{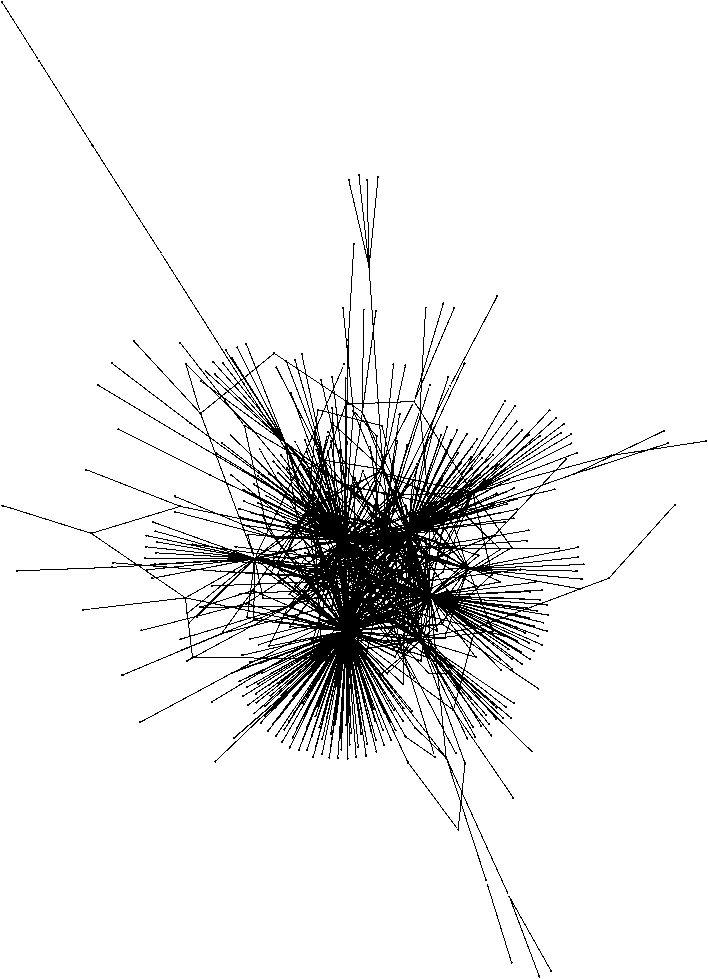} &
\includegraphics[width=0.16\textwidth]{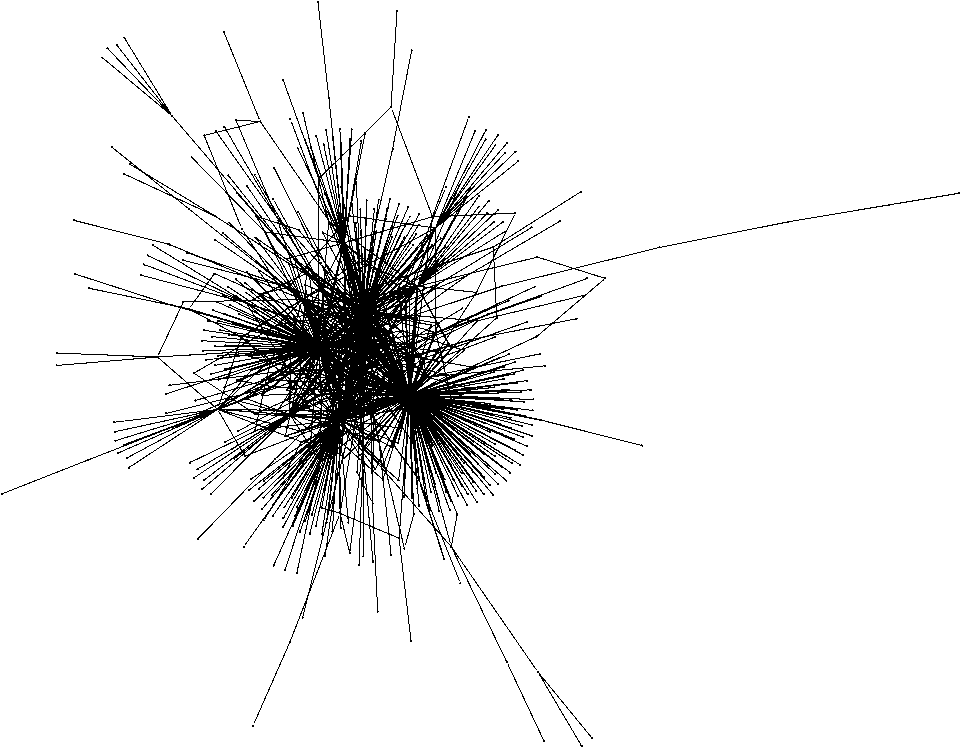} &
\includegraphics[width=0.16\textwidth]{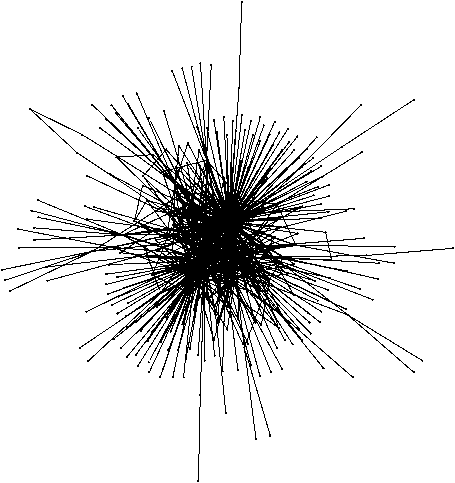} &
\includegraphics[width=0.16\textwidth]{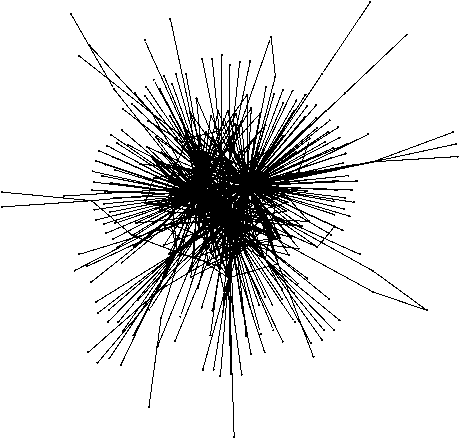} &
\includegraphics[width=0.16\textwidth]{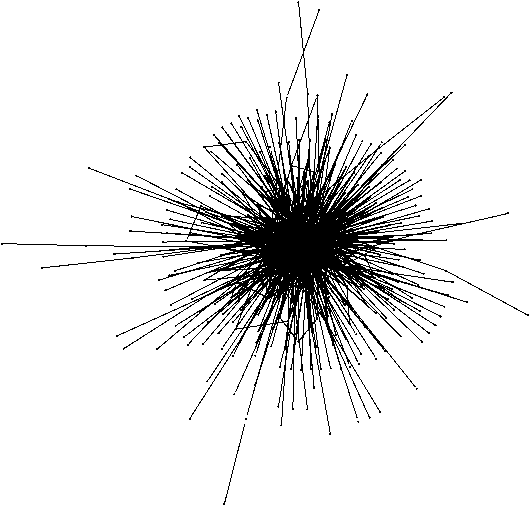} &
\includegraphics[width=0.16\textwidth]{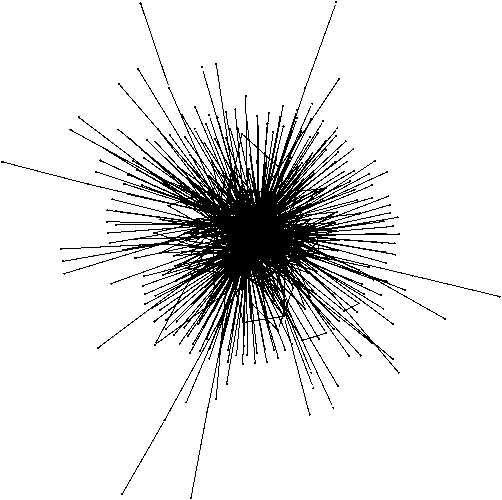} \\ \hline
$SM$ & $ShSM$ &$SM$ & $ShSM$ &$SM$ & $ShSM$ \\ \hline
\multicolumn{2}{|c|}{Max. outerplanar} & \multicolumn{2}{|c|}{Biconn. outerplanar} & \multicolumn{2}{|c|}{Tree} \\ \hline
\includegraphics[width=0.16\textwidth]{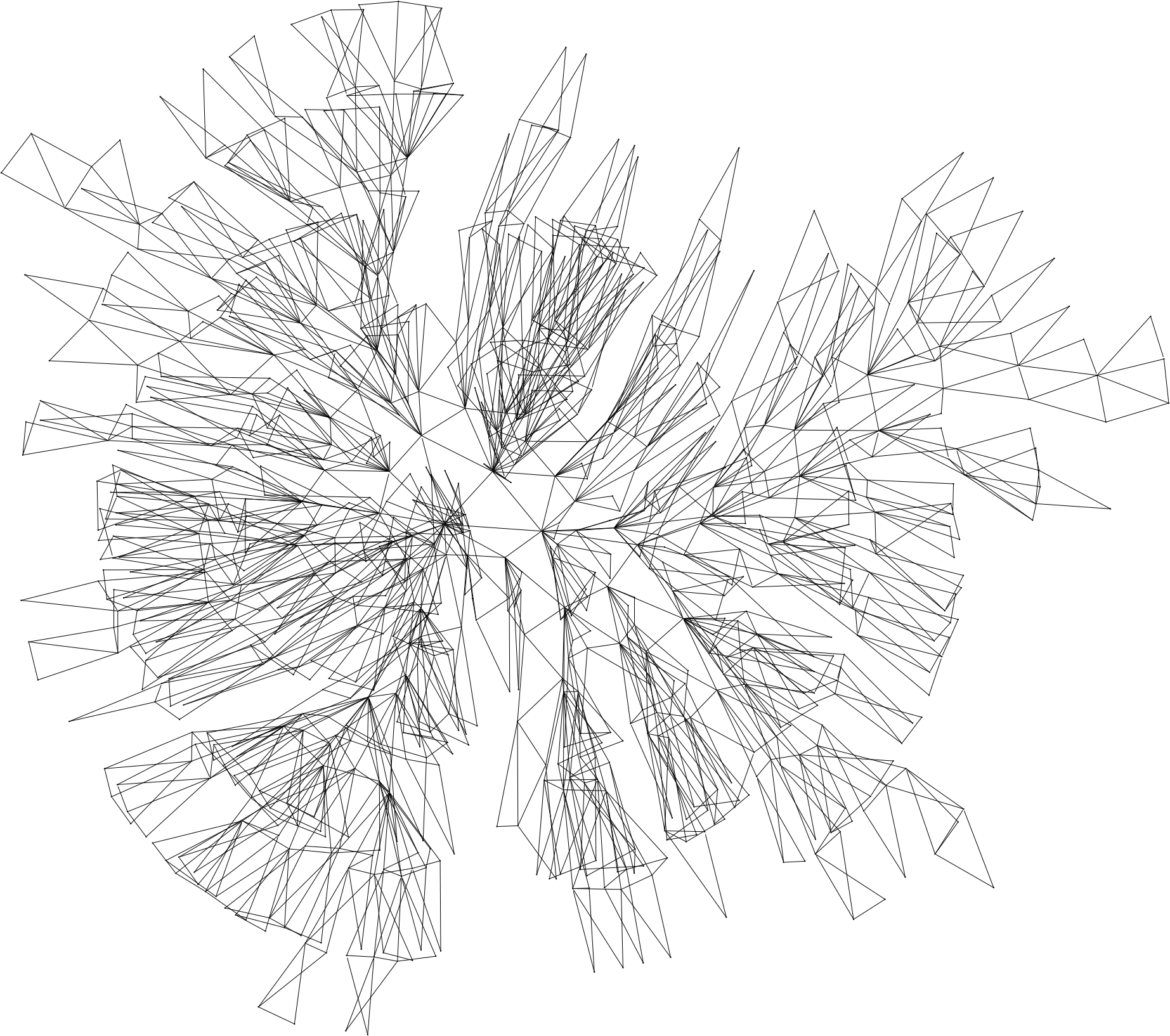} &
\includegraphics[width=0.16\textwidth]{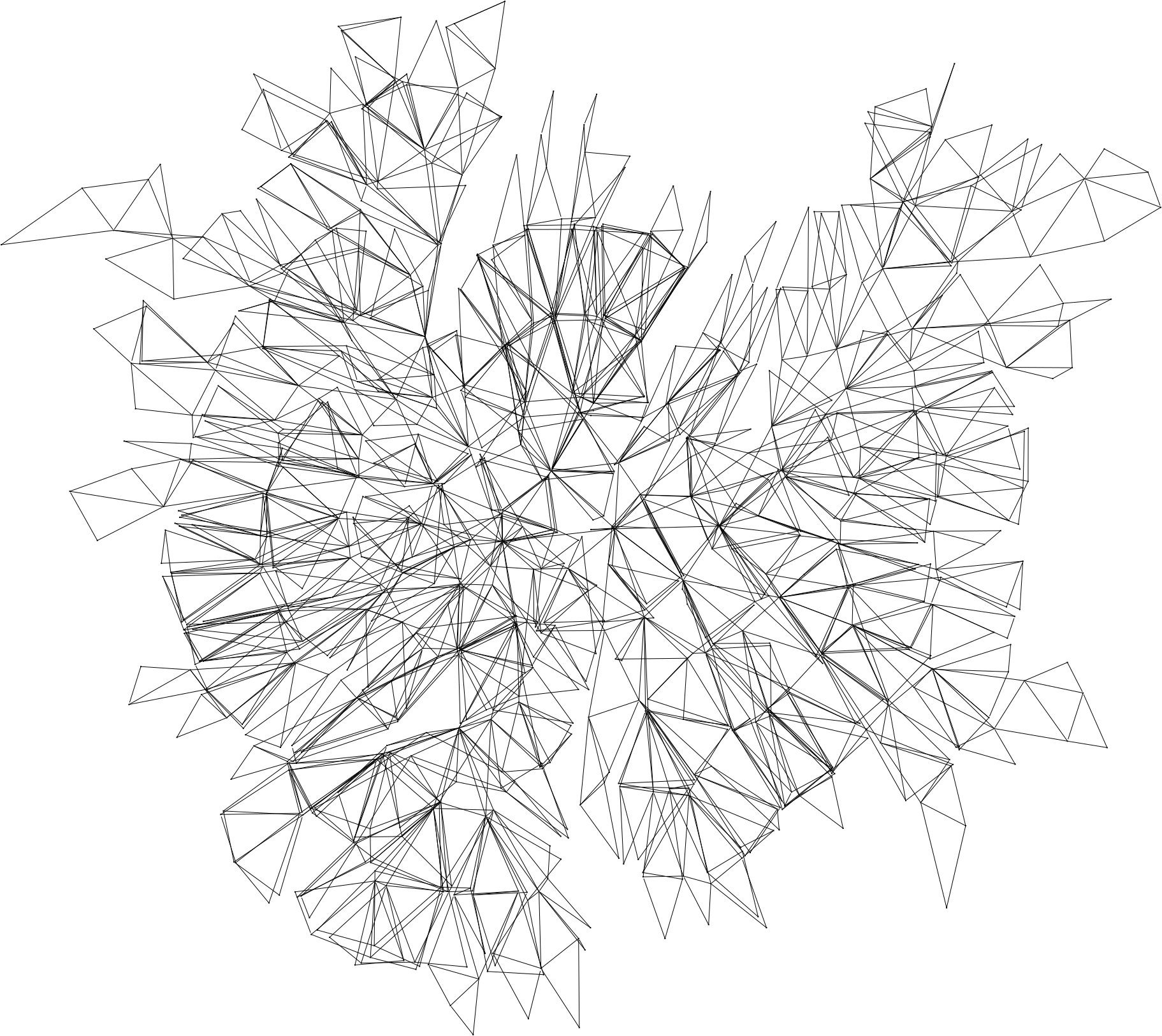} &
\includegraphics[width=0.16\textwidth]{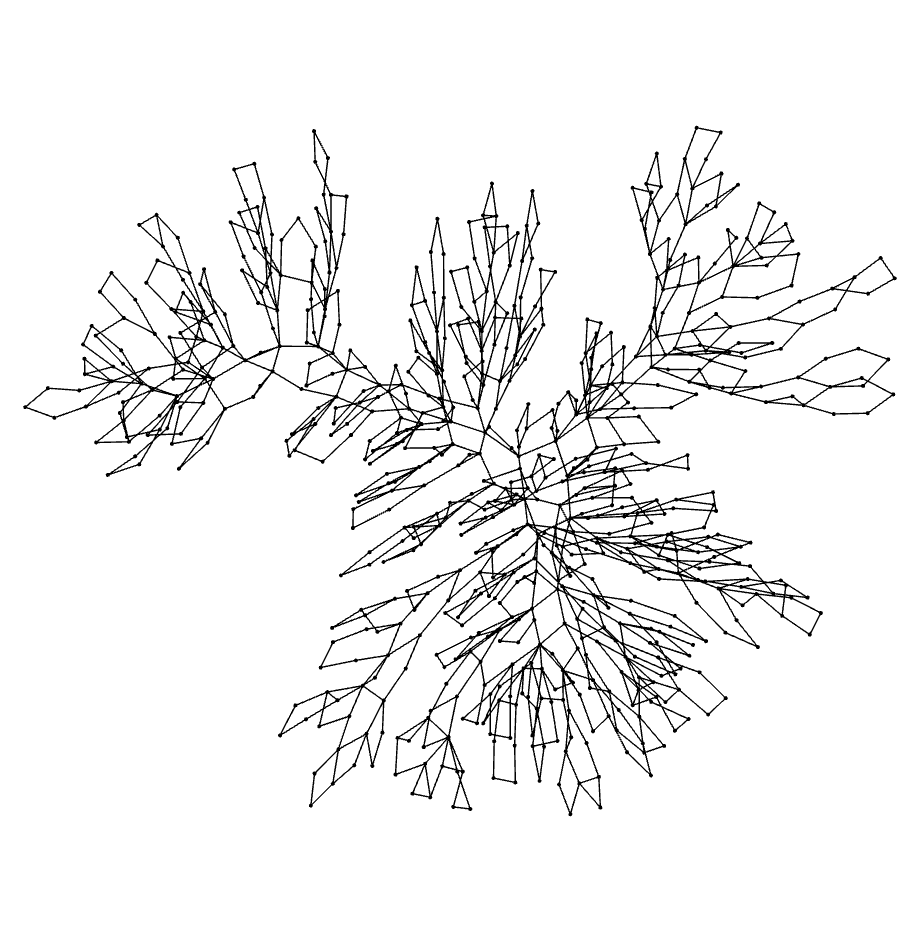} &
\includegraphics[width=0.16\textwidth]{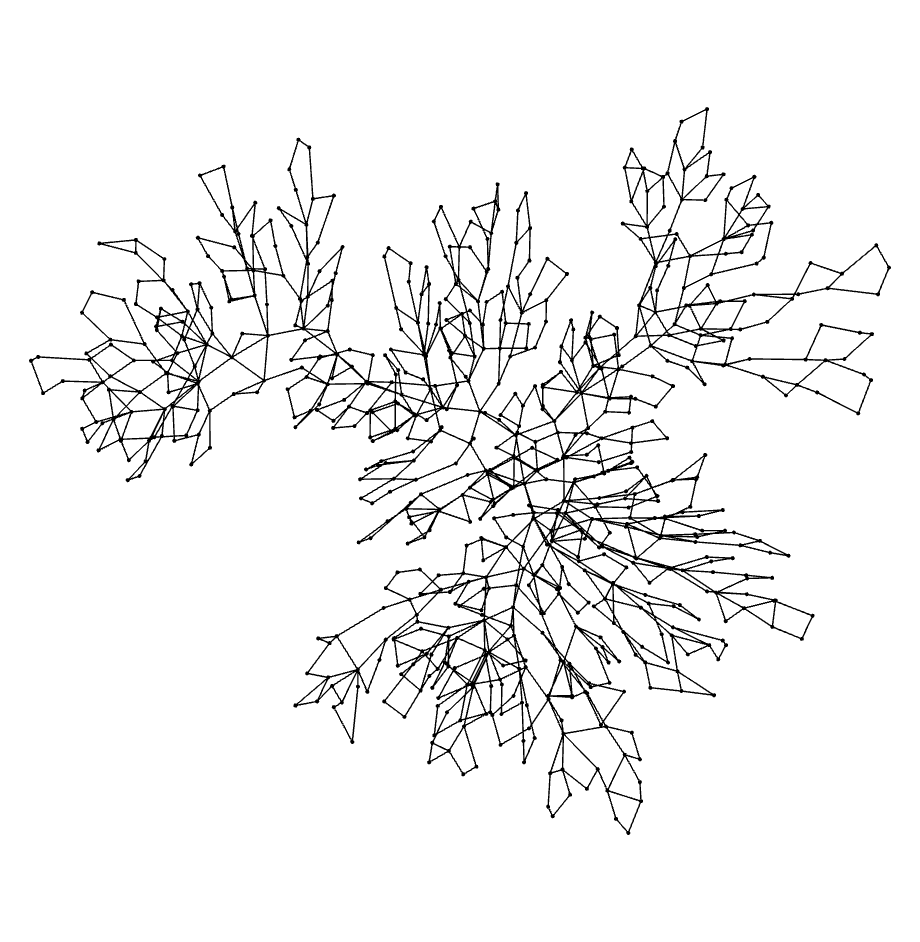} &
\includegraphics[width=0.16\textwidth]{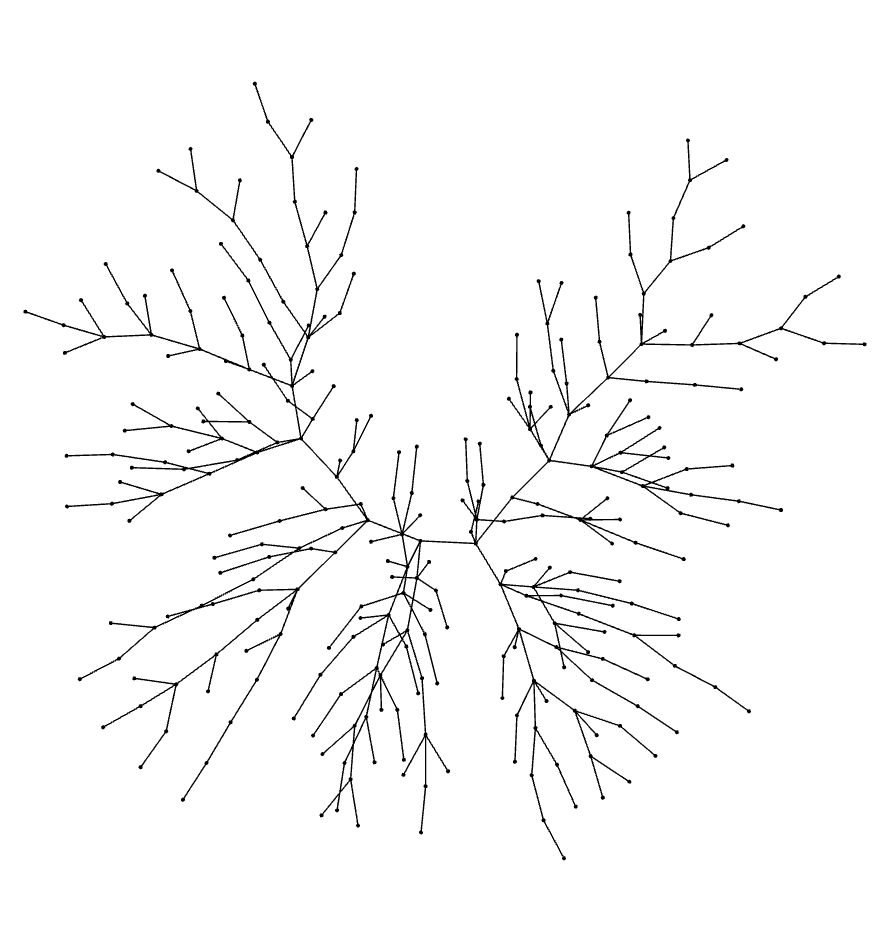} &
\includegraphics[width=0.16\textwidth]{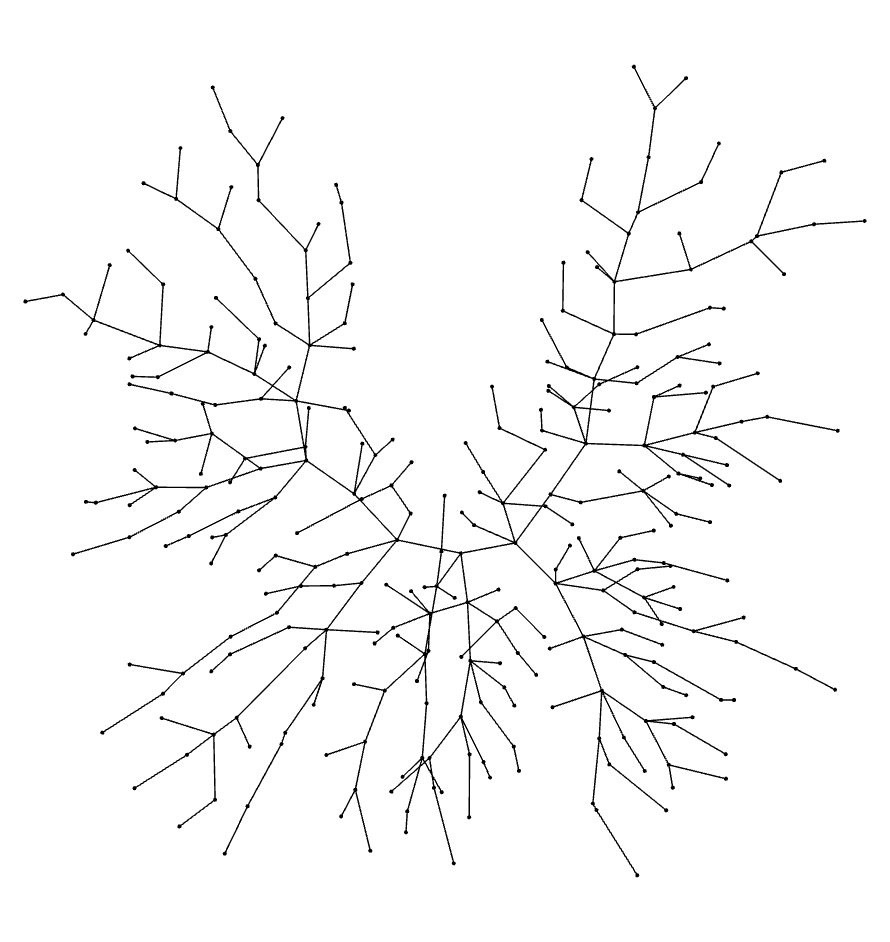} \\ \hline
\multicolumn{2}{|c|}{Scale-free ($d = 2$)} & \multicolumn{2}{|c|}{Scale-free ($d = 3$)} & \multicolumn{2}{|c|}{Scale-free ($d = 5$)} \\ \hline
\includegraphics[width=0.16\textwidth]{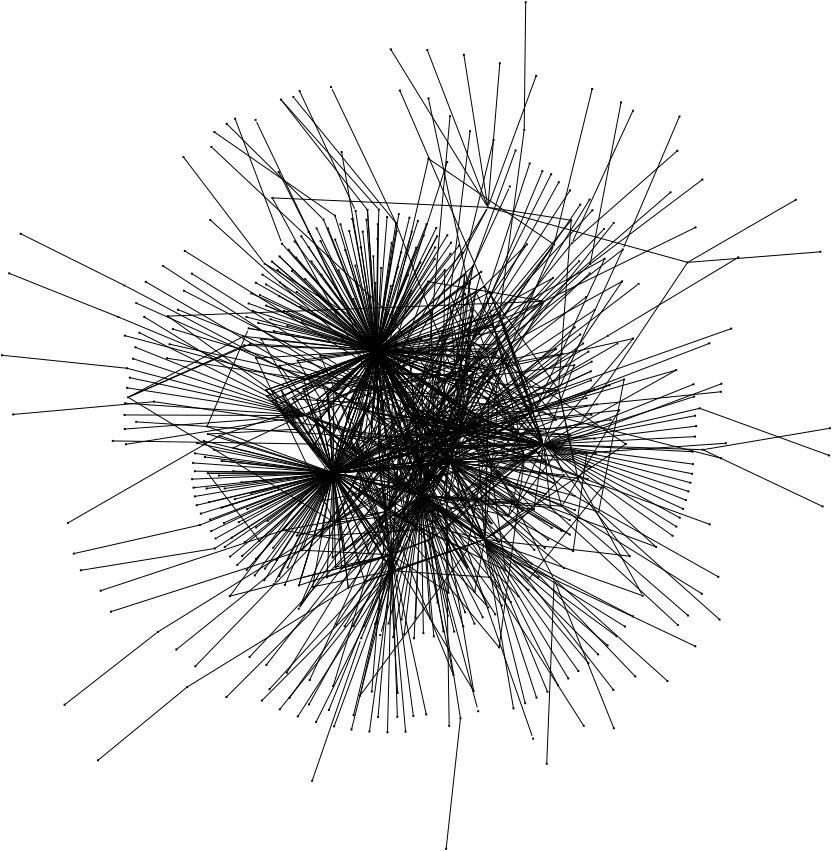} &
\includegraphics[width=0.16\textwidth]{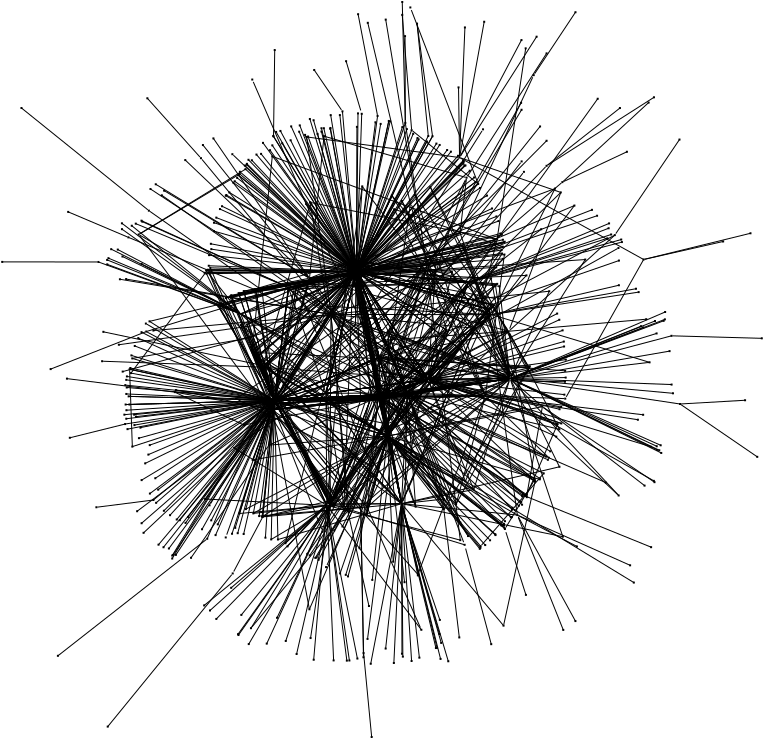} &
\includegraphics[width=0.16\textwidth]{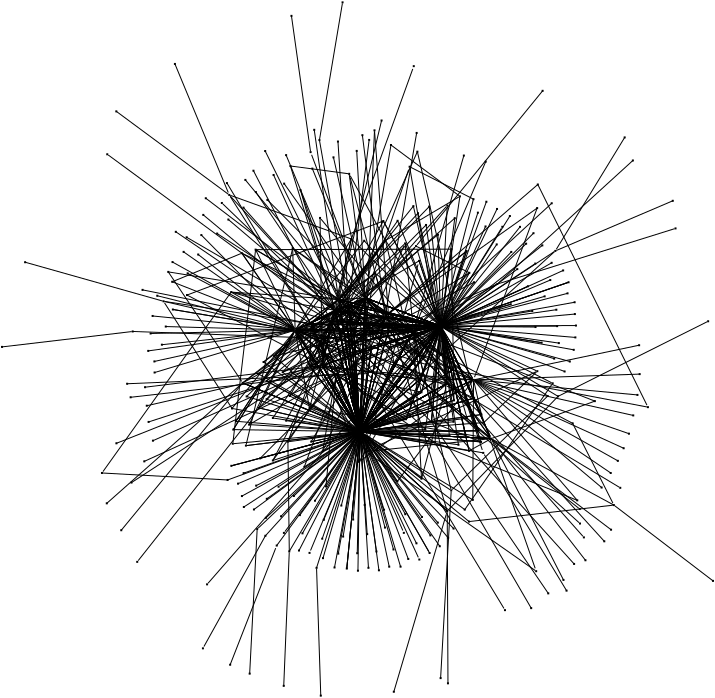} &
\includegraphics[width=0.16\textwidth]{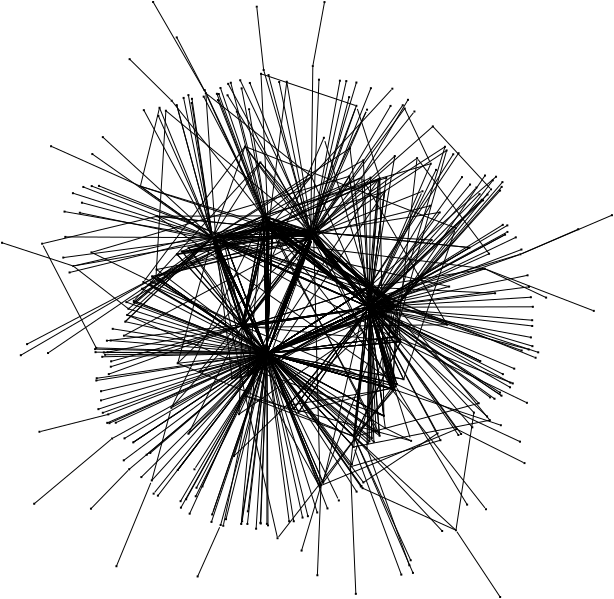} &
\includegraphics[width=0.16\textwidth]{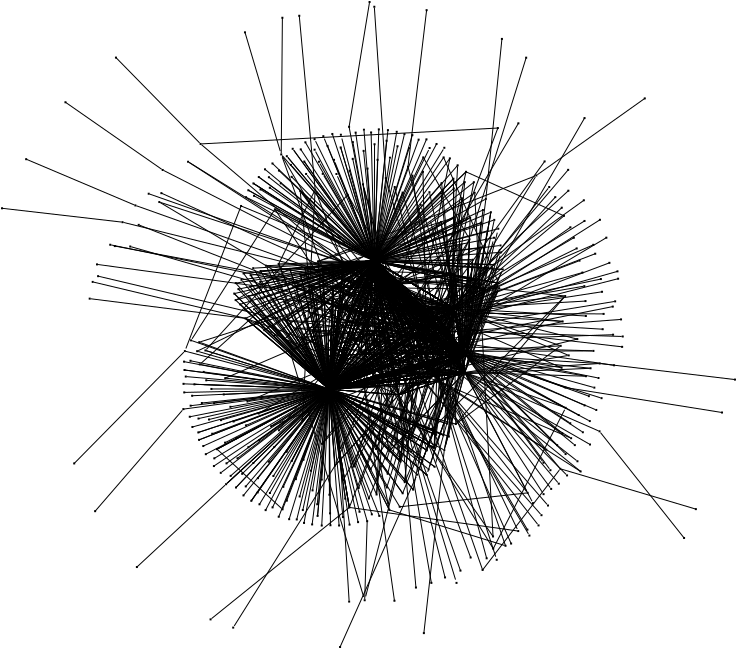} &
\includegraphics[width=0.16\textwidth]{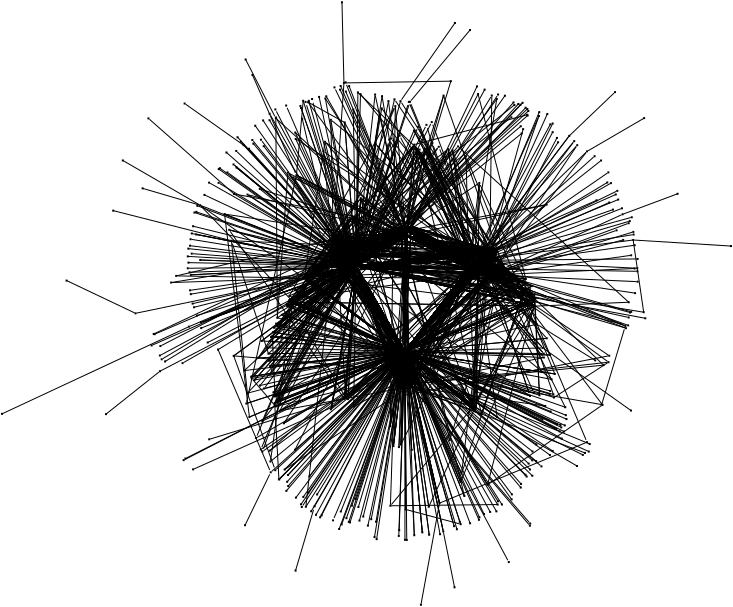} \\ \hline
\end{tabular}
\end{table}

\end{subappendices}

\end{document}